\documentclass[12pt,superscriptaddress,showpacs,nobibnotes,longbibliography]{iopart}

\usepackage[margin=1in]{geometry}
\usepackage[colorlinks,urlcolor=blue,citecolor=blue,linkcolor=blue]{hyperref}
\usepackage[pdftex]{graphicx}
\usepackage[usenames,dvipsnames]{xcolor}
\usepackage{url}
\usepackage{tabularx}
\usepackage{tabulary}
\usepackage{multirow} 
\usepackage{comment}
\usepackage{multicol}
\usepackage{graphicx}
\usepackage{mathrsfs}
\usepackage{booktabs}
\usepackage{gensymb}

\usepackage{makecell}

\expandafter\let\csname equation*\endcsname\relax
\expandafter\let\csname endequation*\endcsname\relax
\usepackage{amssymb,amsmath}

\usepackage{pdfpages}
\usepackage{abstract}
\usepackage{wrapfig}
\usepackage{fancyhdr}
\usepackage{parskip}
\usepackage{color}
\usepackage{enumitem}
\usepackage{bm}
\usepackage{lipsum}
\usepackage[sc]{mathpazo}
\usepackage{subcaption}
\usepackage{titlesec}
\usepackage{lineno}
\usepackage{nicefrac}
\usepackage[titletoc]{appendix}
\usepackage[sort&compress,numbers]{natbib}
\usepackage{doi}%<----------
\usepackage{nameref}
\usepackage{comment}
\usepackage[draft, inline, nomargin]{fixme}

\usepackage[normalem]{ulem}

\fxsetup{theme=color, mode=multiuser}
\FXRegisterAuthor{rc}{1}{\color{blue}RC}
\FXRegisterAuthor{bl}{2}{\color{red}BL}
\FXRegisterAuthor{pm}{3}{\color{purple}PM}

\newcommand{\nuebar}{\ensuremath{\overline{\nu}_{e}} }
\newcommand{\uFive}{$^{235}$U}
\newcommand{\uEight}{$^{238}$U}
\newcommand{\pNine}{$^{239}$Pu}
\newcommand{\pOne}{$^{241}$Pu}
\newcommand{\cevns}{CE$\nu$NS }

\begin{document}

\title[]{HEP Physics Opportunities Using Reactor Antineutrinos: A Snowmass 2021 White Paper Submission}
\vspace{20pt}
%\input{AuthorListJun2021}
%\pagebreak

\begin{abstract}
Nuclear reactors are uniquely powerful, abundant, and flavor-pure sources of antineutrinos that continue to play a vital role in the US neutrino physics program.  
The US reactor antineutrino physics community is a diverse interest group encompassing many detection technologies and many particle physics topics, including Standard Model and short-baseline oscillations, BSM physics searches, and reactor flux and spectrum modelling.  
The community's aims offer strong complimentary with numerous aspects of the wider US neutrino program and have direct relevance to most of the topical sub-groups composing the Snowmass 2021 Neutrino Frontier.  
Reactor neutrino experiments also have a direct societal impact and have become a strong workforce and technology development pipeline for DOE National Laboratories and universities.  

This white paper, prepared as a submission to the Snowmass 2021 community organizing exercise, will survey the state of the reactor antineutrino physics field and summarize the ways in which current and future reactor antineutrino experiments can play a critical role in advancing the field of particle physics in the next decade.  
As it is directed towards the Snowmass 2021 Neutrino Frontier, Sections~\ref{sec:lbl} through~\ref{sec:applications} are organized around specific Topical Groups within that Frontier, with the relevant Topical Group specified in each Section's title.  
Finally, to enable quick reference to the document's main themes, two to four `Key Takeaways' are provided at the beginning of each Section.  
\end{abstract}

\clearpage



O. A. Akindele,[1] J. M. Berryman,[2,3] N. S. Bowden,[1] R. Carr,[4] A. J. Conant,[5] P. Huber,[6]
T. J. Langford,[7] J.M. Link,[6] B. R. Littlejohn,[8] G. Fernandez-Moroni,[9,10] J. P. Ochoa-Ricoux,[11]
C. Roca,[1] S. Schoppmann,[12,3] L. Strigari,[13] J. Xu,[1] C. Zhang,[14] and X. Zhang[1]
(White Paper Editors)

C. Awe,[15,16] P.S. Barbeau,[15,16] A. Haghighat,[6,17] S. C. Li,[6]
J.M. Link,[6] V. Mascolino,[17] T. Subedi,[6] and K. Walkup[6]
(The CHANDLER Collaboration)

P. S. Barbeau,[15,16] E. P. Bernard,[1] N. S. Bowden,[1] I. Jovanovic,[18] J. W. Kingston,[19,1] E. Mizrachi,[20,1]
E. Pantic,[19] T. J. Pershing,[1] S. V. Pereverzev,[1] C. G. Prior,[15,16] R. Smith,[21] J. Xu,[1] and A. Bernstein[1]
(The CHILLAX development team)

A. Aguilar-Arevalo,[22] N. Avalos,[23] J. Bernal,[24] X. Bertou,[23] C. Bonifazi,[25,26] G. Cancelo,[9]
V. G. P. B. de Carvalho,[25] B. A. Cervantes-Vergara,[22] C. Chavez,[24] J. C. D'Olivo,[22] J. M. De
Egea,[24] J. C. dos Anjos,[27] J. Estrada,[9] A. R. Fernandez Neto,[28] G. Fernandez-Moroni,[9,10]
A. Foguel,[25] R. Ford,[9] J. Gasanego,[29] J. Gonzalez-Cuevas,[24] A. Haghighat,[6,17] S. Hernandez,[9]
F. Izraelevitch,[29] B. Kilminster,[30] K. Kuk,[9] P. Lemos,[25] H. P. Lima Jr,[27] M. Makler,[27,26]
F. Marinho,[31] K. Maslova,[25] L. H. Mendes,[25] J. Molina,[24] M. M. Montero,[22] P. Mota,[27]
I. Nasteva,[25] A.C. Oliveira,[25] E. Paolini,[10] L. Paulucci,[32] D. Rodrigues,[33] C. Romero,[24]
Y. Sarkis,[22] M. S. Haro,[23] A. Soto,[10] D. Stalder,[24] J. Tiffenberg,[9] C. Torres,[24] and P. Z. Ventura[25]
(The CONNIE Collaboration)

A. Bonhomme,[34] C. Buck,[34] J. Hakenmüller,[34] M. Lindner,[34]
W. Maneschg,[34] E. Sanchez Garcia,[34] and T. Rink[34]
(on behalf of The CONUS Collaboration)

F. P. An,[35] W. D. Bai,[36] A. B. Balantekin,[37] M. Bishai,[14] S. Blyth,[38] G. F. Cao,[39] J. Cao,[39] J. F. Chang,[39]
Y. Chang,[40] H. S. Chen,[39] H. Y. Chen,[41] S. M. Chen,[41] Y. Chen,[42,36] Y. X. Chen,[43] J. Cheng,[39]
Z. K. Cheng,[36] J. J. Cherwinka,[37] M. C. Chu,[44] J. P. Cummings,[45] O. Dalager,[11] F. S. Deng,[46] Y. Y. Ding,[39]
M. V. Diwan,[14] T. Dohnal,[47] D. Dolzhikov,[48] J. Dove,[49] D. A. Dwyer,[50] J. P. Gallo,[8] M. Gonchar,[48]
G. H. Gong,[41] H. Gong,[41] W. Q. Gu,[14] J. Y. Guo,[36] L. Guo,[41] L. Guo,[41] X. H. Guo,[51] Y. H. Guo,[52] Z. Guo,[41]
R. W. Hackenburg,[14] A. Haghighat,[6,17] S. Hans,[14,∗] M. He,[39] K. M. Heeger,[7] Y. K. Heng,[39] Y. K. Hor,[36]
Y. B. Hsiung,[38] B. Z. Hu,[38] J. R. Hu,[39] T. Hu,[39] Z. J. Hu,[36] H. X. Huang,[53] J. H. Huang,[39] X. T. Huang,[54]
Y. B. Huang,[55] P. Huber,[6] D. E. Jaffe,[14] K. L. Jen,[56] X. L. Ji,[39] X. P. Ji,[14] D. Jones,[57] L. Kang,[58]
S. H. Kettell,[14] S. Kohn,[21] M. Kramer,[50,21] T. J. Langford,[7] J. Lee,[50] J. H. C. Lee,[59] R. T. Lei,[58]
R. Leitner,[47] J. K. C. Leung,[59] F. Li,[39] H. L. Li,[39] J. J. Li,[41] Q. J. Li,[39] R. H. Li,[39] S. Li,[58] S. C. Li,[6]
W. D. Li,[39] X. N. Li,[39] X. Q. Li,[60] Y. F. Li,[39] Z. B. Li,[36] H. Liang,[46] C. J. Lin,[50] G. L. Lin,[56] S. Lin,[58]
J. J. Ling,[36] J. M. Link,[6] L. Littenberg,[14] B. R. Littlejohn,[8] J. C. Liu,[39] J. L. Liu,[14] J. X. Liu,[39] C. Lu,[62]
H. Q. Lu,[39] K. B. Luk,[21,50] B. Z. Ma,[54] X. B. Ma,[43] X. Y. Ma,[39] Y. Q. Ma,[39] R. C. Mandujano,[11]
C. Marshall,[50] K. T. McDonald,[62] R. D. McKeown,[63,64] Y. Meng,[61] J. Napolitano,[57] D. Naumov,[48]
E. Naumova,[48] T. M. T. Nguyen,[56] J. P. Ochoa-Ricoux,[11] A. Olshevskiy,[48] H.-R. Pan,[38] J. Park,[6]
S. Patton,[50] J. C. Peng,[49] C. S. J. Pun,[59] F. Z. Qi,[39] M. Qi,[65] X. Qian,[14] N. Raper,[36] J. Ren,[53]
C. Morales Reveco,[11] R. Rosero,[14] B. Roskovec,[11] X. C. Ruan,[53] J. L. Sun,[66] T. Tmej,[47] K. Treskov,[48]
W.-H. Tse,[44] C. E. Tull,[50] B. Viren,[14] V. Vorobel,[47] C. H. Wang,[40] J. Wang,[36] M. Wang,[54]
N. Y. Wang,[51] R. G. Wang,[39] W. Wang,[36,64] X. Wang,[67] Y. Wang,[65] Y. F. Wang,[39] Z. Wang,[39]
Z. Wang,[41] Z. M. Wang,[39] H. Y. Wei,[14] L. H. Wei,[39] L. J. Wen,[39] K. Whisnant,[68] C. G. White,[8]
H. L. H. Wong,[21,50] E. Worcester,[14] D. R. Wu,[39] Q. Wu,[54] W. J. Wu,[39] D. M. Xia,[69] Z. Q. Xie,[39]

Z. Z. Xing,[39] H. K. Xu,[39] J. L. Xu,[39] T. Xu,[41] T. Xue,[41] C. G. Yang,[39] L. Yang,[58] Y. Z. Yang,[41] H. F. Yao,[39] M. Ye,[39] M. Yeh,[14] B. L. Young,[68] H. Z. Yu,[36] Z. Y. Yu,[39] B. B. Yue,[36] V. Zavadskyi,[48] S. Zeng,[39] Y. Zeng,[36] L. Zhan,[39] C. Zhang,[14] F. Y. Zhang,[61] H. H. Zhang,[36] J. L. Zhang,[65] J. W. Zhang,[39] Q. M. Zhang,[52] S. Q. Zhang,[36] X. T. Zhang,[39] Y. M. Zhang,[36] Y. X. Zhang,[66] Y. Y. Zhang,[61] Z. J. Zhang,[58] Z. P. Zhang,[46] Z. Y. Zhang,[39] J. Zhao,[39] R. Z. Zhao,[39] L. Zhou,[39] H. L. Zhuang,[39] and J. H. Zou[39]

(The Daya Bay Collaboration)

T. Abrahão,[70, 71] H. Almazan,[72] J.C. dos Anjos,[70] S. Appel,[73] J.C. Barriere,[74] I. Bekman,[75] T.J.C. Bezerra,[76] L. Bezrukov,[77] E. Blucher,[78] T. Brugière,[79] C. Buck,[72] J. Busenitz,[80] A. Cabrera,[71, 81, 82] M. Cerrada,[83] E. Chauveau,[84] P. Chimenti,[70] O. Corpace,[74] J.V. Dawson,[71] Z. Djurcic,[85] A. Etenko,[86] H. Furuta,[87] I. Gil-Botella,[83] A. Givaudan,[71] H. Gomez,[71, 74] L.F.G. Gonzalez,[88] M.C. Goodman,[85] T. Hara,[89] J. Haser,[72] D. Hellwig,[75] A. Hourlier,[71] M. Ishitsuka,[90] J. Jochum,[91] C. Jollet,[84] K. Kale,[84, 79] M. Kaneda,[90] M. Karakac,[71] T. Kawasaki,[90] E. Kemp,[88] D. Kryn,[71] M. Kuze,[90] T. Lachenmaier,[91] C.E. Lane,[93] T. Lasserre,[74, 71] C. Lastoria,[83] D. Lhuillier,[74] H.P. Lima Jr,[70] M. Lindner,[72] J.M. López-Castaño,[83] J.M. LoSecco,[94] B. Lubsandorzhiev,[77] J. Maeda,[95, 89] C. Mariani,[96] J. Maricic,[93] J. Martino,[76] T. Matsubara,[95] G. Mention,[74] A. Meregaglia,[84] T. Miletic,[93] R. Milincic,[93] A. Minotti,[74] D. Navas-Nicolás,[71, 83, 82] P. Novella,[83] L. Oberauer,[73] M. Obolensky,[71] A. Onillon,[74] A. Oralbaev,[86] C. Palomares,[83] I.M. Pepe,[70] G. Pronost,[76] J. Reichenbacher,[80] B. Reinhold,[72] S. Schönert,[73] S. Schoppmann,[72] L. Scola,[74] R. Sharankova,[90] V. Sibille,[74] V. Sinev,[77] M. Skorokhvatov,[86] P. Soldin,[75] A. Stahl,[75] I. Stancu,[80] L.F.F. Stokes,[91] F. Suekane,[87, 71] S. Sukhotin,[86] T. Sumiyoshi,[95] Y. Sun,[80] C. Veyssiere,[74] B. Viaud,[76] M. Vivier,[74] S. Wagner,[71, 70] C. Wiebusch,[75] G. Yang,[85] and F. Yermia[76]

(The Double Chooz Collaboration)

T. Anderson,[97] E. Anderssen,[12] M. Askins,[12, 3] A. J. Bacon,[98] Z. Bagdasarian,[12, 3] A. Baldoni,[97] N. Barros,[99, 100] L. Bartoszek,[101] A. Bat,[102] M. Bergevin,[1] A. Bernstein,[1] E. Blucher,[78] J. Boissevain,[101] R. Bonventre,[12] D. Brown,[12] E. J. Callaghan,[12, 3] D. F. Cowen,[97] S. Dazeley,[1] M. Diwan,[14] K. Frankiewicz,[103] C. Grant,[103] T. Kaptanoglu,[12, 3] J.R. Klein,[98] C. Kraus,[104, 105] T. Kroupa,[98] B. Land,[12, 3, 98] L. Lebanowski,[98] V. Lozza,[99, 100] A. Marino,[106] A. Mastbaum,[107] C. Mauger,[98] S. Naugle,[98] M. Newcomer,[98] A. Nikolica,[98] G. D. Orebi Gann,[12, 3] L. Pickard,[108] J. Saba,[12] S. Schoppmann,[12, 3] J. Sensenig,[98] M. Smiley,[12, 3] H. Steiger,[109, 110] R. Svoboda,[108] E. Tiras,[102, 111] W. H. Trzaska,[112] R. van Berg,[98, 101] G. Wendel,[97] M. Wetstein,[113] M. Wurm,[109] M. Yeh,[14] and E.D. Zimmerman[106]

(The EOS team)

A. Abusleme,[114] T. Adam,[115] S. Ahmad,[116] R. Ahmed,[116] S. Aiello,[117] M. Akram,[116] A. Aleem,[116] T. Alexandros,[118] F. P. An,[119] Q. An,[46] G. Andronico,[117] N. Anfimov,[48] V. Antonelli,[120] T. Antoshkina,[48] B. Asavapibhop,[121] J. P. A. M. de André,[115] D. Auguste,[122] W. Bai,[36] N. Balashov,[48] W. Baldini,[123] A. Barresi,[124] D. Basilico,[120] E. Baussan,[115] M. Bellato,[125] A. Bergnoli,[125] T. Birkenfeld,[118] S. Blin,[122] D. Blum,[126] S. Blyth,[38] A. Bolshakova,[48] M. Bongrand,[122] C. Bordereau,[127, 40] D. Breton,[122] A. Brigatti,[120] R. Brugnera,[128] R. Bruno,[117] A. Budano,[129] Jose Busto,[130] A. Cabrera,[122] B. Caccianiga,[120] H. Cai,[131] X. Cai,[39] Y. K. Cai,[39] Z. Y. Cai,[39] R. Callegari,[128] A. Cammi,[132] A. Campeny,[114] C. Y. Cao,[39] G. F. Cao,[39] J. Cao,[39] R. Caruso,[117] C. Cerna,[127] C. Chan,[56] J. F. Chang,[39] Y. Chang,[40] P. P. Chen,[133] P. A. Chen,[38] S. Chen,[134] X. Chen,[135] Y. Chen,[43] Y. Chen,[36] Z. Chen,[39] Z. Chen,[36] J. Cheng,[43] Y. Cheng,[136] Y. Z. Cheng,[38] A. Chetverikov,[48] D. Chiesa,[124] P. Chimenti,[137] Z. Chu,[39] A. Chukanov,[48] G. Claverie,[127] C. Clementi,[138] B. Clerbaux,[139] S. Di Lorenzo,[122] D. Corti,[125] F. D. Corso,[125] O. Dalager,[11] C. De La Taille,[122] Z. Deng,[134] Z. Y. Deng,[39] W. Depnering,[140] M. Diaz,[114] X. F. Ding,[120] Y. Y. Ding,[39] B. Dirgantara,[141] S. Dmitrievsky,[48] T. Dohnal,[47] D. Dolzhikov,[48] G. Donchenko,[142] J. M. Dong,[134] E. Doroshkevich,[143] W. Dou,[134] M. Dracos,[115] F. Druillole,[127] R. Du,[39] S. X. Du,[144]

S. Dusini,[125] M. Dvorak,[47] T. Enqvist,[145] A. Fabbri,[129] U. Fahrendholz,[146] D. H. Fan,[147] L. Fan,[39]
J. Fang,[39] W. X. Fang,[39] M. Fargetta,[117] D. Fedoseev,[48] Z. Fei,[39] L. C. Feng,[56] Q. C. Feng,[148]
F. Ferraro,[120] R. Ford,[120] A. Fournier,[127] H. N. Gan,[149] F. Gao,[118] A. Garfagnini,[128] A. Gavrikov,[48]
M. Giammarchi,[120] N. Giudice,[117] M. Gonchar,[48] G. Gong,[134] H. Gong,[134] Y. Gornushkin,[48]
A. Göttel,[150, 118] M. Grassi,[128] V. Gromov,[48] M. Gu,[39] X. Gu,[144] Y. Gu,[151] M. Y. Guan,[39] Y. D. Guan,[39]
N. Guardone,[117] C. Guo,[39] J. Y. Guo,[36] W. L. Guo,[39] X. H. Guo,[51] Y. H. Guo,[150] C. Hagner,[152] R. Han,[36]
Y. Han,[36] M. He,[39] W. He,[39] T. Heinz,[126] P. Hellmuth,[127] Y. K. Heng,[39] R. Herrera,[114] Y. K. Hor,[36]
S. J. Hou,[39] Y. Hsiung,[38] B. Z. Hu,[38] H. Hu,[36] J. R. Hu,[39] J. Hu,[39] S. Y. Hu,[53] T. Hu,[39] Y. X. Hu,[39]
Z. J. Hu,[36] G. H. Huang,[39] H. X. Huang,[53] K. X. Huang,[36] W. H. Huang,[54] X. Huang,[39] X. T. Huang,[54]
Y. B. Huang,[55] J. Q. Hui,[153] L. Huo,[148] W. Huo,[46] C. Huss,[127] S. Hussain,[116] A. Ioannisian,[154]
R. Isocrate,[125] B. Jelmini,[128] I. Jeria,[114] X. L. Ji,[39] H. H. Jia,[60] J. J. Jia,[131] S. Y. Jian,[53] D. Jiang,[46]
W. Jiang,[39] X. S. Jiang,[39] X. P. Jing,[39] C. Jollet,[127] J. Joutsenvaara,[145] L. Kalousis,[115] P. Kampmann,[150, 118]
L. Kang,[133] R. Karaparambil,[155] N. Kazarian,[154] A. Khatun,[156] K. Khosonthongkee,[141] D. Korablev,[48]
K. Kouzakov,[142] A. Krasnoperov,[48] N. Kutovskiy,[48] P. Kuusiniemi,[145] T. Lachenmaier,[126] C. Landini,[120]
S. Leblanc,[127] V. Lebrin,[155] F. Lefevre,[155] R. Lei,[133] R. Leitner,[47] J. Leung,[56] D. Z. Li,[39] D. Li,[144] F. Li,[39]
F. Li,[134] H. Li,[36] G. S. Li,[39] H. L. Li,[39] M. Z. Li,[39] M. Li,[43] N. Li,[67] N. Li,[39] Q. J. Li,[67] R. H. Li,[39]
R. Li,[153] S. F. Li,[133] T. Li,[36] T. Li,[54] W. D. Li,[39] W. G. Li,[39] X. M. Li,[53] X. N. Li,[39] X. L. Li,[53]
Y. Li,[133] Y. C. Li,[39] Y. F. Li,[39] Z. Li,[39] Z. H. Li,[39] Z. B. Li,[36] Z. Y. Li,[36] Z. H. Li,[131] H. Liang,[46]
H. Liang,[53] J. J. Liao,[36] A. Limphirat,[141] G. L. Lin,[56] S. X. Lin,[133] T. Lin,[39] J. J. Ling,[36] I. Lippi,[125]
F. Liu,[43] H. D. Liu,[144] H. T. Liu,[131] H. B. Liu,[55] H. J. Liu,[157] H. T. Liu,[36] H. Liu,[151] J. L. Liu,[153, 158]
J. C. Liu,[39] M. Liu,[157] Q. Liu,[159] Q. Liu,[46] R. X. Liu,[150, 118] S. B. Liu,[46] S. L. Liu,[39] X. W. Liu,[36]
X. Liu,[55] Y. Liu,[39] Y. Liu,[39] A. Lokhov,[142] P. Lombardi,[120] C. Lombardo,[160] K. Loo,[140] C. Lu,[161]
H. Q. Lu,[39] J. B. Lu,[162] J. G. Lu,[39] P. Z. Lu,[36] S. X. Lu,[144] B. Lubsandorzhiev,[143] S. Lubsandorzhiev,[143]
L. Ludhova,[150, 118] A. Lukanov,[143] D. B. Luo,[39] F. J. Luo,[39] G. Luo,[36] S. Luo,[163] W. M. Luo,[39]
X. J. Luo,[39] V. Lyashuk,[143] B. Z. Ma,[54] B. Ma,[144] Q. M. Ma,[39] S. Ma,[39] X. Y. Ma,[39] X. B. Ma,[43]
J. Maalmi,[122] M. Magoni,[120] J. Y. Mai,[36] Y. Malyshkin,[48] R. C. Mandujano,[11] F. Mantovani,[123]
X. Mao,[136] Y. J. Mao,[164] S. M. Mari,[129] F. Marini,[128] C. Martellini,[129] G. Martin-Chassard,[122]
A. Martini,[165] M. Mayer,[146] D. Mayilyan,[154] I. Mednieks,[166] A. Meinusch,[140] Y. Meng,[153]
A. Meregaglia,[127] E. Meroni,[120] D. Meyhöfer,[152] M. Mezzetto,[125] J. Miller,[149] L. Miramonti,[120]
M. C. Molla,[139] P. Montini,[129] M. Montuschi,[123] A. Müller,[126] M. Nastasi,[124] D. V. Naumov,[48]
E. Naumova,[48] D. Navas-Nicolas,[122] I. Nemchenok,[48] T. M. T. Nguyen,[56] F. P. Ning,[39] Z. Ning,[39]
H. Nunokawa,[167] L. Oberauer,[146] J. P. Ochoa-Ricoux,[11, 114] A. Olshevskiy,[48] D. Orestano,[129] F. Ortica,[138]
R. Othegraven,[140] A. Paoloni,[165] S. Parmeggiano,[120] Y. Pei,[39] L. Pelicci,[150, 118] N. Pelliccia,[138] A. Peng,[157]
H. Peng,[46] Y. Peng,[39] Z. Y. Peng,[39] F. Perrot,[127] P. A. Petitjean,[139] F. Petrucci,[129] O. Pilarczyk,[140]
L. F. Piñeres Rico,[115] A. Popov,[142] P. Poussot,[115] E. Previtali,[124] F. Qi,[39] M. Qi,[65] S. Qian,[39] X. H. Qian,[39]
Z. Qian,[39] H. Qiao,[164] Z. H. Qin,[39] S. K. Qiu,[157] G. Ranucci,[120] N. Raper,[36] R. Rasheed,[127] A. Re,[120]
H. Rebber,[152] A. Rebii,[127] M. Redchuk,[125, 128] B. Ren,[133] J. Ren,[53] B. Ricci,[123] M. Rifai,[150, 118]
M. Roche,[127] N. Rodphai,[121] A. Romani,[138] B. Roskovec,[47] X. Ruan,[53] S. Rujirawat,[141] A. Rybnikov,[48]
A. Sadovsky,[48] P. Saggese,[120] G. Salamanna,[129] S. Sanfilippo,[129] A. Sangka,[168] U. Sawangwit,[168]
J. Sawatzki,[146] M. Schever,[150, 118] C. Schwab,[115] K. Schweizer,[146] A. Selyunin,[48] A. Serafini,[123]
G. Settanta,[150] M. Settimo,[155] Z. Shao,[169] V. Sharov,[48] A. Shaydurova,[48] J. Shi,[39] Y. Shi,[39] V. Shutov,[48]
A. Sidorenkov,[143] F. Šimkovic,[156] C. Sirignano,[128] J. Siripak,[141] M. Sisti,[124] M. Slupecki,[145]
M. Smirnov,[36] O. Smirnov,[48] T. Sogo-Bezerra,[155] J. Songwadhana,[141] B. Soonthornthum,[168]
A. Sotnikov,[48] O. Sramek,[47] W. Sreethawong,[141] A. Stahl,[118] L. Stanco,[125] K. Stankevich,[142]
D. Stefanik,[156] H. Steiger,[146] J. Steinmann,[118] T. Sterr,[126] M. R. Stock,[146] V. Strati,[123] A. Studenikin,[142]
J. Su,[36] S. F. Sun,[43] X. L. Sun,[39] Y. J. Sun,[46] Y. Z. Sun,[39] N. Suwonjandee,[121] M. Szelezniak,[115] J. Tang,[36]
Q. Tang,[36] Q. Tang,[157] X. Tang,[39] V. T. Hariharan,[152] E. Theisen,[140] A. Tietzsch,[126] I. Tkachev,[143]

T. Tmej,[47] M. D. C. Torri,[120] K. Treskov,[48] A. Triossi,[115] G. Troni,[114] W. Trzaska,[145] C. Tuve,[117] N. Ushakov,[143] V. Vedin,[166] G. Verde,[117] M. Vialkov,[142] B. Viaud,[155] C. M. Vollbrecht,[118,150] C. Volpe,[122] K. Von Sturm,[128] V. Vorobel,[47] D. Voronin,[143] L. Votano,[165] P. Walker,[114] C. Wang,[133] C. H. Wang,[40] E. Wang,[144] G. Wang,[148] J. Wang,[46] J. Wang,[36] L. Wang,[39] M. F. Wang,[39] M. Wang,[54] M. Wang,[157] R. G. Wang,[39] S. G. Wang,[164] W. Wang,[65] W. Wang,[36] W. S. Wang,[39] X. Wang,[167] X. Y. Wang,[36] Y. F. Wang,[39] W. Wang,[131] Y. Wang,[134] Y. Wang,[147] Y. Wang,[39] Y. F. Wang,[39] Y. Q. Wang,[134] Y. Wang,[65] Z. Wang,[134] Z. Wang,[39] Z. M. Wang,[39] Z. Y. Wang,[134] A. Watcharangkool,[168] W. Wei,[39] W. Wei,[144] W. L. Wei,[39] Y. D. Wei,[133] K. Wen,[39] L. J. Wen,[39] J. Weng,[134] C. Wiebusch,[118] S. C. F. Wong,[36] B. Wonsak,[152] D. Wu,[39] Q. Wu,[54] Z. Wu,[39] M. Wurm,[140] J. Wurtz,[115] C. Wysotzki,[118] Y. F. Xi,[161] D. M. Xia,[69] X. Xiao,[36] X. C. Xie,[55] Y. G. Xie,[39] Z. Q. Xie,[39] Z. Xin,[39] Z. Z. Xing,[39] B. Xu,[134] C. Xu,[157] D. L. Xu,[158,153] F. R. Xu,[151] H. K. Xu,[39] J. L. Xu,[39] J. Xu,[51] M. H. Xu,[39] Y. Xu,[60] B. J. Yan,[39] Q. Yan,[159] T. Yan,[141] W. Q. Yan,[39] X. B. Yan,[39] Y. P. Yan,[141] C. G. Yang,[39] C. F. Yang,[55] H. Yang,[39] J. Yang,[144] L. Yang,[133] X. Y. Yang,[39] Y. . Yang,[139] Y. F. Yang,[39] H. F. Yao,[39] J. X. Ye,[39] M. Ye,[39] Z. P. Ye,[158] F. Yermia,[155] Z. Y. You,[36] B. X. Yu,[39] C. Y. Yu,[133] C. X. Yu,[60] H. Z. Yu,[36] M. Yu,[131] X. H. Yu,[60] Z. Y. Yu,[39] Z. Z. Yu,[39] C. Yuan,[36] C. Z. Yuan,[39] Y. Yuan,[164] Z. X. Yuan,[134] B. B. Yue,[36] N. Zafar,[116] V. Zavadskyi,[48] S. Zeng,[39] T. X. Zeng,[39] Y. D. Zeng,[36] L. Zhan,[39] A. Q. Zhang,[134] B. Zhang,[144] B. T. Zhang,[39] F. Y. Zhang,[153] G. Q. Zhang,[39] H. H. Zhang,[36] J. L. Zhang,[65] J. W. Zhang,[39] J. Zhang,[55] J. Zhang,[39] J. B. Zhang,[148] J. Zhang,[39] M. Zhang,[39] P. Zhang,[39] Q. M. Zhang,[169] S. Zhang,[36] S. Zhang,[36] T. Zhang,[153] X. M. Zhang,[39] X. Zhang,[39] X. T. Zhang,[39] Y. Zhang,[39] Y. H. Zhang,[39] Y. Y. Zhang,[39] Y. P. Zhang,[39] Y. Y. Zhang,[153] Y. M. Zhang,[36] Z. Y. Zhang,[131] Z. J. Zhang,[133] F. Y. Zhao,[135] J. Zhao,[39] R. Zhao,[36] R. Zhao,[39] S. J. Zhao,[144] D. Q. Zheng,[151] H. Zheng,[133] Y. H. Zheng,[159] W. R. Zhong,[151] J. Zhou,[53] L. Zhou,[39] N. Zhou,[46] S. Zhou,[39] T. Zhou,[39] X. Zhou,[131] J. Zhu,[36] J. S. Zhu,[119] K. F. Zhu,[169] K. J. Zhu,[39] Z. H. Zhu,[39] B. Zhuang,[39] H. L. Zhuang,[39] L. Zong,[134] J. H. Zou,[39] and S. Zwickel[146]

(The JUNO Collaboration)

J. dos Anjos,[70] L. Asquith,[170] J.-L. Beney,[171] T. J. C. Bezerra,[170] M. Bongrand,[171] C. Bourgeois,[172] D. Brasse,[79] D. Breton,[172] M. Briere,[172] J. Busto,[173] A. Cabrera,[172] A. Cadiou,[171] E. Calvo,[83] H. Carduner,[171] V. Chaumat,[172] E. Chauveau,[174] M. Chen,[175] P. Chimenti,[176] F. Dal Corso,[177] A. Dahmane,[79] S. Dusini,[177] A.D. Earle,[170] C. Frigerio-Martins,[176] J. Galán,[178] J.A. García,[178] R. Gazzini,[172] A. Gibson,[170] D. Giovagnoli,[79] P. Govoni,[179,180] M. Grassi,[181] W.C. Griffith,[170] F. Haddad,[171] J. Hartnell,[170] A. Hourlier,[79] G. Hull,[172] I. Irastorza,[178] L. Koch,[182] P. Laniéce,[172] P. Lasorak,[170] C. Lefevre,[175] F. Lefevre,[171] P. Loaiza,[172] G. Luzon,[178] J. Maalmi,[172] F. Mantovani,[183,184] C. Marquet,[174] M. Martinez,[178] J. Martino,[171] L. Menard,[172] D. Navas-Nicolás,[172] H. Nunokawa,[185] M. Obolensky,[172] J. P. Ochoa-Ricoux,[11] C. Palomares,[83] P. Pillot,[171] J. C. C. Porter,[170] M. S. Pravikoff,[174] M. Roche,[174] B. Roskovec,[186] M.L. Sarsa,[178] S. Schoppmann,[182] A. Serafini,[177,181] L. Simard,[172] M. Sisti,[179] D. Stocco,[171] V. Strati,[183,184] J.-S. Stutzmann,[171] F. Suekane,[187] M.-A. Verdier,[172] A. Verdugo,[83] B. Viaud,[171] A. Weber,[182] and F. Yermia[171]

(The LiquidO Consortium)

B. C. Rasco[188] and K. P. Rykaczewski[188]
(on behalf of The MTAS Collaboration)

B. Y. Han,[189] E. J. Jeon,[190] Y. Jeong,[191] H. S. Jo,[192] D. K. Kim,[192] J. Y. Kim,[193] J. G. Kim,[194] Y. D. Kim,[190,193,195] Y. J. Ko,[190] H. M. Lee,[189] M. H. Lee,[190,195] J. Lee,[190] C. S. Moon,[192] Y. M. Oh,[190] H. K. Park,[196] K. S. Park,[190] S. H. Seo,[190] K. Siyeon,[191] G. M. Sun,[189] Y. S. Yoon,[197] and I. Yu[194]

(The NEOS Collaboration)

G. Angloher,[198] A. Bento,[198,199] L. Canonica,[198] F. Cappella,[200] L. Cardani,[200] N. Casali,[200]

R. Cerulli,[201, 202] I. Colantoni,[203, 200] A. Cruciani,[200] G. del Castello,[204, 200] A. Erhart,[205] M. Friedl,[206] A. Garai,[198] V.M. Ghete,[206] C. Goupy,[74] V. Guidi,[207, 208] D. Hauff,[198, 209] M. Kaznacheeva,[205] A. Kinast,[205] L. Klinkenberg,[205] H. Kluck,[206] A. Langenkämper,[205] T. Lasserre,[74, 71] D. Lhuillier,[74] M. Mancuso,[198] B. Mauri,[74] A. Mazzolari,[208] E. Mazzucato,[74] H. Neyrial,[74] C. Nones,[74] L. Oberauer,[205] A. Onillon,[74] T. Ortmann,[205] L. Pattavina,[210, 205] F. Petricca,[198] W. Potzel,[205] F. Pröbst,[198] F. Pucci,[198] F. Reindl,[206, 211] R. Rogly,[74] J. Rothe,[205] V. Savu,[74] N. Schermer,[205] J. Schieck,[206, 211] S. Schönert,[205] C. Schwertner,[206, 211] L. Scola,[74] L. Stodolsky,[198] R. Strauss,[205] C. Tomei,[200] K. v. Mirbach,[205] M. Vignati,[204, 200] M. Vivier,[74] V. Wagner,[205] and A. Wex[205]

(The NUCLEUS Collaboration)

H. Abele,[211] M. del Gallo Roccagiovine,[204, 200] S. Dorer,[211] A. Doblhammer,[211] E. Jericha,[211] L. Peters,[205] and G. Soum[74]

(Associated to the NUCLEUS Collaboration)

M. Andriamirado,[8] A. B. Balantekin,[37] C. D. Bass,[212] D. E. Bergeron,[213] E. P. Bernard,[1] N. S. Bowden,[1] C. D. Bryan,[214] R. Carr,[4] T. Classen,[1] A. J. Conant,[5] G. Deichert,[214] A. Delgado,[188, 215] M. V. Diwan,[14] M. J. Dolinski,[93] A. Erickson,[216] B. T. Foust,[7] J. K. Gaison,[7] A. Galindo-Uribarri,[188, 215] C. E. Gilbert,[188] S. Gokhale,[14] C. Grant,[217] S. Hans,[14] A. B. Hansell,[218] K. M. Heeger,[7] B. Heffron,[188, 215] D. E. Jaffe,[14] S. Jayakumar,[93] X. Ji,[14] D. C. Jones,[57] O. Kyzylova,[93] P. Kunkle,[217] C. E. Lane,[93] T. J. Langford,[7] J. LaRosa,[213] B. R. Littlejohn,[8] X. Lu,[188, 215] J. Maricic,[219] M. P. Mendenhall,[1] A. M. Meyer,[219] R. Milincic,[219] P. E. Mueller,[188] H. P. Mumm,[213] J. Napolitano,[57] R. Neilson,[93] J. A. Nikkel,[7] S. Nour,[213] J. L. Palomino,[8] D. A. Pushin,[220] X. Qian,[14] C. Roca,[14] R. Rosero,[14] M. Searles,[214] P. T. Surukuchi,[7] F. Sutanto,[1] M. A. Tyra,[213] R. L. Varner,[188] D. Venegas-Vargas,[188, 215] P. B. Weatherly,[93] J. Wilhelmi,[7] A. Woolverton,[220] M. Yeh,[14] C. Zhang,[14] and X. Zhang[1]

(The PROSPECT Collaboration)

H. I. Jang,[221] J. S. Jang,[222] S. H. Jeon,[194] K. K. Joo,[223] D. E. Jung,[194] J. Y. Kim,[223] S. B. Kim,[194] S. Y. Kim,[224] W. Kim,[192] E. Kwon,[194] H. G. Lee,[224] W. J. Lee,[224] I. T. Lim,[223] D. H. Moon,[223] M. Y. Pac,[225] R. G. Park,[223] H. Seo,[224] J. W. Seo,[194] C. D. Shin,[223] B. S. Yang,[224] J. Yoo,[224] S. G. Yoon,[226] I. S. Yeo,[225] and I. Yu[194]

(The RENO Collaboration)

C. Augier,[227] J. Billard,[227] A. Broniatowski,[228] M. Calvo,[229] A. Cazes,[227] C. Chang,[230] M. Chapellier,[228] G. Chemin,[231] M. De Jesus,[227] L. Dumoulin,[228] E. Figueroa-Feliciano,[232] J. A. Formaggio,[233] J. Gascon,[227] A. Giuiani,[228] J. Goupy,[229] C. Goy,[231] P. Harrington,[233] S. T. Heine,[233] S. A. Hertel,[234] Z. Hong,[235] Y. Jin,[236] A. Juillard,[227] J. Lamblin,[231] H. Lattaud,[227] A. Lubashevskiy,[237] P. de Marcillac,[228] S. Marnieros,[228] A. Monfardini,[229] W. Oliver,[233] E. Olivieri,[228] K. J. Palladino,[238] D. Poda,[228] J. S. Real,[231] J. S. Ricol,[231] A. Robert,[239] S. Rozov,[237] V. Sanglard,[227] T. Salagnac,[227] B. Schmidt,[232] V. Sibille,[233, 227] J. Stachurska,[233] A. Stutz,[231] W. Van De Ponteseele,[233] L. Winslow,[233] E. Yakushev,[237] and D. Zinatulina[237]

(The Ricochet Collaboration)

O. A. Akindele,[1] N. S. Bowden,[1] L. Carman,[1] T. Classen,[1] S. Dazeley,[1] M. Ford,[1] V. A. Li,[1] M. P. Mendenhall,[1] C. Roca,[1] F. Sutanto,[1] N. Zaitseva,[1] and X. Zhang[1]

(The ROADSTR Near-Field Working Group)

W. Beaumont,[240] S. Binet,[241] I. Bolognino,[242] M. Bongrand,[242] J. Borg,[243] V. Buridon,[244] H. Chanal,[241] B. Coupé,[245] P. Crochet,[241] D. Cussans,[246] A. De Roeck,[240, 247] D. Durand,[244] M. Fallot,[242] D. Galbinski,[243] S. Gallego,[244] L. Giot,[242] B. Guillon,[244] D. Henaff,[242]

S. Hayashida,[243] B. Hosseini,[243] S. Kalcheva,[245] G. Lehaut,[244] I. Michiels,[248] S. Monteil,[241] D. Newbold,[246, 249] N. Roy,[250] D. Ryckbosch,[248] H. Rejeb Sfar,[240] L. Simard,[250, 251] A. Vacheret,[243] G. Vandierendonck,[248] S. Van Dyck,[245] N. van Remortel,[240] S. Vercaemer,[240] M. Verstraeten,[240] B. Viaud,[242] A. Weber,[252, 249] M. Yeresko,[241] and and F. Yermia.[242]

(The SoLid Collaboration)


A. Bonhomme,[34, 74] C. Buck,[34] P. del Amo Sanchez,[253] I. El Atmani,[74] L. Labit,[253] J. Lamblin,[254] A. Letourneau,[74] D. Lhuillier,[74] M. Licciardi,[254] M. Lindner,[34] T. Materna,[74] H. Pessard,[253] J.-S. Réal,[254] J.-S. Ricol,[254] R. Rogly,[74] V. Savu,[74] S. Schoppmann,[34] T. Soldner,[255] A. Stutz,[254] and M. Vialat[255]

(The STEREO Collaboration)


M. Askins,[3, 12] Z. Bagdasarian,[3, 12] N. Barros,[98, 99, 100] E.W. Beier,[98] A. Bernstein,[1] M. Böhles,[109] E. Blucher,[78] R. Bonventre,[12] E. Bourret,[12] E. J. Callaghan,[3, 12] J. Caravaca,[3, 12] M. Diwan,[14] S.T. Dye,[256] J. Eisch,[257] A. Elagin,[78] T. Enqvist,[112] U. Fahrendholz,[110] V. Fischer,[108] K. Frankiewicz,[103] C. Grant,[103] D. Guffanti,[109] C. Hagner,[258] A. Hallin,[259] C. M. Jackson,[260] R. Jiang,[78] T. Kaptanoglu,[3, 12] J.R. Klein,[98] Yu. G. Kolomensky,[3, 12] C. Kraus,[104, 105] F. Krennrich,[113] T. Kutter,[261] T. Lachenmaier,[262] B. Land,[3, 12, 98] K. Lande,[98] L. Lebanowski,[98] J.G. Learned,[256] V.A. Li,[1] V. Lozza,[99, 100] L. Ludhova,[263, 75] M. Malek,[264] S. Manecki,[105, 265, 104] J. Maneira,[99, 100] J. Maricic,[256] J. Martyn,[109] A. Mastbaum,[107] C. Mauger,[98] M. Mayer,[110] J. Migenda,[266] F. Moretti,[12] J. Napolitano,[267] B. Naranjo,[268] S. Naugle,[98] M. Nieslony,[109] L. Oberauer,[110] G. D. Orebi Gann,[3, 12] J. Ouellet,[269] T. Pershing,[108] S.T. Petcov,[270] L. Pickard,[108] R. Rosero,[14] M. C. Sanchez,[113] J. Sawatzki,[110] S. Schoppmann,[3, 12] S.H. Seo,[190] M. Smiley,[3, 12] M. Smy,[271] A. Stahl,[75] H. Steiger,[109, 110] M. R. Stock,[110] H. Sunej,[14] R. Svoboda,[108] E. Tiras,[102, 111] W. H. Trzaska,[112] M. Tzanov,[261] M. Vagins,[271] C. Vilela,[272] Z. Wang,[273] J. Wang,[274] M. Wetstein,[113] M.J. Wilking,[272] L. Winslow,[269] P. Wittich,[275] B. Wonsak,[258] E. Worcester,[14, 272] M. Wurm,[109] G. Yang,[272] M. Yeh,[14] E.D. Zimmerman,[106] S. Zsoldos,[3, 12] and K. Zuber[276]

(The THEIA Collaboration)


A. Algora,[277] A. Beloeuvre,[242] M. Estienne,[242] M. Fallot,[242] L. Giot,[242] R. Kean,[242] A. Porta,[242] and J. L. Tain[277]

(The Valencia-Nantes TAGS Collaboration)


J. I. Collar,[278] Z. Djurcic,[279] A. Erlandson,[280] M. Foxe,[260] S. Gariazzo,[281] M.V. Garzelli,[282] C. Giunti,[283] S. Hedges,[1] I. Jovanovic,[18] J.Learned,[219] C. M. Lewis,[278] J.LoSecco,[284] X. Mougeot,[285] J. Newby,[5] K. Ni,[286] R. Pestes,[6] and L. Périssé[74]

[1] *Lawrence Livermore National Laboratory, Livermore, CA 94550, USA*

[2] *Institute for Nuclear Theory, University of Washington, Seattle, WA 98195, USA*

[3] *Department of Physics, University of California at Berkeley, Berkeley, CA 94720-7300, USA*

[4] *Department of Physics, United States Naval Academy, Annapolis, MD, USA*

[5] *Oak Ridge National Laboratory, Oak Ridge, TN, USA*

[6] *Center for Neutrino Physics, Virginia Tech, Blacksburg, Virginia 24061*

[7] *Wright Laboratory and Department of Physics, Yale University, New Haven, Connecticut 06520, USA*

[8] *Department of Physics, Illinois Institute of Technology, Chicago, Illinois 60616*

[9] *Fermi National Accelerator Laboratory, Batavia, IL, United States*

[10] *Instituto de Investigaciones en Ingeniería Eléctrica, Departamento de Ingeniería Eléctrica y Computadoras, Universidad Nacional del Sur (UNS) - CONICET, Bahía Blanca, Argentina*

[11]Department of Physics and Astronomy, University of California, Irvine, California 92697

[12]Lawrence Berkeley National Laboratory, 1 Cyclotron Road, Berkeley, CA 94720-8153, USA

[13]Texas A&M University, College Station, TX 77843, USA

[14]Brookhaven National Laboratory, Upton, New York 11973, USA

[15]Department of Physics, Duke University, Durham, NC 27708

[16]Triangle Universities Nuclear Laboratory, Durham, NC 27708

[17]Nuclear Engineering Program, Virginia Tech, Falls Church, Virginia 22043

[18]University of Michigan, Ann Arbor, MI, USA

[19]Department of Physics, University of California at Davis, Davis, California 95616, USA

[20]University of Maryland, Department of Physics, College Park, MD 20742

[21]Department of Physics, University of California, Berkeley, California 94720

[22]Universidad Nacional Autónoma de México, Ciudad de México, México

[23]Centro Atómico Bariloche and Instituto Balseiro, Comisión Nacional de Energía Atómica (CNEA), Consejo Nacional de Investigaciones Científicas y Técnicas (CONICET), Universidad Nacional de Cuyo (UNCUYO), San Carlos de Bariloche, Argentina.

[24]Facultad de Ingeniería - Universidad Nacional de Asunción, Asunción, Paraguay

[25]Universidade Federal do Rio de Janeiro, Instituto de Física, Rio de Janeiro, RJ, Brazil

[26]International Center for Advanced Studies & Instituto de Ciencias Físicas, ECyT-UNSAM and CONICET, Buenos Aires, Argentina

[27]Centro Brasileiro de Pesquisas Físicas, Rio de Janeiro, RJ, Brazil

[28]Centro Federal de Educação Tecnológica Celso Suckow da Fonseca, Angra dos Reis, RJ, Brazil

[29]Universidad Nacional de San Martín (UNSAM), Comisión Nacional de Energía Atómica (CNEA), Consejo Nacional de Investigaciones Científicas y Técnicas (CONICET), Argentina

[30]Universität Zürich Physik Institut, Zurich, Switzerland

[31]Universidade Federal de São Carlos, Araras, SP, Brazil

[32]Universidade Federal do ABC, Santo André, SP, Brazil

[33]Departamento de Física, FCEN, Universidad de Buenos Aires and IFIBA, CONICET, Buenos Aires, Argentina

[34]Max-Planck-Institut für Kernphysik, Saupfercheckweg 1, 69117 Heidelberg, Germany

[35]Institute of Modern Physics, East China University of Science and Technology, Shanghai

[36]Sun Yat-Sen (Zhongshan) University, Guangzhou

[37]University of Wisconsin, Madison, Wisconsin 53706

[38]Department of Physics, National Taiwan University, Taipei

[39]Institute of High Energy Physics, Beijing

[40]National United University, Miao-Li

[41]Department of Engineering Physics, Tsinghua University, Beijing

[42]Shenzhen University, Shenzhen

[43]North China Electric Power University, Beijing

[44]Chinese University of Hong Kong, Hong Kong

[45]Siena College, Loudonville, New York 12211

[46]University of Science and Technology of China, Hefei

[47] *Charles University, Faculty of Mathematics and Physics, Prague*

[48] *Joint Institute for Nuclear Research, Dubna, Moscow Region*

[49] *Department of Physics, University of Illinois at Urbana-Champaign, Urbana, Illinois 61801*

[50] *Lawrence Berkeley National Laboratory, Berkeley, California 94720*

[51] *Beijing Normal University, Beijing*

[52] *Department of Nuclear Science and Technology, School of Energy and Power Engineering, Xi'an Jiaotong University, Xi'an*

[53] *China Institute of Atomic Energy, Beijing*

[54] *Shandong University, Jinan*

[55] *Guangxi University, No.100 Daxue East Road, Nanning*

[56] *Institute of Physics, National Chiao-Tung University, Hsinchu*

[57] *Department of Physics, College of Science and Technology, Temple University, Philadelphia, Pennsylvania 19122*

[58] *Dongguan University of Technology, Dongguan*

[59] *Department of Physics, The University of Hong Kong, Pokfulam, Hong Kong*

[60] *School of Physics, Nankai University, Tianjin*

[61] *Department of Physics and Astronomy, Shanghai Jiao Tong University, Shanghai Laboratory for Particle Physics and Cosmology, Shanghai*

[62] *Joseph Henry Laboratories, Princeton University, Princeton, New Jersey 08544*

[63] *California Institute of Technology, Pasadena, California 91125*

[64] *College of William and Mary, Williamsburg, Virginia 23187*

[65] *Nanjing University, Nanjing*

[66] *China General Nuclear Power Group, Shenzhen*

[67] *College of Electronic Science and Engineering, National University of Defense Technology, Changsha*

[68] *Iowa State University, Ames, Iowa 50011*

[69] *Chongqing University, Chongqing*

[70] *Centro Brasileiro de Pesquisas Físicas, Rio de Janeiro, RJ, 22290-180, Brazil*

[71] *APC, Université de Paris, CNRS, Astroparticule et Cosmologie, F-75006, Paris*

[72] *Max-Planck-Institut für Kernphysik, 69117 Heidelberg, Germany*

[73] *Physik Department, Technische Universität München, 85748 Garching, Germany*

[74] *IRFU, CEA, Université Paris-Saclay, 91191 Gif-sur-Yvette, France*

[75] *III. Physikalisches Institut, RWTH Aachen University, 52056 Aachen, Germany*

[76] *SUBATECH, CNRS/IN2P3, Université de Nantes, IMT-Atlantique, 44307 Nantes, France*

[77] *Institute of Nuclear Research of the Russian Academy of Sciences, Moscow 117312, Russia*

[78] *The Enrico Fermi Institute, The University of Chicago, Chicago, Illinois 60637, USA*

[79] *IPHC, CNRS/IN2P3, Université de Strasbourg, 67037 Strasbourg, France*

[80] *Department of Physics and Astronomy, University of Alabama, Tuscaloosa, Alabama 35487, USA*

[81] *LNCA Underground Laboratory, CNRS/IN2P3-CEA, Chooz, France*

[82] *IJC Laboratory, CNRS/IN2P3, Université Paris-Saclay, Orsay, France*

[83] *Centro de Investigaciones Energéticas, Medioambientales y Tecnológicas, CIEMAT, 28040, Madrid, Spain*

[84] *Université de Bordeaux, CNRS/IN2P3, CENBG, F-33175 Gradignan, France*

[85]Argonne National Laboratory, Argonne, Illinois 60439, USA

[86]NRC Kurchatov Institute, 123182 Moscow, Russia

[87]Research Center for Neutrino Science, Tohoku University, Sendai 980-8578, Japan

[88]Universidade Estadual de Campinas-UNICAMP, Campinas, SP, 13083-970, Brazil

[89]Department of Physics, Kobe University, Kobe, 657-8501, Japan

[90]Department of Physics, Tokyo Institute of Technology, Tokyo, 152-8551, Japan

[91]Kepler Center for Astro and Particle Physics, Universität Tübingen, 72076 Tübingen, Germany

[92]Department of Physics, Kitasato University, Sagamihara, 252-0373, Japan

[93]Department of Physics, Drexel University, Philadelphia, Pennsylvania 19104, USA

[94]University of Notre Dame, Notre Dame, Indiana 46556, USA

[95]Department of Physics, Tokyo Metropolitan University, Tokyo, 192-0397, Japan

[96]Center for Neutrino Physics, Virginia Tech, Blacksburg, Virginia 24061, USA

[97]Pennsylvania State University, University Park, PA 16802, USA

[98]Department of Physics and Astronomy, University of Pennsylvania, Philadelphia, PA 19104-6396

[99]Universidade de Lisboa, Faculdade de Ciências (FCUL), Departamento
de Física, Campo Grande, Edifício C8, 1749-016 Lisboa, Portugal

[100]Laboratório de Instrumentação e Física Experimental de Partículas
(LIP), Av. Prof. Gama Pinto, 2, 1649-003, Lisboa, Portugal

[101]Bartoszek Engineering, Aurora, IL 60506, USA

[102]Department of Physics, Erciyes University, 38030, Kayseri, Turkey

[103]Boston University, Department of Physics, Boston, MA 02215, USA

[104]SNOLAB, Creighton Mine 9, 1039 Regional Road 24, Sudbury, ON P3Y 1N2, Canada

[105]Laurentian University, Department of Physics, 935 Ramsey Lake Road, Sudbury, ON P3E 2C6, Canada

[106]University of Colorado at Boulder, Department of Physics, Boulder, Colorado, USA

[107]Department of Physics and Astronomy, Rutgers, The State University
of New Jersey, 136 Frelinghuysen Road, Piscataway, NJ 08854-8019 USA

[108]University of California, Davis, 1 Shields Avenue, Davis, CA 95616, USA

[109]Institute of Physics and Excellence Cluster PRISMA,
Johannes Gutenberg-Universität Mainz, 55099 Mainz, Germany

[110]Physics Department, Technische Universität München, 85748 Garching, Germany

[111]Department of Physics and Astronomy, The University of Iowa, Iowa City, Iowa, USA

[112]Department of Physics, University of Jyväskylä, Finland

[113]Department of Physics and Astronomy, Iowa State University, Ames, IA 50011, USA

[114]Pontificia Universidad Católica de Chile, Santiago, Chile

[115]IPHC, Université de Strasbourg, CNRS/IN2P3, F-67037 Strasbourg, France

[116]Pakistan Institute of Nuclear Science and Technology, Islamabad, Pakistan

[117]INFN Catania and Dipartimento di Fisica e Astronomia dell'Università di Catania, Catania, Italy

[118]III. Physikalisches Institut B, RWTH Aachen University, Aachen, Germany

[119]East China University of Science and Technology, Shanghai, China

[120]INFN Sezione di Milano and Dipartimento di Fisica dell'Università di Milano, Milano, Italy

[121]*Department of Physics, Faculty of Science, Chulalongkorn University, Bangkok, Thailand*

[122]*IJCLab, Université Paris-Saclay, CNRS/IN2P3, 91405 Orsay, France*

[123]*Department of Physics and Earth Science, University of Ferrara and INFN Sezione di Ferrara, Ferrara, Italy*

[124]*INFN Milano Bicocca and University of Milano Bicocca, Milano, Italy*

[125]*INFN Sezione di Padova, Padova, Italy*

[126]*Eberhard Karls Universität Tübingen, Physikalisches Institut, Tübingen, Germany*

[127]*Université de Bordeaux, CNRS, CENBG-IN2P3, F-33170 Gradignan, France*

[128]*Dipartimento di Fisica e Astronomia dell'Universita' di Padova and INFN Sezione di Padova, Padova, Italy*

[129]*University of Roma Tre and INFN Sezione Roma Tre, Roma, Italy*

[130]*Centre de Physique des Particules de Marseille, Marseille, France*

[131]*Wuhan University, Wuhan, China*

[132]*INFN Milano Bicocca and Politecnico di Milano, Milano, Italy*

[133]*Dongguan University of Technology, Dongguan, China*

[134]*Tsinghua University, Beijing, China*

[135]*Institute of Modern Physics, Chinese Academy of Sciences, Lanzhou, China*

[136]*Beijing Institute of Spacecraft Environment Engineering, Beijing, China*

[137]*Universidade Estadual de Londrina, Londrina, Brazil*

[138]*INFN Sezione di Perugia and Dipartimento di Chimica,
Biologia e Biotecnologie dell'Università di Perugia, Perugia, Italy*

[139]*Université Libre de Bruxelles, Brussels, Belgium*

[140]*Institute of Physics and EC PRISMA$^+$, Johannes-Gutenberg Universität Mainz, Mainz, Germany*

[141]*Suranaree University of Technology, Nakhon Ratchasima, Thailand*

[142]*Lomonosov Moscow State University, Moscow, Russia*

[143]*Institute for Nuclear Research of the Russian Academy of Sciences, Moscow, Russia*

[144]*School of Physics and Microelectronics, Zhengzhou University, Zhengzhou, China*

[145]*University of Jyvaskyla, Department of Physics, Jyvaskyla, Finland*

[146]*Technische Universität München, München, Germany*

[147]*Wuyi University, Jiangmen, China*

[148]*Harbin Institute of Technology, Harbin, China*

[149]*Universidad Tecnica Federico Santa Maria, Valparaiso, Chile*

[150]*Forschungszentrum Jülich GmbH, Nuclear Physics Institute IKP-2, Jülich, Germany*

[151]*Jinan University, Guangzhou, China*

[152]*Institute of Experimental Physics, University of Hamburg, Hamburg, Germany*

[153]*School of Physics and Astronomy, Shanghai Jiao Tong University, Shanghai, China*

[154]*Yerevan Physics Institute, Yerevan, Armenia*

[155]*SUBATECH, Université de Nantes, IMT Atlantique, CNRS-IN2P3, Nantes, France*

[156]*Comenius University Bratislava, Faculty of Mathematics, Physics and Informatics, Bratislava, Slovakia*

[157]*The Radiochemistry and Nuclear Chemistry Group in University of South China, Hengyang, China*

[158]*Tsung-Dao Lee Institute, Shanghai Jiao Tong University, Shanghai, China*

[159]*University of Chinese Academy of Sciences, Beijing, China*

[160] INFN Catania and Centro Siciliano di Fisica Nucleare e Struttura della Materia, Catania, Italy

[161] Institute of Hydrogeology and Environmental Geology,
Chinese Academy of Geological Sciences, Shijiazhuang, China

[162] Jilin University, Changchun, China

[163] Xiamen University, Xiamen, China

[164] School of Physics, Peking University, Beijing, China

[165] Laboratori Nazionali di Frascati dell'INFN, Roma, Italy

[166] Institute of Electronics and Computer Science, Riga, Latvia

[167] Pontificia Universidade Catolica do Rio de Janeiro, Rio, Brazil

[168] National Astronomical Research Institute of Thailand, Chiang Mai, Thailand

[169] Xi'an Jiaotong University, Xi'an, China

[170] Department of Physics and Astronomy, University of Sussex, Falmer, Brighton BN1 9QH, United Kingdom

[171] Subatech, CNRS/IN2P3, Nantes Université, IMT-Atlantique, 44307 Nantes, France

[172] IJCLab, CNRS/IN2P3, Université Paris-Saclay, Université
de Paris, 15 rue Georges Clémenceau, 91400 Orsay, France

[173] Aix Marseille Univ, CNRS/IN2P3, CPPM, Marseille, France

[174] Université de Bordeaux, CNRS, LP2i, F-33170 Gradignan, France

[175] Department of Physics, Engineering Physics & Astronomy,
Queen's University, Kingston, Ontario K7L3N6, Canada

[176] Departamento de Física, Universidade Estadual de Londrina, Rodovia Celso Garcia Cid, PR
445 Km 380, Campus Universitário Cx. Postal 10.011, CEP 86.057-970, Londrina – PR, Brazil

[177] INFN, Sezione di Padova, via Marzolo 8, I-35131 Padova, Italy

[178] University of Zaragoza, Facultad de Ciencias, Departamento de Física, Pedro Cerbuna 12, 50009, Zaragoza

[179] INFN, Sezione di Milano-Bicocca, I-20126 Milano, Italy

[180] Dipartimento di Fisica, Università di Milano-Bicocca, I-20126 Milano, Italy

[181] Dipartimento di Fisica e Astronomia, Università di Padova, via Marzolo 8, I-35131 Padova, Italy

[182] Johannes Gutenberg Universität Mainz, Institut für Physik, Staudingerweg 7, 55128 Mainz, Germany

[183] INFN, Ferrara Section, Via Saragat 1, 44122 Ferrara, Italy

[184] Department of Physics and Earth Sciences, University of Ferrara, Via Saragat 1, 44122 Ferrara, Italy

[185] Department of Physics, Pontifícia Universidade Católica do
Rio de Janeiro, C.P. 38097, 22451-900, Rio de Janeiro, Brazil

[186] Institute of Particle and Nuclear Physics Faculty of Mathematics and Physics,
Charles University, V Holešovičkách 2 180 00 Prague 8, Czech Republic

[187] RCNS, Tohoku University, 6-3 AzaAoba, Aramaki, Aoba-ku, 980-8578, Sendai, Japan

[188] Physics Division, Oak Ridge National Laboratory, Oak Ridge, TN, USA

[189] Neutron Science Division, Korea Atomic Energy Research Institute, Deajeon, 34057, Korea

[190] Center for Underground Physics, Institute for Basic Science (IBS), Daejeon, 34126, Korea

[191] Department of Physics, Chung-Ang University, Seoul, 06974, Korea

[192] Department of Physics, Kyungpook National University, Daegu 41566, Korea

[193] Department of Physics and Astronomy, Sejong University, Seoul, 05006, Korea

[194] Department of Physics, SungKyun Kwan University, Suwon, 16419, Korea

[195] *IBS School, University of Science and Technology (UST), Daejeon, 34113, Korea*

[196] *Department of Accelerator Science, Korea University, Sejong, 30019, Korea*

[197] *Center for Ionizing Radiation, Korea Research Institute of Standards and Science, Daejeon, 34113, Korea*

[198] *Max-Planck-Institut für Physik, D-80805 München, Germany*

[199] *CIUC, Departamento de Fisica, Universidade de Coimbra, P3004 516 Coimbra, Portugal*

[200] *Istituto Nazionale di Fisica Nucleare – Sezione di Roma, Roma I-00185, Italy*

[201] *Istituto Nazionale di Fisica Nucleare – Sezione di Roma "Tor Vergata", Roma I-00133, Italy*

[202] *Dipartimento di Fisica, Università di Roma "Tor Vergata", Roma I-00133, Italy*

[203] *Consiglio Nazionale delle Ricerche, Istituto di Nanotecnologia, Roma I-00185, Italy*

[204] *Dipartimento di Fisica, Sapienza Università di Roma, Roma I-00185, Italy*

[205] *Physik-Department and Excellence Cluster ORIGINS, Technische Universität München, D-85748 Garching, Germany*

[206] *Institut für Hochenergiephysik der Österreichischen Akademie der Wissenschaften, A-1050 Wien, Austria*

[207] *Dipartimento di Fisica, Università di Ferrara, I-44122 Ferrara, Italy*

[208] *Istituto Nazionale di Fisica Nucleare – Sezione di Ferrara, I-44122 Ferrara, Italy*

[209] *Present address: Eberhard-Karls-Universität Tübingen, D-72076 Tübingen, Germany*

[210] *Istituto Nazionale di Fisica Nucleare – Laboratori Nazionali del Gran Sasso, Assergi (L'Aquila) I-67100, Italy*

[211] *Atominstitut, Technische Universität Wien, A-1020 Wien, Austria*

[212] *Department of Physics, Le Moyne College, Syracuse, NY, USA*

[213] *National Institute of Standards and Technology, Gaithersburg, MD, USA*

[214] *High Flux Isotope Reactor, Oak Ridge National Laboratory, Oak Ridge, TN, USA*

[215] *Department of Physics and Astronomy, University of Tennessee, Knoxville, TN, USA*

[216] *George W. Woodruff School of Mechanical Engineering, Georgia Institute of Technology, Atlanta, GA USA*

[217] *Department of Physics, Boston University, Boston, MA, USA*

[218] *Department of Physics, Susquehanna University, Selinsgrove, PA, USA*

[219] *Department of Physics & Astronomy, University of Hawaii, Honolulu, HI, USA*

[220] *Institute for Quantum Computing and Department of Physics and Astronomy, University of Waterloo, Waterloo, ON, Canada*

[221] *Department of Fire Safety, Seoyeong University, Gwangju 61268, Korea*

[222] *GIST College, Gwangju Institute of Science and Technology, Gwangju 61005, Korea*

[223] *Institute for Universe and Elementary Particles, Chonnam National University, Gwangju 61186, Korea*

[224] *Department of Physics and Astronomy, Seoul National University, Seoul 08826, Korea*

[225] *Institute for High Energy Physics, Dongshin University, Naju 58245, Korea*

[226] *Department of Physics, KAIST, Daejeon 34141, Korea*

[227] *Institut de Physique Nucléaire de Lyon (IPNL), France*

[228] *Centre de Sciences Nucléaires et de Sciences de la Matière (CSNSM), France*

[229] *Institut Neel PI at Neel Institute Grenoble, France*

[230] *Argonne National Laboratory, USA*

[231] *Laboratoire de Physique Subatomique et de Cosmologie (LPSC), France*

[232] *Northwestern University, USA*

[233] *Massachusetts Institute of Technology, USA*

[234] *University of Massachusetts at Amherst, USA*

[235] *University of Toronto, Canada*

[236] *CNRS, Univ. Paris-Saclay, France*

[237] *Joint Institute for Nuclear Research (JINR), Russian Federation*

[238] *University of Oxford, United Kingdom*

[239] *Institut Laue Langevin, Grenoble, France*

[240] *Universiteit Antwerpen, Antwerpen, Belgium*

[241] *Université Clermont Auvergne, CNRS/IN2P3, LPC, Clermont-Ferrand, France*

[242] *SUBATECH, CNRS/IN2P3, Université de Nantes, Ecole des Mines de Nantes, Nantes, France*

[243] *Imperial College London, Department of Physics, London, United Kingdom*

[244] *Normandie Univ, ENSICAEN, UNICAEN, CNRS/IN2P3, LPC Caen, 14000 Caen, France*

[245] *SCK-CEN, Belgian Nuclear Research Centre, Mol, Belgium*

[246] *University of Bristol, Bristol, UK*

[247] *CERN, 1211 Geneva 23, Switzerland*

[248] *Universiteit Gent, Gent, Belgium*

[249] *STFC, Rutherford Appleton Laboratory, Harwell Oxford, and Daresbury Laboratory*

[250] *LAL, Univ Paris-Sud, CNRS/IN2P3, Université Paris-Saclay, Orsay, France*

[251] *Institut Universitaire de France, F-75005 Paris, France*

[252] *University of Oxford, Oxford, UK*

[253] *Univ. Grenoble Alpes, Université Savoie Mont Blanc, CNRS/IN2P3, LAPP, 74000 Annecy, France*

[254] *Univ. Grenoble Alpes, CNRS, Grenoble INP, LPSC-IN2P3, 38000 Grenoble, France*

[255] *Institut Laue-Langevin, CS 20156, 38042 Grenoble Cedex 9, France*

[256] *University of Hawai'i at Manoa, Honolulu, Hawai'i 96822, USA*

[257] *Fermi National Accelerator Laboratory, Batavia, IL 60510, USA*

[258] *Institut für Experimentalphysik, Universität Hamburg, 22761 Hamburg, Germany*

[259] *University of Alberta, Department of Physics, 4-181 CCIS, Edmonton, AB T6G 2E1, Canada*

[260] *Pacific Northwest National Laboratory, Richland, WA 99352, USA*

[261] *Department of Physics and Astronomy, Louisiana State University, Baton Rouge, LA 70803*

[262] *Kepler Center for Astro and Particle Physics, Universität Tübingen, 72076 Tübingen, Germany*

[263] *Forschungszentrum Jülich, Institute for Nuclear Physics, Wilhelm-Johnen-Straße 52425 Jülich, Germany*

[264] *University of Sheffield, Physics & Astronomy, Western Bank, Sheffield S10 2TN, UK*

[265] *Queen's University, Department of Physics, Engineering Physics & Astronomy, Kingston, ON K7L 3N6, Canada*

[266] *King's College London, Department of Physics, Strand Building, Strand, London WC2R 2LS, UK*

[267] *Department of Physics, Temple University, Philadelphia, PA, USA*

[268] *University of California Los Angeles, Department of Physics & Astronomy, 475 Portola Plaza, Los Angeles, CA 90095-1547, USA*

[269] *Massachusetts Institute of Technology, Department of Physics and Laboratory for Nuclear Science, 77 Massachusetts Ave Cambridge, MA 02139, USA*

[270] *SISSA/INFN, Via Bonomea 265, I-34136 Trieste, Italy ,*

*Kavli IPMU (WPI), University of Tokyo, 5-1-5 Kashiwanoha, 277-8583 Kashiwa, Japan*

[271]*University of California, Irvine, Department of Physics, CA 92697, Irvine, USA*

[272]*State University of New York at Stony Brook, Department of Physics and Astronomy, Stony Brook, New York, USA*

[273]*Department of Engineering Physics, Tsinghua University, Beijing 100084, China*

[274]*Department of Physics, South Dakota School of Mines and Technology, Rapid City, SD 57701, USA*

[275]*Cornell University, Ithaca, NY, USA*

[276]*Institut für Kern und Teilchenphysik, TU Dresden, Zellescher Weg 19, 01069, Dresden, Germany*

[277]*Instituto de Física Corpuscular, CSIC-Universitat de Valencia, E-46071 Valencia, Spain*

[278]*University of Chicago, Chicago, IL 60637, USA*

[279]*Argonne National Laboratory, Lemont, IL, United States*

[280]*Canadian Nuclear Laboratories Ltd., Chalk River, ON, K0J 1J0, Canada*

[281]*Istituto Nazionale di Fisica Nucleare (INFN),*
*Sezione di Torino, Via P. Giuria 1, I–10125 Torino, Italy*

[282]*University of Hamburg, II Institute for Theoretical Physics,*
*Luruper Chaussee 149, D-22761 Hamburg, Germany*

[283]*Istituto Nazionale di Fisica Nucleare (INFN),*
*Sezione di Torino, Via P. Giuria 1, I-10125 Torino, Italy*

[284]*University of Notre Dame du Lac, Notre Dame, IN 46556-5670*

[285]*Université Paris-Saclay, CEA, List, Laboratoire National*
*Henri Becquerel (LNE-LNHB), F-91120 Palaiseau, France*

[286]*University of California San Diego, La Jolla, 92093, CA*

---

[*] Now at Department of Chemistry and Chemical Technology, Bronx Community College, Bronx, New York 10453


%, offset=75 -75

\clearpage
\tableofcontents{}
\clearpage
\markboth{}{}

\section{Introduction}
\label{sec:intro}

%Editors: Bryce Littlejohn, Pedro Ochoa, Jon Link, Nathaniel Bowden, Patrick Huber

%\begin{comment}
\subsection{Key Takeaways}
\begin{itemize}
    \item Nuclear reactors are a uniquely powerful, abundant and flavor-pure source of antineutrinos that  continue to play a vital role in the US neutrino physics program.  
    \item Neutrino physics opportunities at reactors offer strong complimentarity with numerous aspects of the wider US neutrino program, including Standard Model and BSM oscillations studies and technology development.  
    \item Reactor neutrino experiments have a direct societal impact and have become a strong workforce and technology development pipeline for DOE National Laboratories and universities.
\end{itemize}

\subsection{Narrative}

%Nuclear reactors are a uniquely powerful, abundant, flavour-pure, and well-understood source of antineutrinos that have advanced and will continue to advance the US neutrino physics program.  

Nuclear reactors are a uniquely powerful, abundant, and flavor-pure source of MeV-scale antineutrinos.  Electron-flavored antineutrinos (\nuebar) are produced in reactors as the unstable, neutron-rich products of nuclear fission undergo beta decay reactions: 
\begin{equation}
^{A}_{Z}B_{N} \rightarrow ^{A}_{Z+1}C_{N-1} + e^- + \overline{\nu}_e
\label{eq:betadecay}
\end{equation}
While only a few percent of the the roughly 200~MeV of excess rest mass energy from one nuclear fission is ultimately expressed as \nuebar kinetic energy, this equates to a total release of $2\times10^{20}$ \nuebar per~GW$_{\textrm{th}}$ power generated.  
The energy spectrum of \nuebar emitted by an operating reactor core reflects the decay schemes of the decaying isotopes, whose endpoints roughly range from the sub-MeV to the 10~MeV scale, as well as the relative abundance of these isotopes in the nuclear fuel, which is driven primarily by the likelihood of their production (or yield) in the core's fission reactions~\cite{Wigner,bib:vogel,vogel_review,sonzogni_insights}.  

\begin{figure}[!h]
  \centering
  \includegraphics[width=0.95\textwidth]{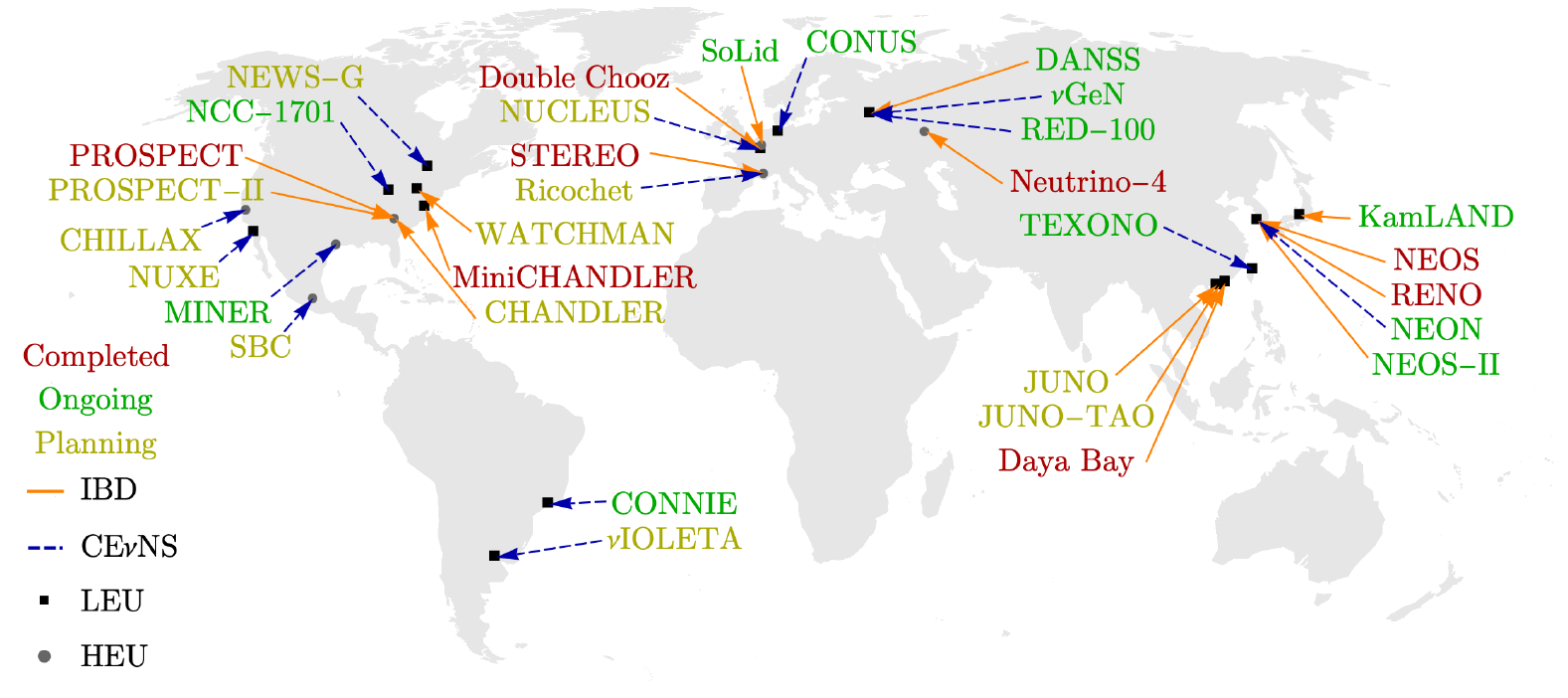}
  \caption{Map of planned, current, and completed reactor antineutrino experiments.  Text color indicates experimental status, while arrow color indicates the interaction channel used by the experiment.  Only completed experiments taking data after 2010 are included.  Further description of these experiments are given in Tables~\ref{tab:ibd_experiments} and~\ref{tab:cevns_experiments}.}
  \label{fig:RxMap}
\end{figure}

The antineutrino emissions of dozens of nuclear reactors across three different continents have been observed with neutrino detectors.  
Locations of current and recent past experiments are illustrated in Figure~\ref{fig:RxMap}; detailed reviews of these and other past experiments can be found in  Refs~\cite{bib:mention2011,AnomalyWhite,huber_berryman}.  
Most of these have been commercial power reactors, which operate in the $\sim$GW$_{\textrm{th}}$ regime and burn fuel with a relatively low level of~\uFive~enrichment (low enriched, or LEU).  
These reactors' neutrino emissions are produced by a mixture of fissionable isotopes, with the dominant isotopes \uFive~and \pNine~providing $>$80\% of all fissions, and \uEight~and \pOne~each providing less than 10\%.  
A substantial number of experiments have been performed at research reactors operating at substantially lower power, $\sim$10-100~MW$_{\textrm{th}}$, than commercial LEU cores.  
These cores have generally been smaller in spatial extent ($<$1~m dimensions) than commercial ones ($>$m dimensions), and have used fuel of substantially higher \uFive~enrichment (highly enriched, or HEU), leading to \nuebar emissions overwhelmingly dominated by \uFive~fission products.  
While other reactor types exist that contain substantially different fuel content than these two options, such as mixed oxide~\cite{Jaffke:2016xdt,Bernstein:2016ayp,Behera:2020qwf} or natural uranium reactors~\cite{Carroll:2018kad}, no successful measurements of these reactor types have been performed.  

As differing fission isotopes have differing yet overlapping fission product yields, HEU and LEU reactors modestly differ in the mean number and energy spectrum of neutrinos they release per fission.  
Considering the decay production mechanism in Eq.~\ref{eq:betadecay}, predictions of HEU and LEU reactor \nuebar emissions can be composed by relying either primarily on knowledge of the produced parent and daughter nuclei, referred to as the \textit{summation} or \textit{ab initio} approach~\cite{bib:vogel,bib:mueller2011,bib:fallot,sonzogni_insights,bib:fallot2}, or primarily on knowledge of the properties of the decay electron produced in concert with each \nuebar, referred to as the \textit{conversion} approach~\cite{bib:ILL_1,bib:ILL_2,bib:ILL_3,bib:huber}.  
These prediction methods are described in further detail in Section~\ref{sec:source}.  

Reactor \nuebar can be detected via multiple detection channels, including inverse beta decay (IBD) on protons or other nuclei, neutral current inelastic nuclear scattering, neutrino-electron elastic scattering, and coherent elastic neutrino-nucleus scattering~\cite{qian_review}.  
The proton IBD interaction, $p + $\nuebar$\rightarrow n + e^+$, represents the vast majority of all observed  interactions to date.  
The presence of two final-state particles that can be individually and coincidentally detected in organic scintillator detectors is  advantageous in achieving excellent background reduction; this channel also facilitates high-fidelity determination of \nuebar energies via reconstruction of e$^+$ energies.  
Detectors with some combination of very low background contamination, very low energy detection thresholds, and specialized materials are required for detection of reactor \nuebar using other detection channels.  
As an example, detection via coherent neutrino-nucleus scattering  require cryogenic detectors using semiconductors or bolometric crystals as targets, with energy detection thresholds well below 1~keV$_{nr}$.  
Figure~\ref{fig:RxMap} also indicates the exploited interaction channel in recent and future reactor \nuebar experiments.  

\begin{wrapfigure}{r}{0.45\textwidth}

\vspace{-8mm}
\includegraphics[width = 0.45\textwidth]{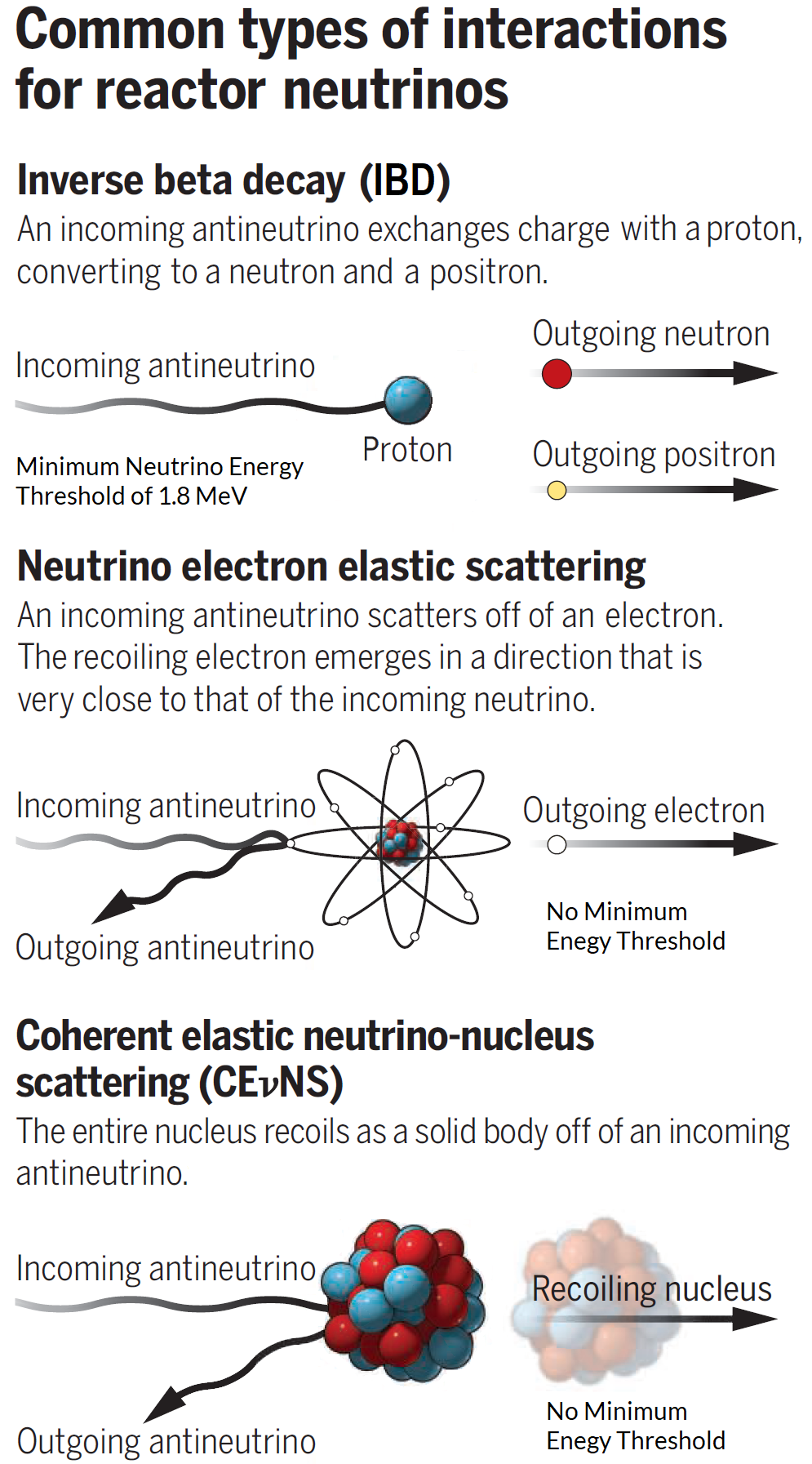}

\vspace{2mm}

\begin{minipage}[c]{0.5\textwidth}
\caption{\label{interactions} {Illustrations of reactor \nuebar interaction mechanisms.  Adapted from~\cite{doi:10.1126/science.aao4050} }}
\end{minipage}

\vspace{-8mm}
\end{wrapfigure}

Past reactor antineutrino experiments have been critically important in the elucidation of the contemporary view of the Standard Model of particle physics (SM).  
Proton IBD-based reactor measurements were the first to verify the existence of neutrinos~\cite{bib:reines_cc}, and have yielded world-leading or competitive precision on three of the six SM neutrino mixing parameters~\cite{KamLAND_rate,KamLAND_shape,bib:prl_rate,bib:reno,bib:dc,bib:prl_shape}.  
Deuteron IBD-based reactor measurements provided early validations of weak interaction theory~\cite{bib:reines_nc}.  
Reactor \nuebar-electron scattering measurements have enabled measurement of the Weinberg mixing angle and competitive limits on measurements of the magnetic moment of the neutrino~\cite{bib:reines_escat,TEXONO:2009knm}.  
Reactor experiments have also enabled world-leading probes of new beyond-the-Standard-Model (BSM) physics.  
Short-baseline proton IBD experiments have been used to set new limits on active-sterile neutrino mixing in the eV-scale range and below~\cite{bib:Bugey3_osc,DayaBay:2016qvc,bib:neos,danss_osc,stereo_2019,prospect_prd,neutrino4_prd}.  
Efforts to measure reactor-based coherent neutrino scattering, while so far unsuccessful in detecting a statistically significant quantity of neutrino interactions, have nonetheless established world-leading limits on some prospective hidden sector couplings to neutrinos~\cite{connie_bsm,collar_bsm}.  
All of these measurements have been performed with fairly imprecise knowledge regarding the true underlying flux and spectrum of reactor \nuebar emissions.  
%A summary pictorial overview of previous reactor \nuebar experiments are indicated in Table~\ref{tab:blah} and Figure~\ref{fig:blah}.  

In the coming decade, reactor-based neutrino measurements can continue to provide crucial new insights into the nature of the Standard Model and beyond.  
New reactor-based oscillation experiments can continue extending the boundaries of our understanding of key SM mixing parameters~\cite{DYB_LOI,JUNO_LOI,juno2}, while also pushing active-sterile mixing parameter space coverage in the electron flavor sector close to the few-percent level over a wide range of mass splittings from $\Delta m^2_{13}$ to the 10s of eV$^2$ scale~\cite{PROSPECT-I,PROSPECT-II,Andriamirado:2021qjc,TAO_LOI,juno_tao}.  
Both SM and BSM oscillation measurements are highly synergistic with other aspects of the US neutrino physics program, including its long-baseline and short-baseline accelerator neutrino efforts.  
Future high-statistics \nuebar measurements at differing reactor types can greatly improve our understanding of the absolute flux and spectrum of reactor \nuebar produced by all reactor core types, both above and below the 1.8~MeV IBD interaction threshold~\cite{RxFlux}.  
In addition to improving the achievable precision of some reactor-based BSM measurements, such as those performed by CEvNS experiments, these improvements are clearly synergistic with facets of the applied reactor physics, nuclear safeguards, and nuclear data communities~\cite{PROSPECT-App,RxApp,bib:IAEA,Akindele:2021sbh,bib:WoNDRAM}.  
The reactor \nuebar field's comparatively low barrier to entry and small experiment scales enable it to serve as a valuable workforce and technology development pipeline while concurrently delivering world-class physics results.  

%Due to the relatively low costs and comparatively plentiful reactor \nuebar sources, 
%A visual summary of impending and future experiments is also overview in Figure~\ref{} and Table~\ref{fig:blah}.  

The purpose of this white paper, written as part of the Snowmass 2021 community organizing exercise, is to survey the impressive range of high-impact physics that can be achieved in the coming decade with current and prospective future reactor antineutrino experiments, to highlight the large degree of synergy between this future reactor \nuebar measurement program, and to emphasize direct societal impacts of near-term investments of the HEP community in the reactor \nuebar sector.  
% as well as the broader nuclear science community.  
In this context, Section organization and content will be generally aligned with the boundaries of specific Neutrino Frontier Topical Groups.  
Sections~\ref{sec:synergies} and~\ref{sec:synergies_broad} will begin by summarizing  synergies between future reactor \nuebar efforts and the US neutrino program and the broader US science community, respectively.  
Sections~\ref{sec:lbl} (directed towards Topical Group NF01) and~\ref{sec:sbl} (towards NF02) highlights potential improvements in understanding of SM oscillations and current short-baseline neutrino anomalies, respectively.  
Section~\ref{sec:bsm} (towards NF03 and NF04) discusses how future reactor measurements can improve knowledge of other SM neutrino properties and possible hidden-sector couplings.  
Section~\ref{sec:source} (towards NF09) overviews potential advancements in global understanding of \nuebar emissions from various reactor types and the ability to accurately model these emissions.  
Finally, Sections~\ref{sec:detectors} and~\ref{sec:applications}  (towards NF10 and NF07, respectively) will focus on applications and detector technology developments relevant to reactor \nuebar.  

%Outline:
%\begin{itemize}

%\item Overview rich history of neutrino physics discoveries with reactor antineutrinos

%\item Overview reactor neutrino detection methods and relevant detector technologies

%\item Overview the broad strokes of future physics to be done.  
%\end{itemize}

%TABLE: Detector Technologies and thresholds, etc, similar to Qian/Peng Review?

%TABLE with checks: Matching Future Experiments to different physics goals?  

\section{Synergies with the US Neutrino Program}
\label{sec:synergies}

%\textcolor{red}{Editors: Bryce Littlejohn, Pedro Ochoa, Jon Link, Nathaniel Bowden, Patrick Huber}

\subsection{Key Takeaways}
\begin{itemize}
    \item Due to their comparatively low energies and high electron flavor purity, reactor \nuebar experiments play a necessary role in a multi-faceted global effort to probe the potential BSM origins of existing short-baseline neutrino anomalies.  
    \item High-precision reactor-based probes of active-sterile neutrino couplings are important for ensuring clear interpretations of DUNE's long-baseline oscillation physics results.  
    \item Reactor-based measurements of $\theta_{13}$, $\theta_{12}$, $\Delta m^2_{31}$, $\Delta m^2_{21}$ and the mass hierarchy serve to expand the physics deliverables and ultimate sensitivity of DUNE and the US long-baseline neutrino program.  
    \item Reactor \nuebar experiments continue to develop  technologies well-suited for application to other areas in neutrino and particle physics, such as detection of light dark matter, geoneutrinos, solar neutrinos, and neutrinoless double beta decay.
\end{itemize}

\subsection{Narrative}

\begin{figure}[!h]
  \centering
  \includegraphics[width=0.7\textwidth]{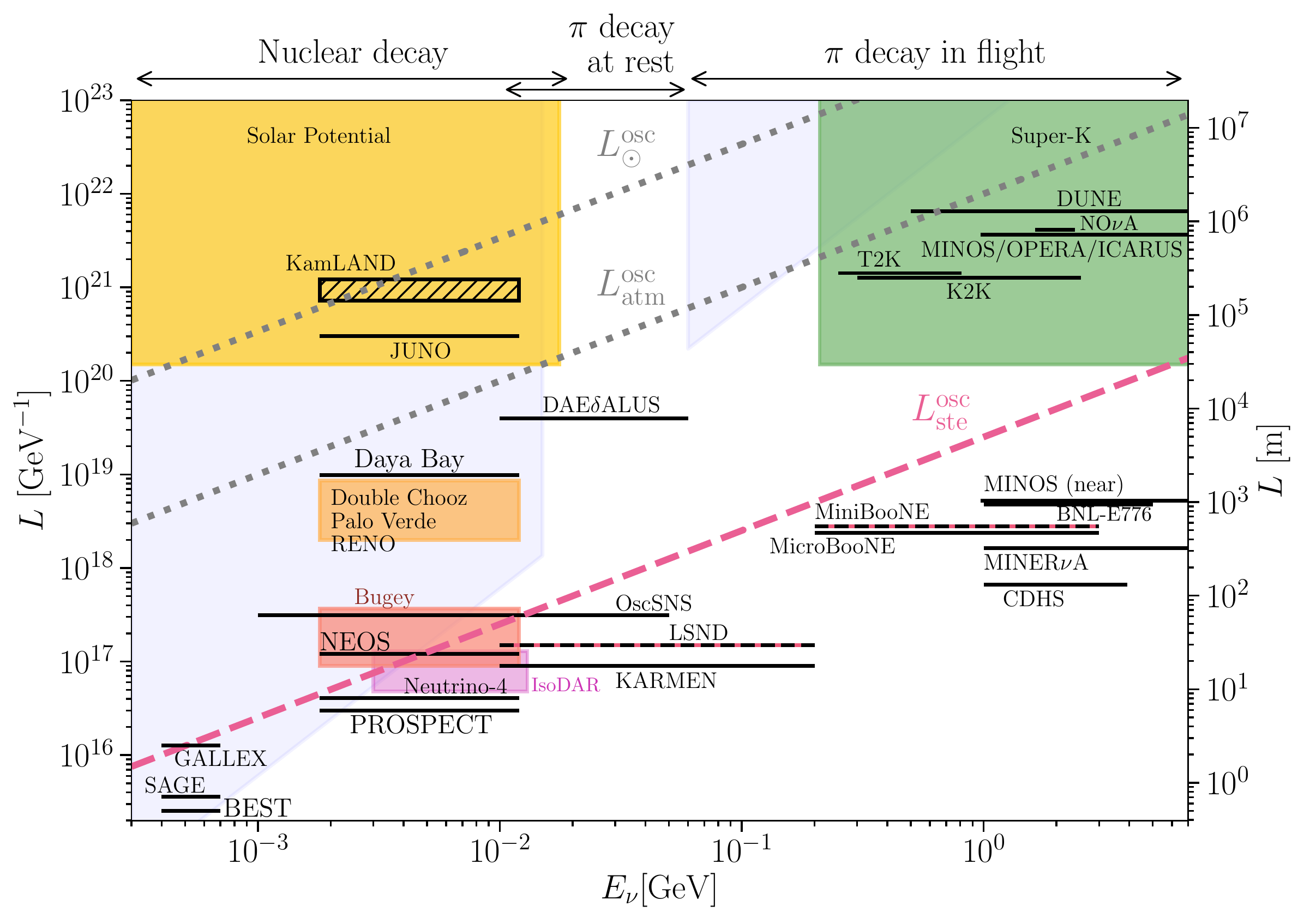}

  \caption{Overview of experimental source-detector baselines (L) and neutrino energies (E) sampled by neutrino experiments worldwide; adapted from Ref~\cite{Arguelles:2022bvt}.}
  \label{fig:LEFig}
\end{figure}

\begin{figure}[!h]
  \centering
  \includegraphics[trim=0.3cm 10.1cm 40.0cm 8.0cm, clip=true, width=0.65\textwidth]{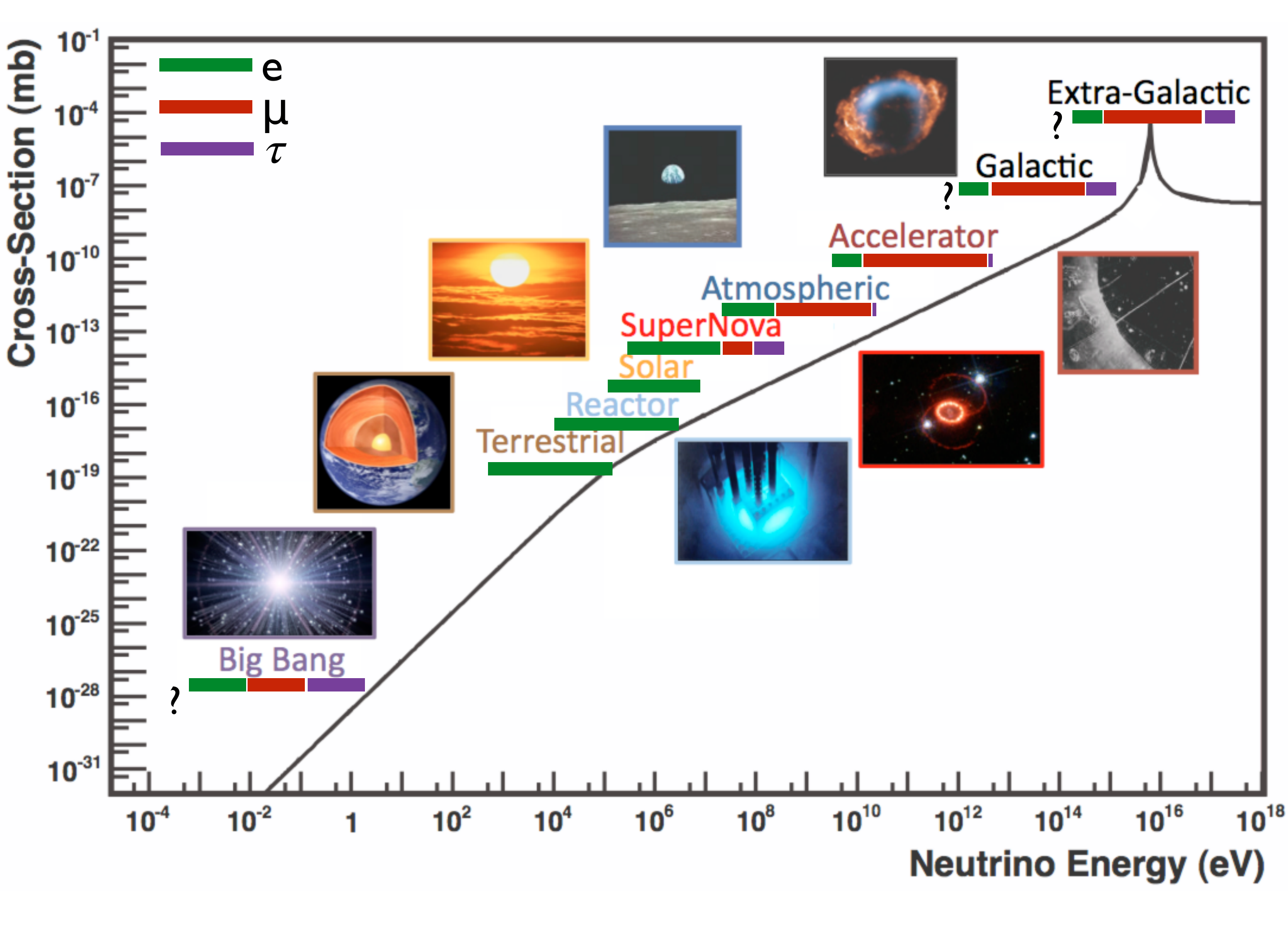}

  \caption{Approximate flavor composition of commonly discussed neutrino sources; adapted from~\cite{Formaggio:2012cpf}.  
  Reactor experiments are notable in their use of lower energy neutrinos, their access to very short baselines, and their extreme electron flavor purity.}
  \label{fig:FlavorFig}
\end{figure}

Reactor \nuebar data plays a variety of essential roles in performing future Standard Model and BSM oscillation measurements vital to the US neutrino community.  
Their power and complimentary position in the global landscape is well illustrated in Figures~\ref{fig:LEFig} and~\ref{fig:FlavorFig}.  
As they sample lower neutrino energies than most other efforts (Figure~\ref{fig:LEFig}), reactor experiments can feasibly access all $\Delta m^2$ ranges of interest in current oscillation studies with a single source type.  
They also sample a pure flux of electron-flavor neutrinos (Figure~\ref{fig:FlavorFig}), enabling  particularly clean tests of specific mixing parameters.  
Since lower energies in reactor experiments are also accompanied by shorter baselines, reactor-based oscillation tests are also less influenced by some commonly-studied neutrino sector BSM effects, such as non-standard matter interactions or heavy-mediator couplings between neutrinos and hidden sectors.  

In the context of the today's US neutrino program, one of reactor experiments' most prominent roles is in testing the origin of anomalies observed by short-baseline neutrino experiments.  
This topic addresses two of the five Science Drivers identified in the 2014 P5 report~\cite{P5}.  
Many of these persistent anomalies rest in the electron flavor realm, where reactor experiments, in particular, excel.  
For example, the BEST experiment recently confirmed the robustness of the so-called 'Gallium Anomaly' by detecting a $\sim$20\% deficit in observed interactions of sub-MeV $\nu_e$ generated by an intense radioactive source~\cite{Barinov:2021asz}.  
Even MicroBooNE, which primarily samples $\nu_{\mu}$ from Fermilab's BNB beamline, has attracted attention with weak hints of a deficit of $\nu_e$ interactions~\cite{MicroBooNE:2021rmx}, prompting further theoretical examination of sterile-mediated electron-flavor disappearance~\cite{Denton:2021czb,Arguelles:2021meu}.  
This recent result contrasts with long-standing MiniBooNE results showing an excess of $\nu_e$-like events in the same beamline~\cite{MiniBooNE:2020pnu}.  
%, while in the reactor sector, the Neutrino-4 experiment provided moderate, albeit controversial~\cite{}, hints for electron-flavor disappearance with an L/E characteristics.
%BEST is an excellent example.  
%Neutrino-4 is another.  
%MiniBooNE and LSND provide nue and nuebar-like appearance indications in numu sources of varying flavor purity.  
%Some are even arguing for hints of nue DISappearance in the BNB from MicroBooNE, although others disagree about whether such a hint is statistically significant.  

Future aspects of the US neutrino program, such as Fermilab SBN~\cite{sbn}, will certainly fight in the following decade to elucidate the causes of these and other anomalies.  
However, when viewed in the canonical BSM framework of oscillations between three active neutrino states and one additional sterile state (3+1), it seems likely that conclusively demonstrating the consonance or dissonance of these varied datasets will be a challenging task.  
Datasets that test, with maximal clarity, the 3+1 oscillation interpretation in specific channels and suggested phase space regions are a particularly important ingredient in this effort.  
Short-baseline reactor neutrino experiments, with their well-defined flavor profile, straightforward energy reconstruction, and purely relative analysis methods, offer an ideal experimental arrangement for  targeted, clear tests.  
For these reasons, reactor experiments are a crucial piece of a diverse future global effort capable of elucidating whether or not a 3+1 model is an acceptable solution to the various short-baseline anomalies.  

It is also possible that the observed short-baseline anomalies are instead explained by a hidden sector physics scenario more complex than the canonical 3+1 model, such as one with multiple sterile neutrinos (3+N)~\cite{bib:kopp}, sterile neutrino decay~\cite{decay_machado,decay_deg}, NSI~\cite{nsi}, hidden sector couplings~\cite{Batell:2009di}, or some combination of effects~\cite{decay_conrad}.  
If this is the case, data from diverse channels, energies, and sources will be even more crucial for disentangling the different contributing effects, as each effect may or may not manifest itself differently in specific experimental regimes.  
In the stable of global measurements, short-baseline reactor oscillation measurements are unique in their capability to very purely probe sterile oscillation effects.  
As mentioned above, this is due to the lower energies involved in interactions and decays in the reactor, which prohibits production and decay of heavier hidden-sector particles, and their very short baselines, which minimize the impact of NSI.  

Short-baseline reactor experiment results also have particular relevance to upcoming measurements of Standard Model neutrino properties.  
Theoretical studies have pointed to specific regions of 3+1 phase space that could complicate interpretation of DUNE and other future US long-baseline neutrino measurements~\cite{Klop:2014ima,deGouvea:2014aoa}.  
For example, a sterile sector with specific combinations of non-zero active and sterile CP violating phases could mimic CP-conserved signatures in DUNE \cite{Kayser}.  
Parameter degeneracies can be avoided for DUNE if separate measurements are used to constrain the level of active-sterile mixing; scenarios like the one above can be avoided if limits on $\theta_{14}$ and $\theta_{24}$ can be improved to approximately the $5^{\circ}$ (sin$^2$2$\theta$ = 0.03) level~\cite{KayserVal}.  
$\theta_{14}$ limits meeting this stringent requirement are only accessible with intense electron-flavor sources, such as reactors and tritium decay facilities like KATRIN and Project-8~\cite{katrin_sterile}.  
Thus, reactor \nuebar experiments play a synergistic role in enabling clear interpretations of the neutrino community's centerpiece experiment, DUNE, and its physics centerpiece, measurement of leptonic CP-violation.  

Medium- and long-baseline reactor oscillation measurements are also crucial in extending the US Standard Model neutrino oscillation  measurements program.  
It should first be emphasized that reactor-based measurements of a large $\theta_{13}$ value paved the way for DUNE by demonstrating that CP-violation measurements are feasible with conventional neutrino beams.  
In the near future, Daya Bay's still-improving limits on $\theta_{13}$ remain essential in current accelerator-based probes of CP-violation with T2K and NOvA~\cite{NOvA:2021nfi,T2K:2021xwb}, and later, when included in DUNE fits, they will modestly enhance DUNE's oscillation parameter measurement precision~\cite{Bass:2013vcg}.  
Approached from a different perspective, comparisons of Daya Bay's and DUNE's independently-measured $\theta_{13}$ values can be directly compared to yield tests of unitary in the PMNS mixing matrix~\cite{unitarity_qian,unitarity_ellis}.  
In the solar sector, JUNO, along with DUNE, are the primary pieces in a future program for sharpening our view of tensions in solar- and reactor-derived measurements of $\Delta m^2_{21}$~\cite{Capozzi:2018dat}; if such a discrepancy persists in these higher-precision experiments, it could provide the first clear evidence for non-standard neutrino interactions~\cite{nsi}.  
Last but not least, JUNO will measure the mass hierarchy independently of other experiments~\cite{JUNO:2022hxd}, providing unique information on a parameter that is extremely important across many branches of neutrino physics, including neutrinoless double beta decay and neutrino mass experiments, as well as DUNE's long-baseline oscillation~\cite{DUNE:2020jqi,DUNE:2021mtg} and supernova neutrino burst~\cite{DUNE:2020zfm} physics programs.  

%At the lowest reactor neutrino energies, Rx coherent scattering experiments exhibit synergies with similar measurements using decay-at-rest sources.  
%BLAH: sentence about the specific physics topics.  
%Relationship between RxCEvNS nuclear physics measurements and modelling of some neutrino production sources, like supernovae?  

\begin{comment}
\textcolor{blue}{Key takeaways:}
\begin{itemize}
    \item{Compelling synergies exist that strengthen other parts of the US program, in addition to pursuing good physics in own right.}

    \item{HEP: Reactors provide an artificial source of neutrinos that is unique in its electron-like purity, in its intensity, and in its low energy.  These featues enable it to play a unique or complementary role in standard and BSM physics.  Be more explicit.}
    
    \end{itemize}
\end{comment}

\section{Synergies with the Broader US Science Program}
\label{sec:synergies_broad}

\subsection{Key Takeaways}
\begin{itemize}
    \item \emph{Nuclear physics}: Nuclear reactors' antineutrinos provide a novel source of information regarding short-lived, high-Q isotopes whose properties are in some cases poorly understood.  

    \item \emph{Applications}: HEP-oriented reactor neutrino detector technologies and techniques are highly relevant to future antineutrino-oriented safeguards, reactor exclusion and reactor monitoring use cases, as well as neutron and gamma-ray detection.

    \item \emph{US Workforce Development}: Compared with large international experiments, the relatively small size and fast timescale from design to data-taking of reactor experiments provide an inviting workforce development opportunity, enabling the realization of versatile skillsets for new generations of young nuclear and particle physicists. Additionally, antineutrino-based applications at all scales offer a unique opportunity for collaborative engagement between applications-focused and basic science-focused community members.

    \item \emph{US Facility Enhancement}: Reactor neutrino experiments enable the use of crucial US facilities for purposes beyond their initially intended or envisioned scope, which strengthens the scientific interest and vitality of these facilities.  

%     \item{ Use of unique US-facilities; vitality of US-based physics program.}

\end{itemize}

\subsection{Narrative}

%     \item{ Workforce development.}

% \end{itemize}

Efforts pursing neutrino physics goals using reactor \nuebar{} have many benefits to and synergies with the broader scientific community of the United States and the World. These range across the direct contribution of important scientific knowledge, development of cross-cutting technologies and facilities, enabling applications with significant societal impact, and the development of a highly skilled workforce. 

Beyond the high energy physics  topics mentioned in the previous section, reactor \nuebar{} measurements can contribute to other fields of scientific enquiry. As described in Sec.~\ref{sec:source}, the \nuebar{} emissions from a reactor provide a probe of the nuclear fission process that is complementary to other techniques that measure more readily accessible particles like fission fragments, gamma-rays and neutrons. Specifically, the reactor \nuebar{} energy spectrum encodes information about fission product yields and the energy spectrum of beta-decays of those fission daughters. Included in the total \nuebar{} spectrum are contributions from short-lived, high Q-value isotopes, some of which have received limited experimental investigation.  High statistics and high precision \nuebar{} spectrum measurements therefore have the potential to test the nuclear data evaluations that underlie many areas of nuclear physics, nuclear energy, and nuclear security. Nuclear data needs and benefits that can be addressed with reactor \nuebar{} have been described in recent workshops and reports~\cite{bib:IAEA,bib:WoNDRAM}.

Advances in scientific knowledge regarding neutrino production in nuclear reactors and characterizing such nuclear systems themselves also underlie another significant societal benefit of reactor \nuebar{} studies. As described in Sec.~\ref{sec:applications} and Ref.~\cite{Bernstein:2019hix}, the \nuebar{} emitted by operating nuclear reactors
%spent nuclear fuel may be useful in cooperative nonproliferation applications such as monitoring fissile material production in reactors, discovery or exclusion of undeclared reactors, and monitoring of spent fuel and reprocessing facilities.  
% and spent nuclear fuel and reprocessing facilities can be used to monitor the operation of these facilities to verify that they are being used in a manner consistent with a nation's international obligations. 
and spent nuclear fuel may be useful for cooperative nonproliferation applications such as monitoring fissile material production in reactors, exclusion of undeclared reactors, and monitoring of spent fuel and reprocessing facilities.

A recent study focused on the potential utility of \nuebar{} for nuclear energy and nuclear security applications elucidates some of the relevant characteristics of these particles and potential use cases for them~\cite{Akindele:2021sbh}. The highly penetrating nature of neutrinos poses detection and implementation challenges in the context of monitoring applications, but also holds promise as a non-intrusive technique that does not require direct access to complex and/or sensitive facilities. Considering user need and constraints, forthcoming advanced reactor types for which nuclear safeguards techniques are still be developed and nuclear security deals between nations were found to be promising use cases for \nuebar{} monitoring measurements. 

Of course, potential applications of \nuebar{} depend heavily upon the detection tools and techniques developed by neutrino physics experiments. All application oriented demonstrations of reactor \nuebar{} have been enabled by the multi-decade succession of reactor \nuebar{} scientific experiments that have preceded them~\cite{Bernstein:2019hix}. Recent advances like aboveground  \nuebar{} detection without substantial  overburden~\cite{prospect_osc,Haghighat:2018mve} have greatly broadened the range of applications that can be considered. Since neutron identification is central to detection of the IBD interactions, materials and techniques developed for reactor \nuebar{} also have significant potential for neutron detection in support a wide range of nuclear security applications~\cite{doi:10.1063/1.3503495}.

Beyond the scope of the US neutrino oscillation physics program and potential applications, reactor \nuebar experiments continue to develop technologies well-suited for other areas in neutrino and particle physics.  
For example, technology being developed to enable detection of low-energy signals from coherent neutral current nuclear scattering of reactor \nuebar addresses similar challenges  to those needed to seek dark matter interactions with electrons and nuclei~\cite{connie,ricochet,NF10:MINER:2016igy}.  
Doped aqueous, plastic, or opaque scintillator technology used for reactor IBD detection may offer value in other sectors of the US neutrino physics program, such as in neutrinoless double beta decay experiments~\cite{SNO:2021xpa,NF10:LiquidO:2019}, measurements of neutrino-induced neutron production~\cite{ANNIE:2017nng}, and future water-based DUNE far detector modules~\cite{Theia:2019non}. 
For these and other cases described in Section~\ref{sec:detectors}, synergies clearly exist between the pursuit of reactor neutrino detection and other aspects of the US particle physics program.  

The training and mentoring of a skilled, creative, and diverse workforce is not only essential to the future of HEP, but it is also one of the primary societal benefits that justifies public investment in this field. Reactor \nuebar{} experiments are an especially effective training ground for producing highly skilled and well rounded scientists. Due to their relatively small size, they offer young scientists the rare opportunity to experience the experimental process from the idea and design stage to data taking and analysis. Experiments of order of 5 years duration offer invaluable training opportunities matched to the research timescale of postdocs (3 years) and graduate students (4-6 years). In addition, the relatively smaller size of collaborations of these experiments offer supportive and nurturing environments that are complementary to the opportunities found in large, international collaborations and reduce the threshold for early career scientists to get involved.  

More broadly, antineutrino-based applications offer a unique opportunity for collaborative engagement between applications-focused and basic science-focused community members.  
Since a key programmatic goal of the US HEP program is workforce training for US National Laboratories, opportunities for bridging between applied and fundamental science research should be fostered wherever it is possible to do so.  
By engaging in antineutrino applications research, applications and fundamental physics researchers can work side by side in performing technology development, engineering and deployment, and data analysis.  

%The final point to be noted in this section is that reactor \nuebar{} experiments typically broaden the research program of their host facilities, thereby contributing to the vitality of the overall U.S. scientific enterprise. By drawing on the skills of reactor staff and nuclear engineers,  

\section{Three-Neutrino Oscillation Physics with Reactors (NF01)}
\label{sec:lbl}

\subsection{Key Takeaways}
\begin{itemize}
    \item Nuclear reactors are a powerful, abundant, cost-effective, and well-understood source of antineutrinos, and have been used to make some seminal measurements in neutrino oscillations.
    \item In the next half-decade, reactor antineutrino experiments are expected to provide the world’s best estimates for the foreseeable future of 4 out of 6 oscillation parameters: \begin{itemize}
        \item $\sin^2 \theta_{13}$ with Daya Bay, Double Chooz and RENO. 
        \item $\Delta m^2_{31}$, $\Delta m^2_{21}$, and $\sin^2 \theta_{12}$ with JUNO (with sub-half-percent precision).
    \end{itemize}
    \item Reactor antineutrinos in JUNO will also enable an independent measurement of the mass ordering with very different baseline, energy, backgrounds, and detector systematic uncertainties to what other experiments will do. 
\end{itemize}

\begin{comment}
Outline:
\begin{itemize}

\item give PMNS picture, and describe oscillations.

\item Summarize previous things learned: KamLAND and theta13 measurements via IBD-based measurement.

\item Summarize wrap-up of Daya Bay, and ultimate sensitivity prospects.

\item Summarize JUNO's prospects for MH measurement

\item Summarize JUNO's prospect for theta12 / dm2 measurement.  

\item (maybe make a short mention of how sterile couplings could affect DUNE and other experiments' ability to measure $\Delta_{CP}$, and point readers to the next section?)

\item TABLE: Current mixing angle / dm2 precisions; future precision versus calendar year?

\end{itemize}
\end{comment}

\subsection{Narrative}

%Nuclear reactors are one of the most powerful, abundant, cost-effective, and well-understood sources of neutrinos. Pure electron antineutrinos are emitted in a reactor core from the beta decays of the nuclear fission products. On average, about $6\times10^{20}$ $ \bar\nu_e$'s are produced every second from a typical 3~GW$_\textrm{th}$ commercial reactor core. 
%This intense human-controllable neutrino source enabled the first experimental detection of neutrinos in 1959 by Reines and Cowan~\cite{cowan1956}. 

In recent years, reactors have played a major role in the study of neutrino oscillations and helped establish the three-neutrino oscillation framework that still stands as the leading paradigm of this phenomenon~\cite{Vogel:2015wua,ParticleDataGroup:2020ssz}. In this section, we review the theory, experiments, and prospects of three-neutrino oscillation physics with reactors.

In the Standard Model of particle physics, three neutrino flavors, $\nu_e$, $\nu_\mu$, and $\nu_\tau$, participate in the weak interaction. However, if neutrinos have a non-zero mass, the flavor composition of a neutrino beam could change as the neutrinos propagate in space. This phenomenon is called neutrino oscillations and is a quantum mechanical effect stemming from the fact that a neutrino with a definite flavor need not have a definite mass. In fact, a neutrino flavor eigenstate can be viewed as a linear superposition of the neutrino mass eigenstates, $\nu_1$, $\nu_2$, and $\nu_3$:
\begin{equation}\label{eq:pmns_matrix}
    \left( \begin{array}{c} \nu_{e} \\ \nu_{\mu} \\ \nu_{\tau} \end{array} \right) = 
    \left ( \begin{array}{ccc} U_{e1} & U_{e2} & U_{e3} \\
        U_{\mu1} & U_{\mu2} & U_{\mu3} \\
        U_{\tau1} & U_{\tau2} & U_{\tau3} 
    \end{array} \right) \cdot \left( \begin{array}{c} \nu_{1} \\ 
        \nu_{2} \\ \nu_{3} \end{array} \right).
\end{equation}
The unitary $3\times3$ mixing matrix, $U$, is called the Pontecorvo-Maki-Nakagawa-Sakata (PMNS) matrix and can be parameterized by three mixing angles, $\theta_{12}$, $\theta_{13}$, $\theta_{23}$, and one CP-violation phase, $\delta_{CP}$\footnote{There are two additional phases if neutrinos are Majorana particles, but they do not play a role in neutrino oscillation experiments.}:
\begin{equation}\label{eq:pmns_rot}
    U_{\textrm{PMNS}} = 
    \left( \begin{array}{ccc}
        1 & 0 & 0 \\
        0 & c_{23} & s_{23}  \\
        0 & -s_{23} & c_{23} 
    \end{array} \right)
    \left( \begin{array}{ccc}
        c_{13} & 0 & s_{13} e^{-i\delta_{CP}} \\
        0 & 1 & 0 \\
        -s_{13} e^{i\delta_{CP}} & 0 & c_{13} 
    \end{array} \right)
    \left( \begin{array}{ccc}
        c_{12} & s_{12} & 0 \\
        -s_{12} & c_{12} & 0 \\
        0 & 0 & 1 
    \end{array} \right),
\end{equation}
where the notation $c_{ij} = \cos\theta_{ij}$,  $s_{ij} = \sin\theta_{ij}$ is used.

As neutrinos travel a certain distance $L$ in vacuum, their mass eigenstates with energy $E$ develop a phase such that $\nu_i(L) = e^{-i\frac{m_i^2}{2E}L} \cdot \nu_i(0)$. Given the neutrino mixing formula in Eq.~\eqref{eq:pmns_matrix}, the probability of a neutrino with flavor $l$ transforming to a different flavor $l^\prime$ can be written as:
\begin{eqnarray}\label{eq:osc_dis}
    P_{\nu_{l}\rightarrow \nu_{l^\prime}} &=& |<\nu_{l^\prime}(L)|\nu_{l}(0)>|^2  \nonumber  \\
    &=&   \left |\sum_{j} U_{lj}U^{*}_{l'j}e^{-i\frac{m_j^2}{2E}L} \right | ^2 \hfill \nonumber \\ 
    &=&  \sum_{j}|U_{lj}U^*_{l'j}|^2 +  \sum_{j} \sum_{k \neq j} U_{lj} U^{*}_{l'j} U^{*}_{lk} U_{l'k} e^{i\frac{\Delta m^2_{jk} L}{2E}},  
\end{eqnarray}
where $\Delta m^2_{jk} = m^2_j - m^2_k$ are the mass-squared differences between mass eigenstates.

\begin{figure}[!htb]
  \centering
  \includegraphics[width=0.9\textwidth]{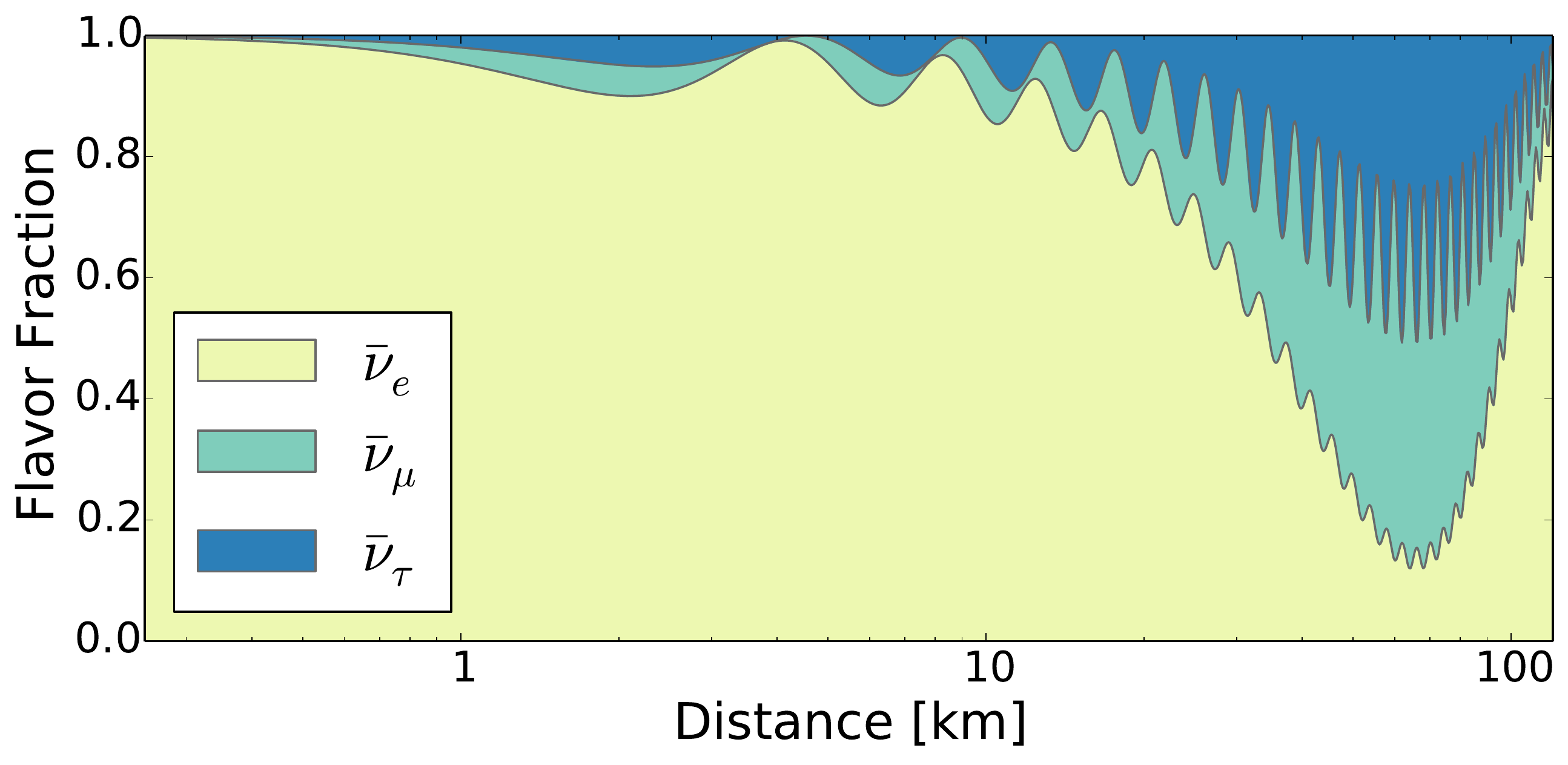}
  \caption{Expected flavor composition of the reactor antineutrino flux as a function of distance to a reactor core for neutrinos of 4~MeV energy. Figure taken from Ref.~\cite{Vogel:2015wua}. The light yellow region corresponds to the survival probability of $\bar\nu_e$ that reactor antineutrino experiments can measure by placing their detectors at different baselines.
  \label{fig:psur_distance_vogel}}
\end{figure}

Since nuclear reactors produce only electron antineutrinos, $\bar\nu_e$, with energy below about 9 MeV that is lower than the production threshold of a muon or a tau lepton, the experimental observation of neutrino oscillations is typically through the disappearance channel. Namely, the $\bar\nu_e$ neutrino flux is measured at some distance $L$ away from the reactor, and the survival probability $ P_{\bar\nu_{e}\rightarrow \bar\nu_{e}}$ is calculated by comparing to the flux near the source. Given Eq.~\eqref{eq:osc_dis}, this survival probability can be expressed as:
\begin{align}\label{eq:dis_osc}
    P_{\bar{\nu}_e \rightarrow \bar{\nu}_e} 
    &= 1 - 4|U^2_{e1}||U^2_{e3}|\sin^2 \Delta_{31} -
    4|U^2_{e2}||U^2_{e3}|\sin^2\Delta_{32} - 4|U^2_{e1}||U^2_{e2}|\sin^2\Delta_{21} \nonumber \\
    &= 1-\sin^{2}2\theta_{13}(\cos^{2}\theta_{12}\sin^{2}\Delta_{31}+\sin^{2}\theta_{12}\sin^{2}{\Delta_{32}}) 
    -\cos^{4}\theta_{13}\sin^{2}2\theta_{12}\sin^{2}\Delta_{21},
\end{align}
where the notation $\Delta_{ij} = \frac{\Delta m^2_{ij}L}{4E}$ is used. From Eq.~\eqref{eq:dis_osc} we see that reactor antineutrino disappearance is a clean channel that is only dependent on $\theta_{12}$, $\theta_{13}$, $\Delta m^2_{21}$, $\Delta m^2_{31}$, and the neutrino mass ordering, making it ideal for precision measurements of these oscillation parameters.
Fig.~\ref{fig:psur_distance_vogel} shows the survival probability as a function of the travel distance $L$ for a typical 4~MeV reactor $\bar\nu_e$. The large disappearance at $\sim$60 kilometers is driven by the solar-mixing mass scale $\Delta m^2_{21}$ and its corresponding large mixing angle $\theta_{12}$, while the smaller disappearance at $\sim$2 kilometers is caused by the atmospheric-mixing mass scale $\Delta m^2_{31} \sim \Delta m^2_{32}$ and the small mixing angle $\theta_{13}$. The two very different $\Delta m^2$ scales benefit designs of reactor antineutrino oscillation experiments, which can isolate the parameters of interest and improve the precision of their determination by placing detectors at strategic baselines. 

The first reactor antineutrino experiment that observed an evidence of neutrino oscillations is the KamLAND experiment~\cite{KamLAND_rate}, built in the early 2000s in Japan.  
The KamLAND experiment was prompted by the ``Solar Neutrino Problem'', which refers to the observation that the $\nu_e$ flux from the Sun is less than a half of the prediction from the Standard Solar Model~\cite{bib:solar}. 
The theory of neutrino oscillations provides an elegant solution to the solar neutrino problem, and can be tested on Earth using reactor antineutrinos assuming \emph{CPT} invariance. 
The KamLAND experiment is located in the middle of Japan, surrounded by 55 Japanese reactor cores with a flux-weighted average baseline of $\sim$180 kilometers. 
As shown in Fig.~\ref{fig:psur_distance_vogel}, at this baseline, the KamLAND experiment is sensitive to the solar-mixing parameters $\Delta m^2_{21}$ and $\theta_{12}$, and benefits from a better understood neutrino source and simpler vacuum oscillation formula compared to solar neutrino experiments. 
The KamLAND detector uses one kiloton of liquid scintillator as the target volume, which is contained in a 13-meter-diameter transparent balloon surrounded by a mineral oil region containing 1,879 photomultiplier tubes (PMTs). 
%The balloon is surrounded by mineral oil housed in a 18-meter-diameter stainless steel sphere, where 1325 17-inch and 554 20-inch PMTs are mounted. 
%The mineral oil shields the target volume from external radioactivity mainly from the PMTs, and this setup provides an excellent energy resolution of 6.5\%/$\sqrt{E\textrm{(MeV)}}$ for measuring the reactor neutrino spectrum. 
%The event rate is very low at KamLAND, less than one reactor neutrino measured per day, because of the long distance to the reactors.   %Nevertheless, the results are indisputable after enough statistics are accumulated. 
The results in 2008 observed a total of 1609 events with a 2.9 kton-year exposure, which was only about 60\% of the predicted signal if there were no oscillations~\cite{KamLAND_2008}. The calculated survival probability shows a clear oscillatory pattern as a function of $L/E_{\nu}$, a smoking gun evidence of the existence of neutrino oscillations. The results were also highly consistent with solar neutrino experiments. When combined with the results from SNO~\cite{SNO}, they provided the most precise measurement of  $\tan^2\theta_{12} = 0.47^{+0.06}_{-0.05}$ and $\Delta m^2_{21} = 7.59^{+0.21}_{-0.21} \times 10^{-5}$ eV$^2$ to date~\cite{KamLAND_2008}.

The first generation of reactor $\theta_{13}$ experiments, CHOOZ~\cite{bib:chooz} and Palo Verde~\cite{bib:paloverde}, did not observe $\bar\nu_e$ disappearance from reactors and only an upper limit of sin$^22\theta_{13}<0.10$ at 90\% C.L. was set. In the 2000s, a new generation consisting of Daya Bay~\cite{DayaBay:2015kir}, Double Chooz~\cite{DoubleChooz:2022ukr}, and RENO~\cite{bib:reno}, was initiated to measure the small mixing angle $\theta_{13}$.

Given the mass-scale $\Delta m^2_{31}$ suggested by the atmospheric neutrino experiments, the corresponding baseline for reactor antineutrino experiments is about 1--2 kilometers, as indicated in Fig.~\ref{fig:psur_distance_vogel}.  
All experiments adopted the strategy of performing a relative measurement between near and far functionally identical detectors to largely suppress the reactor and detector related systematic uncertainties. 
After some early indications in 2011~\cite{bib:dc,T2K:2011ypd,MINOS:2011amj}, all three experiments reported clear evidence of $\bar\nu_{e}$ disappearance in 2012 with a few month's data taking~\cite{DoubleChooz:2012gmf,bib:prl_rate,bib:reno}. 

Among these experiments, the Daya Bay experiment, being the most sensitive one, excluded $\theta_{13}=0$ at $5.2\sigma$~with 55 days of data~\cite{bib:prl_rate}.  
The Daya Bay experiment is located near the six reactors of the Daya Bay nuclear power plant in southern China, with a total reactor power of 17.4 GW$_\mathrm{th}$. Daya Bay uses eight identical antineutrino detectors (ADs), with two ADs at $\sim$360 m from the two Daya Bay reactor cores, two ADs at $\sim$500 m from the four Ling Ao reactor cores, and four ADs at a far site $\sim$1580~m away from the 6-reactor complex. 
Each AD contains 20-tons of gadolinium-loaded liquid scintillator as the target volume. 
%This setup provides approximately 700 reactor neutrino events in each near AD and 70 events in each far AD per day. 
Each AD's target is viewed by 192 8-inch PMTs that yield an energy resolution of 8.5\%/$\sqrt{E\textrm{(MeV)}}$, allowing a precise measurement of the reactor antineutrino energy spectrum that enables the the observation of a spectral distortion between far and near detectors as expected from neutrino oscillations.  
In 2018 results, Daya Bay reported detection of nearly 3.5 million reactor antineutrino events in the near detectors and 500 thousand events in the far detectors over 1958 days of data collection.  
The comparison of relative $\bar\nu_e$ event rates and energy spectra among detectors is consistent with the three-neutrino oscillation formula as introduced in Eq.~\eqref{eq:dis_osc} and yields $\sin^2\theta_{13} = 0.0856\pm0.0029$ and $\Delta m^2_{32} = 2.471^{+0.068}_{-0.070} \times 10^{-3}$ eV$^2$ assuming the normal mass ordering, and $\Delta m^2_{32} = -(2.575^{+0.068}_{-0.070}) \times 10^{-3}$ eV$^2$ assuming the inverted mass ordering~\cite{DayaBay:2018}.  
The remarkable precision makes $\theta_{13}$ the most precisely measured angle among the three neutrino mixing angles in the PMNS matrix, despite being the last known mixing angle to be non-zero. The full data set of Daya Bay from 2012--2020, with over 6 million events, is the largest library of reactor antineutrino events collected in history, and is expected to further improve the precision of $\sin^2\theta_{13}$ and $\Delta m^2_{32}$ to better than 2.5\% and 2\%, respectively. Thanks to the consistent results reported by Double Chooz~\cite{DoubleChooz:2020vtr} and RENO~\cite{RENO:2018dro}, reactor experiments are providing robust and precise constraints to other experiments, including those searching for leptonic CP violation~\cite{NOvA:2021nfi,T2K:2021xwb,DUNE:2020jqi,Hyper-Kamiokande:2018ofw}.

\begin{figure}[!h]
  \centering
  \includegraphics[width=0.55\textwidth]{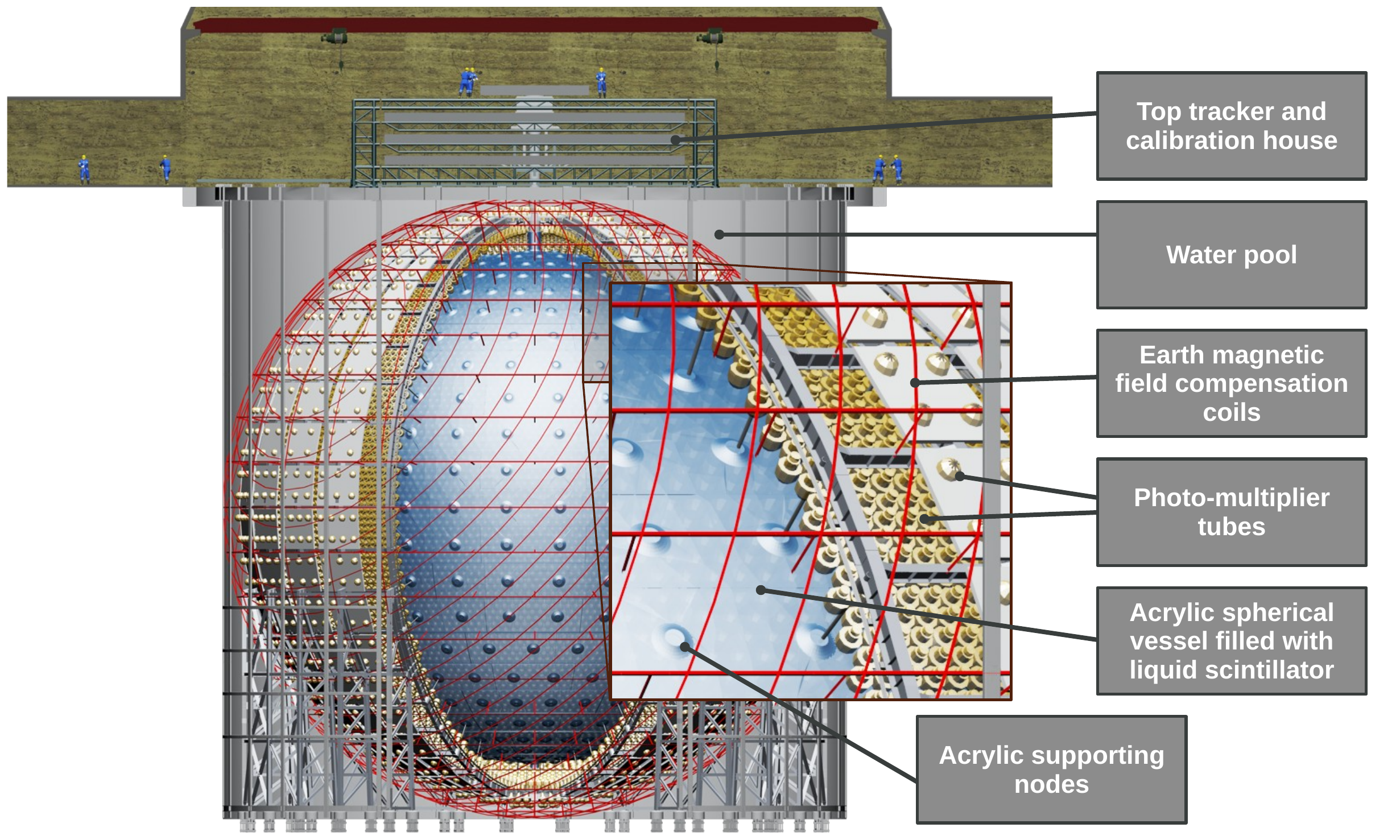}
  \includegraphics[width=0.42\textwidth]{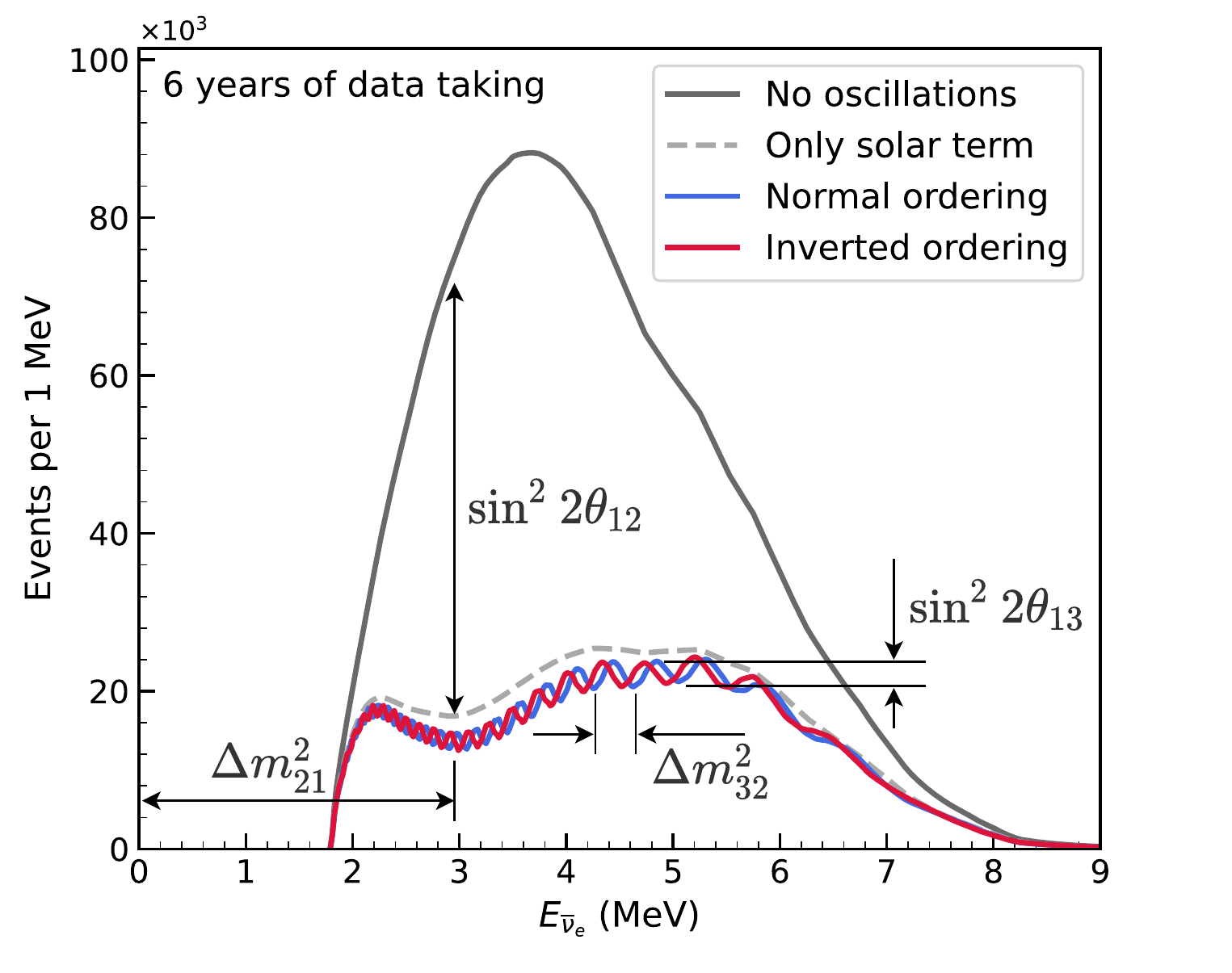}
  \caption{Left: Schematic of the JUNO detector. An acrylic sphere containing 20 kilotons of liquid scintillator serving as the \nuebar detection target is surrounded by 20-inch and 3-inch PMTs. %The sphere is surrounded by an instrumented water pool and field compensating coils that suppress the effects of the Earth's magnetic field on the 20-inch PMTs. 
  Right: JUNO IBD spectrum with and without neutrino oscillation effects. For illustration purposes, a detector with perfect energy resolution is assumed. The gray dashed curve shows the oscillated spectrum when only the term in the disappearance probability that is modulated by $\sin^2 2\theta_{12}$ is included, whereas the blue and red curves show it when the full oscillation probability in vacuum is used assuming the normal and inverted mass orderings, respectively. Some features driven by the $\sin^2 2\theta_{12}$, $\sin^2 2\theta_{13}$, $\Delta m^2_{31}$ and $\Delta m^2_{21}$ oscillation parameters are shown pictorially. Figures obtained from Ref.~\cite{junooscdiana}.}
  %, illustrating the rich information available in a high-resolution measurement of the oscillated spectrum at JUNO's baseline.}
  \label{fig:junodetspectrum}
\end{figure}

%The $\theta_{13}$ mixing angle will not be the only parameter that is best measured by a reactor neutrino experiment. 
Beyond completion of these $\theta_{13}$ experiments, reactor antineutrino experiments continue to be at the forefront of neutrino oscillation physics. The Jiangmen Underground Neutrino Observatory (JUNO) is currently under construction in southern China and is expected to come online in 2023~\cite{JUNO:2022hxd}. JUNO will be located at a baseline of $\sim$52.5~km from six 2.9~GW$_\mathrm{th}$ nuclear reactor cores in the Yangjiang Nuclear Power Plant (NPP) and two 4.6~GW$_\mathrm{th}$ cores in the Taishan NPP. As shown in Fig.~\ref{fig:junodetspectrum}, JUNO's central detector (CD) will consist of 20 kilotons of liquid scintillator contained in an acrylic sphere immersed in water and surrounded by $17,612$ 20-inch and $25,600$ 3-inch PMTs providing more than $75\%$ optical coverage. 
This central region will be supported by an external water Cherenkov veto detector, and a detector-top cosmic veto tracker and calibration house.  
%The CD will be optically decoupled from a surrounding 35 kiloton ultrapure water tank that serves as a Cherenkov veto detector and as a shield against natural radioactivity from the rock and neutrons from cosmic rays. 
%The CD is also surrounded by two sets of large coaxial coils running in different directions whose purpose is to suppress the effects of the Earth's magnetic field on the 20-inch PMTs. 
%The water pool is partially covered by a top tracker consisting of 3 layers of plastic scintillators. A calibration house is used to store and deploy calibration equipment into the detector. 

JUNO will see an unparalleled amount of light for a detector of this type, amounting to over 1,300 photoelectrons per MeV. This, in combination with a comprehensive calibration program that includes the 3-inch PMT system as a handle to assess any instrumental non-linearities in the 20-inch PMT system~\cite{JUNO:2020xtj}, will result in an energy resolution of 3\% at 1~MeV. The unprecedented detector size and energy resolution will allow to simultaneously observe the effects of both the solar and atmospheric oscillations for the first time. As illustrated on the right panel of Fig.~\ref{fig:junodetspectrum}, the former produces a ``slow" oscillation modulated by $\sin^2 2\theta_{12}$ with frequency $\Delta m^2_{21}$, while the latter causes a ``fast" oscillation modulated by $\sin^2 2\theta_{13}$ with frequency $\Delta m^2_{32}$. As also illustrated in Fig.~\ref{fig:junodetspectrum}, the oscillated spectrum changes slightly depending on the neutrino mass ordering, providing sensitivity to this parameter. This difference is caused by the interference effects that occur between the $\Delta m^2_{31}$ and $\Delta m^2_{32}$ terms in the oscillation probability of Eq.~\ref{eq:dis_osc}, which depend on the sign of $\Delta m^2_{31}$. Knowledge of the unoscillated reactor antineutrino spectrum is important for JUNO's physics goals, so the collaboration will deploy a satellite detector at a baseline of $\sim$30~m from one of the Taishan 4.6~GW$_\mathrm{th}$ cores called the Taishan Antineutrino Observatory (JUNO-TAO)~\cite{juno_tao}. JUNO-TAO will be a 1 ton fiducial sphere of liquid scintillator loaded with gadolinium surrounded by silicon photomultipliers providing about 94\% of coverage. It will be able to measure the unoscillated reactor antineutrino spectrum with an unprecedented energy resolution $\lesssim 2\%$ at 1~MeV, thus eliminating any model dependencies in JUNO's oscillation measurements. 

The conventional method to estimate JUNO's median sensitivity to the mass ordering is fitting the oscillated spectrum under the normal and inverted ordering scenarios and considering the difference in the minimum $\chi^2$ values. Using the configuration of Ref.~\cite{juno2}, a value of $\Delta \chi^2 = 10$ with 6 years of data taking is obtained, which corresponds to a sensitivity of about $3\sigma$. This configuration assumes ten nuclear reactors rather than the eight that will actually be built, but also uses lower estimates of the IBD detection efficiency and the PMT detection efficiency, among others. A reassessment of the sensitivity is underway but no significant changes are expected~\cite{JUNO:2022hxd}. 

JUNO's approach to measuring the mass ordering is orthogonal to the one to be carried out by next-generation experiments relying on atmospheric~\cite{IceCube:2016xxt,KM3Net:2016zxf} and accelerator~\cite{DUNE:2020ypp,Hyper-Kamiokande:2018ofw} neutrinos. The latter use neutrinos in the $\sim$GeV energy scale traversing distances of hundreds or thousands of km, while JUNO's neutrinos will be in the $\sim$MeV scale and will only travel for 52.5~km. Likewise, the detection technology and the backgrounds will be completely distinct. Very importantly, JUNO's measurement is completely independent of the $\theta_{23}$ mixing angle and the $\delta_{CP}$ phase. Finally, JUNO's signal arises entirely from vacuum oscillations, whereas all other experiments rely on matter effects. For all these reasons, JUNO's measurement will greatly strengthen the community's confidence in the determination of this critical parameter.% and will enable some of the most stringent tests of the three-neutrino flavor paradigm.  

JUNO's measurement is also complementary to that of other experiments in that it will provide synergistic information beyond the pure statistical addition of results. A combined analysis of JUNO's data with those of ongoing or near term atmospheric~\cite{IceCube-Gen2:2019fet,Chau:2021uqv} or accelerator~\cite{Cabrera:2020own} experiments could yield the first determination of the neutrino mass ordering to $\geq 5\sigma$ significance. This synergy occurs primarily because of a tension in the measured values of $\Delta m^2_{31}$ that arises when the wrong ordering is assumed. As a result, the first unambiguous determination of the neutrino mass ordering could be achieved this decade. 

\begin{figure}[!htb]
  \centering
  \includegraphics[width=0.9\textwidth]{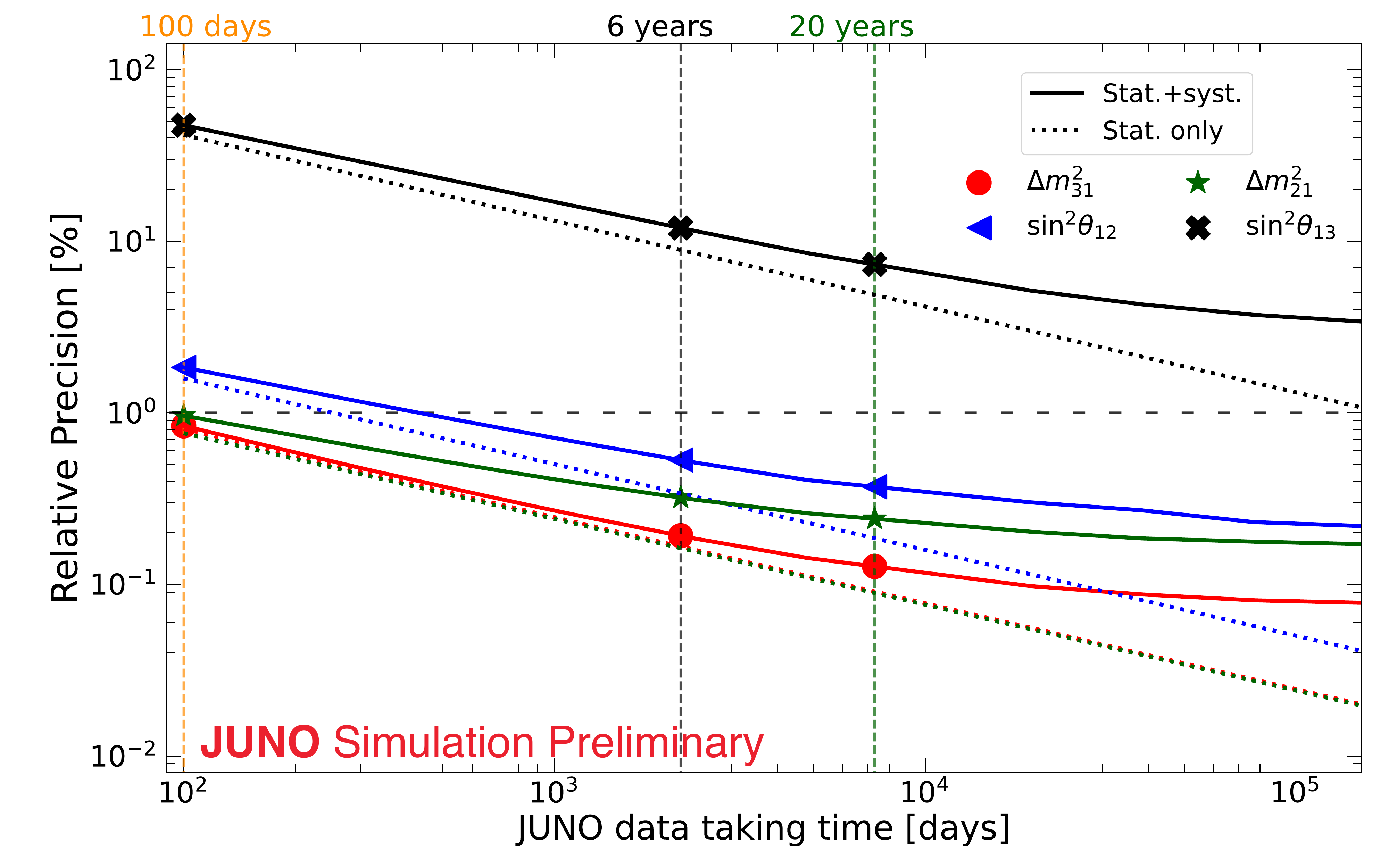}
  \caption{JUNO's relative precision on the oscillation parameters as a function of run time. The markers and vertical lines highlight run times of 100 days, 6 years, and 20 years. The horizontal gray dashed line represents a 1\% relative precision. The green dotted and red dotted lines are indistinguishable from each other since the statistical-only precision is essentially identical for the $\Delta m^2_{31}$ and $\Delta m^2_{21}$ parameters. Figure obtained from Ref.~\cite{junooscdiana}.}
  \label{fig:psur_distance}
\end{figure}

JUNO’s large-statistics measurement of the oscillated spectrum with unprecedented energy resolution will also enable determination of the four oscillation parameters that drive the disappearance of reactor antineutrinos at its $52.5$~km baseline: $\Delta m^2_{31}$, $\Delta m^2_{21}$, $\sin^2 \theta_{12}$, and $\sin^2 \theta_{13}$. The expected sensitivities to these parameters after 6 years of data-taking are shown in Table~\ref{tab:junosens}. The expected relative precision is $\leq 0.5$\% for $\Delta m^2_{31}$, $\Delta m^2_{21}$ and $\sin^2 \theta_{12}$, and the corresponding improvement over current knowledge for those parameters is around an order of magnitude. %Fig.~\ref{fig:junochi2} shows the one-dimensional $\Delta \chi^2$ distributions for the four parameters today and with 6 years of JUNO data, illustrating the dramatic improvement in precision for the first three parameters. 
%For $\sin^2 \theta_{13}$ the precision is worse than what present-day reactor experiments already achieve~\cite{ParticleDataGroup:2020ssz}.%, although JUNO's determination of this parameter will still constitute an important test of the three-neutrino oscillation framework. 
%Using the current knowledge on $\theta_{13}$ as a prior results in a very slight improvement in the precision of the other three parameters due to the small correlation. 
Fig.~\ref{fig:psur_distance} shows the expected precision as a function of running time for the four parameters. As can be seen there, the precision on $\Delta m^2_{21}$ and $\sin^2 \theta_{12}$ will already be world-leading with only $\sim$100 days of data. Moreover, the precision of the four parameters will continue to improve appreciably even after 6 years of data-taking.  
\begin{table}[htbp]
    \centering
	\begin{tabular}{ccccc}
	\hline
	        			&	$\Delta m^{2}_{31}$	&	$\Delta m^{2}_{21}$	&	$\sin^{2}\theta_{12}$ & $\sin^{2}\theta_{13}$ \\
	\hline 
	\hline
	JUNO 6 years		&  $\sim$0.2\%	 & $\sim$0.3\%	 & $\sim$0.5\%	 & $\sim$12\%	\\

	PDG2020				&  $ 1.4  \%$	&  $ 2.4 \%$	&  $ 4.2  \%$	&  $ 3.2  \%$	 \\ 
	\hline
	\end{tabular}%
	\caption{Expected precision of the oscillation parameters after 6~years of JUNO run time. All uncertainties are considered, and no external constraint is applied on $\sin^2 \theta_{13}$.  The precision with which these parameters are currently known is shown for comparison~\cite{ParticleDataGroup:2020ssz}. Numbers obtained from Ref.~\cite{junooscdiana}.}
	\label{tab:junosens}%
\end{table}	

%\begin{figure}[!h]
%  \centering
% \includegraphics[width=0.9\textwidth]{./figures/junochi2.pdf}
%  \caption{One-dimensional $\Delta\chi^2$ distributions of the four oscillation parameters today (PDG2020, dashed black line) and with 6~years of JUNO data (solid red line). The improvement in precision for $\Delta m^2_{31}$, $\sin^2_{12}$, and $\Delta m^2_{21}$ is around an order of magnitude.}
%  \label{fig:junochi2}
%\end{figure}

There is no confirmed experiment on the horizon that will be able to reach this precision on $\Delta m^2_{31}$, $\Delta m^2_{21}$ and $\sin^2 \theta_{12}$, so these measurements are expected to be the best in the world for the foreseeable future. There are several ways in which they are expected to be an important input to the community: 

\begin{itemize}
	\item They will provide important constraints to present and future experiments. 
	\item They will provide stringent inputs for neutrino masses and model building. For instance, the more precise knowledge of $\theta_{12}$ will play a prominent role, since this parameter is more sensitive to quantum corrections due to the fact that $\Delta m^2_{21} << |\Delta m^2_{31}|$ and because the non-zero value of $\theta_{13}$ can induce further corrections for $\theta_{12}$~\cite{Harrison:2002er,Xing:2002sw,He:2003rm}. 
	\item They will narrow down the parameter space of the neutrinoless double beta-decay effective mass $|m_{ee}|$. In inverted mass ordering scenarios where $m_3 < 0.05$~eV, the minimal value of $|m_{ee}|$ is proportional to $\cos ^2 \theta_{12}$ \cite{Lindner:2005kr}. The better knowledge in $\theta_{12}$ will shrink the possible parameter space of $|m_{ee}|$ such that its minimum value can be increased by a factor of 2~\cite{Ge:2015bfa}. This will make a big difference to experiments (roughly a factor of 16 in the combined product of running time, detector mass, background level, and energy resolution for a background dominated experiment) and will thus have a strong impact on when and how a conclusive test of the inverted mass ordering region can be achieved~\cite{juno2}. 
	\item They will play a crucial role in model-independent tests of the three-neutrino oscillation framework, most notably unitarity tests of the PMNS matrix. For example, the combination of JUNO’s results with those of short-baseline reactor experiments like Daya Bay and solar experiments like SNO will enable the first such direct test of 
	${|U_{e1}|}^2+{|U_{e2}|}^2+{|U_{e3}|}^2=1$ to the few percent level~\cite{unitarity_qian,unitarity_ellis,Fong:2016yyh}. Similarly, the combination of JUNO with muon (anti)neutrino disappearance measurements will enable tests of the mass sum rule $\Delta m^2_{13} + \Delta m^2_{21} + \Delta m^2_{32} = 0$, which is another important probe of physics beyond the Standard Model such as the existence of sterile neutrinos.  
\end{itemize}

In summary, reactor antineutrino experiments have played an essential role in unveiling the oscillatory behavior of the neutrino, from the first clear observation of the $L/E_{\nu}$ dependence of this phenomenon with terrestrial neutrinos, to the discovery of the non-zero value of $\theta_{13}$, among other breakthroughs. The precise determination of $\theta_{13}$ by reactor experiments already underpins the world's best knowledge on CP violation, and will continue to do so for the foreseeable future. Reactor antineutrinos will continue to have a prominent role in neutrino oscillation physics, with a measurement within this decade by JUNO of the neutrino mass ordering that is independent and complementary to what atmospheric and accelerator experiments can do. Likewise, by the end of this decade, the most precise knowledge of four out of the six parameters that drive neutrino oscillation will come from reactor antineutrino experiments, namely Daya Bay and JUNO, with three of these determined to $0.5\%$ or better. The United States has traditionally played a leading role in experimental efforts with reactor antineutrinos, but currently has only a small participation in JUNO. This experiment will begin operations soon and thus presents an excellent opportunity to participate in the production of cutting-edge reactor antineutrino physics results before other next-generation neutrino oscillation experiments come online.

% 1300 PEs/MeV + calibration + energy resolution will reach 25,000 
%the detector will have 650~m of overburden. 

\section{Non-Standard Flavor Mixing Searches at Reactors (NF02)}
\label{sec:sbl}

\subsection{Key Takeaways}% of Sec.~\ref{sec:sbl}}

\begin{itemize}
    \item While the 3+1 oscillation scenario is disfavored by a combination of diverse appearance and disappearance results, the desire to explain lingering short-baseline anomalies with new physics has not gone away.  

    \item By performing correlated measurements of the \nuebar spectrum at multiple short baselines, reactor experiments offer a low-cost experimental method for unambiguously probing non-standard neutrino flavor transformation.   
    
    \item There are plausible explanations for the Reactor Antineutrino Anomaly that do not involve sterile neutrinos.  These explanations provide a better match to host of new neutrino and nuclear physics measurements and modelling studies performed in the last decade.

    \item New neutrino mass states enrich studies of $CP$ violation. On one hand, non-standard oscillations can confound inferences of the standard $\delta_{CP}$ at, e.g., DUNE; on the other, they also generically introduce new sources of $CP$ violation.
\end{itemize}

\subsection{The Reactor Antineutrino Anomaly}

%\emph{History of the RAA as invitation to consider nonstandard oscillations --- details of flux models will presumably be covered in Sec.~\ref{sec:source}, so just enough here to give a sense for what's going on.}

In 2011, two independent reevaluations of the reactor antineutrino spectrum were published by Mueller et al.~\cite{bib:mueller2011} and Huber \cite{bib:huber}. Both concluded that the integrated antineutrino flux is $\approx3\%$ larger than previous calculations; we defer a discussion of the details of the flux model to Sec.~\ref{sec:source}. Many of the authors of Ref.~\cite{bib:mueller2011} would then explicitly reanalyze reactor experiments dating back to the early 1980s in Ref.~\cite{bib:mention2011}, finding that observed interaction rates were, on average, $(5.7\pm2.3)\%$ less than what the new `Huber-Mueller' (HM) model predicted; this disagreement was named \emph{the Reactor Antineutrino Anomaly (RAA)}. %By that time, it was understood that SM neutrino oscillations were not operative at the $<100$~m baselines at these experiments. %While the deficit might also indicate shortcomings of the flux model, 
It is pertinent to consider whether modifications to three-neutrino oscillations might be the cause of the RAA.

% \emph{Review 3+$N$ mixing and oscillations; effective mixing angle formalism; new sources of $CP$ violation.}

The SM can be extended by introducing $N$ additional neutrino species. If these are light enough to participate in oscillations, then they must be uncharged under SM interactions, as the invisible decay width of the $Z$ boson is consistent with there being only three light neutrinos \cite{ALEPH:2005ab}. We refer to these as \emph{sterile neutrinos} and denote them $\{\nu_{s_1}, \, \nu_{s_2}, \, \ldots, \, \nu_{s_N}\}$; these are accompanied by new mass eigenstates denoted $\{\nu_{4}, \, \nu_{5}, \, \ldots, \, \nu_{3+N}\}$. The mixing relationship given in Eq.~\eqref{eq:pmns_matrix} can be readily generalized to
\begin{equation}
    \label{eq:pmns_matrix_extended}
    \left( \begin{array}{c} \nu_{e} \\ \nu_{\mu} \\ \nu_{\tau} \\ \nu_{s_1} \\ \vdots \end{array} \right) = 
    \left ( \begin{array}{ccccc} U_{e1} & U_{e2} & U_{e3} & U_{e4} & \ldots\\
        U_{\mu1} & U_{\mu2} & U_{\mu3} & U_{\mu 4} & \ldots \\
        U_{\tau1} & U_{\tau2} & U_{\tau3} & U_{\tau 4} & \ldots \\
        U_{s_1 1} & U_{s_1 2} & U_{s_1 3} & U_{s_1 4} & \ldots \\
        \vdots & \vdots & \vdots & \vdots & \ddots \\
    \end{array} \right) \cdot \left( \begin{array}{c} \nu_{1} \\ 
        \nu_{2} \\ \nu_{3} \\ \nu_{4} \\ \vdots \end{array} \right);
\end{equation}
the $3\times3$ PMNS matrix is replaced by a $(3+N)\times(3+N)$ analog.  The sterile species, by construction, will not interact in a detector; one must infer their existence through their modifications to the oscillation probabilities of the active species. We focus on the case $N=1$ for simplicity and replace $\nu_{s_1} \to \nu_s$. In this case, there are three unique mass-squared differences, $\{ \Delta m_{21}^2, \, \Delta m_{31}^2, \, \Delta m_{41}^2 \}$, and the $4\times4$ extended PMNS matrix may be written in terms of six mixing angles and three $CP$-odd phases. Here, we focus on the survival probability $P(\overline{\nu}_e \to \overline{\nu}_e) \equiv P_{\overline{e}\overline{e}}$ in the limit relevant to SBL reactor experiments. In the three-neutrino scenario, $P_{\overline{e}\overline{e}}$ does not deviate appreciably from unity for baselines $\lesssim\mathcal{O}(100)$ m at reactor energies. Therefore, any oscillations observed on $\mathcal{O}(10-100)$-m length scales would be attributable only to $\Delta m_{41}^2$. In the limit $\Delta_{21}, \, \Delta_{31} \approx 0$, we write
\begin{equation}
    \label{eq:Pee_SBL}
    P_{\overline{e}\overline{e}} \approx 1 - 4 |U_{e4}|^2 (1 - |U_{e4}|^2) \sin^2 \Delta_{41} \equiv 1 - \sin^2 2\theta_{ee} \sin^2 \Delta_{41};
\end{equation}
where $\sin^2 2\theta_{ee}$ is the \emph{effective mixing angle}. If $\sin^2 2\theta_{ee}$ is nonzero, then this can manifest as a deficit of $\overline\nu_e$ relative to prediction --- precisely as indicated by the RAA. 

In Ref.~\cite{bib:mention2011}, rate experiments were explicitly analyzed with respect to the sterile neutrino hypothesis. It was found that the data prefer a sterile neutrino at the level $p\approx3.5\%$; the preferred regions of parameter space were approximately $\sin^2 2\theta_{ee} \in [0.02,0.20]$ and $\Delta m_{41}^2 \gtrsim$ 0.40 eV$^2$. When combined with anomalous $\nu_e$ disappearance results from the radioactive source experiments GALLEX \cite{GALLEX:1997lja, Kaether:2010ag} and SAGE \cite{SAGE:1998fvr, Abdurashitov:2005tb} -- the so-called Gallium Anomaly \cite{Acero:2007su, gallium} -- these become $p\approx0.3\%$, $\sin^2 2\theta_{ee} \in [0.05,0.22]$ and $\Delta m_{41}^2 \gtrsim$ 1.45 eV$^2$. 
%These results have been repeatedly, independently confirmed in the decade since they first appeared \cite{Kopp:2011qd}.

Since Ref.~\cite{bib:mention2011} first appeared, new measurements of the antineutrino rate at HEU reactors were performed at Nucifer \cite{bib:nucifer} and STEREO \cite{stereo_rate}. Moreover, medium-baseline experiments have also become competitive in this endeavor. Double Chooz~\cite{dc_nature}, Daya Bay~\cite{bib:prl_reactor}, and RENO~\cite{bib:reno_shape} all measured time-integrated antineutrino rates consistent with the RAA.  
These results supported the robustness of the suggested data-model flux discrepancy and hint for sterile neutrino oscillations.  
On the other hand, Daya Bay \cite{bib:prl_evol, bib:prl_235239} and RENO \cite{reno_evol} have also exploited a particular feature of LEU reactors: they can track how the \nuebar detection rate evolves with the reactor fuel composition.  
They observe a dependence of the RAA size on fuel content, a clear indication of flux mis-modelling of some sort.  

In parallel to these experimental developments, \nuebar HM flux model has also been the subject of increased scrutiny. While modeling  will be discussed in more depth in Sec.~\ref{sec:source}, we quickly overview  salient details. The HM flux model is largely based on the so-called \emph{conversion method}, whereby one inverts measured isotopic fission $\beta$ spectra \cite{bib:ILL_1, bib:ILL_2, bib:ILL_3} to infer the corresponding $\overline{\nu}_e$ spectra. One could instead calculate the spectrum by direct \emph{summation} of available nuclear data. In 2019, two new, notable flux calculations appeared. The first \cite{bib:fallot2} (hereafter `EF') provided an updated summation calculation, while the second \cite{Hayen:2019eop} (`HKSS') incorporated conversion techniques while accounting for shape alterations contributed by forbidden beta decays.  The EF model predicted a $^{235}$U flux that is $5-10\%$ less than HM, whereas HKSS predicted a modest ($\approx1-2\%$) excess. These models have been compared with reactor rate data in Refs.~\cite{huber_berryman, Berryman:2019hme, giunti_katrin, Giunti:2021kab}; the results of Ref.~\cite{Giunti:2021kab} are given in Table~\ref{tab:raa_rates}. Interestingly, the EF model does not indicate anomalous disappearance, whereas the HKSS model slightly enhances the RAA.

\begin{table*}[t!]
    \centering
    \begin{tabular}{c||c|c||c}
    \hline
    Flux Model & $R$ & Significance & $2\sigma$ Limit on $\sin^2 2\theta_{ee}$ \\ \hline \hline
    HM & $0.930^{+0.024}_{-0.023}$ & 2.8$\sigma$ & [0.031, 0.236] \\ \hline
    EF & $0.975^{+0.032}_{-0.030}$ & 0.8$\sigma$ & $< 0.170$ \\ \hline
    HKSS & $0.922^{+0.024}_{-0.023}$ & 3.0$\sigma$ & [0.039, 0.259] \\ \hline
    KI & $0.970\pm0.021$ & 1.4$\sigma$ & $< 0.144$ \\ \hline
    HKSS-KI & $0.960^{+0.022}_{-0.021}$ & 1.8$\sigma$ & $<0.166$ \\ \hline
    \end{tabular}
    \caption{The ratio $R$ of measured antineutrino rates compared to the predictions from various flux models, adapted from Ref.~\cite{Giunti:2021kab}. Also shown are the corresponding statistical significances and the 2$\sigma$ limit on $\sin^2 2\theta_{ee}$ in the large-$\Delta m_{41}^2$ ($\gtrsim 5$ eV$^2$) region.} 
    %RAA rates \cite{Giunti:2021kab}; focusing on rates + evolution. Last column: limit on $\sin^2 2\theta_{ee}$ in the high-$\Delta m^2_{41}$ region.}
    \label{tab:raa_rates}
\end{table*}

The ratio of the $\beta$ spectra of $^{235}$U and $^{239}$Pu was recently measured at the Kurchatov Institute \cite{Kopeikin:2021rnb, kopeikin2021}. In Ref.~\cite{kopeikin2021}, the $\overline{\nu}_e$ spectrum for $^{235}$U was rederived via $\beta$ conversion assuming that the $^{239}$Pu spectrum is given by the HM model; we call this ``KI.'' Moreover, Ref.~\cite{Giunti:2021kab} derives yet another flux model by rescaling the HKSS prediction for $^{235}$U by the same multiplicative factor ($1.054\pm0.002$) by which the integrated $^{235}$U fluxes for HM and KI disagree; the result is named ``HKSS-KI.'' The experimental deficits with respect to these models are given in Table~\ref{tab:raa_rates}; they are consistent with EF in that they also do not indicate significant, anomalous disappearance. 

It is too soon to consider the RAA definitively resolved by these findings.  
For example, if one had instead assumed that the HM $^{235}$U spectrum is correct and that the $^{239}$Pu one is not, then one would find \emph{increased} evidence for anomalous disappearance~\cite{giunti_evol,giunti_diagnose}. 
Still, these results indicate that the conversion and summation approaches may be converging, which is a decided improvement relative to the past decade.  
However, the RAA is not the only motivating factor for nonstandard oscillation searches at nuclear reactors.  
Anomalous $\nu_e$/$\overline{\nu}_e$ appearance results at LSND \cite{lsnd} and MiniBooNE \cite{MiniBooNE:2013uba, mboone, MiniBooNE:2020pnu} and Gallium Anomaly disappearance results can still be explained in terms of an eV-scale sterile neutrino \cite{Maltoni:2007zf, Giunti:2010jt, Giunti:2011gz, bib:kopp, GiuntiGlobal, schwetz_global, dentler_global, conrad_sterile}.  
If true for LSND and MiniBooNE, then this would require nonstandard contributions to both $\nu_\mu$ and $\nu_e$ disappearance.  
Assuming that the central value model predictions of Table~\ref{tab:raa_rates} accurately reflect reality, flux models can still accomodate a $\sim5-10\%$ change in the antineutrino rate; thus, there is still room for active-sterile mixings of modest size in the reactor sector, even without the RAA.  

\subsection{Reactor Spectrum Ratio Experiments and the Complex Current Landscape}

%\emph{Summarize the past decade of experiments --- the advent of spectral ratio measurements puts the RAA on notice. New $\beta$-spectrum measurements ostensibly bring conversion and summation fluxes in line and spell the end of the RAA --- balance between here and Sec.~\ref{sec:source}?}

\begin{figure}[!t]
  \centering
  \includegraphics[width=0.45\textwidth]{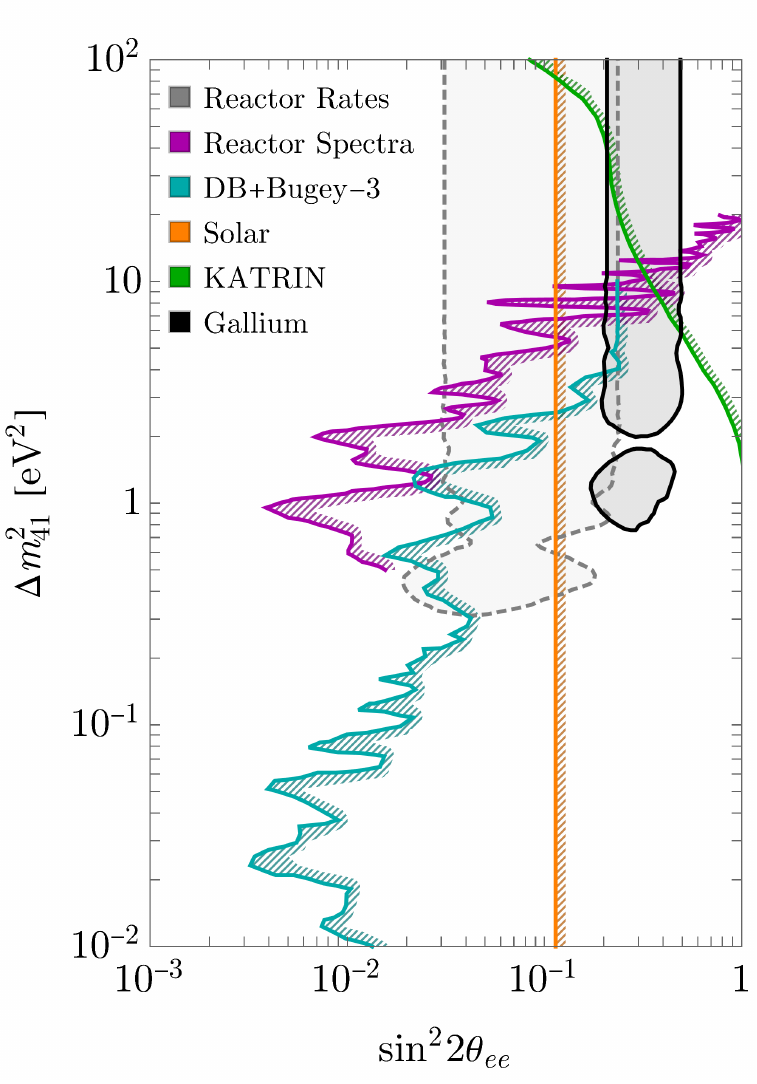}
  \includegraphics[width=0.45\textwidth]{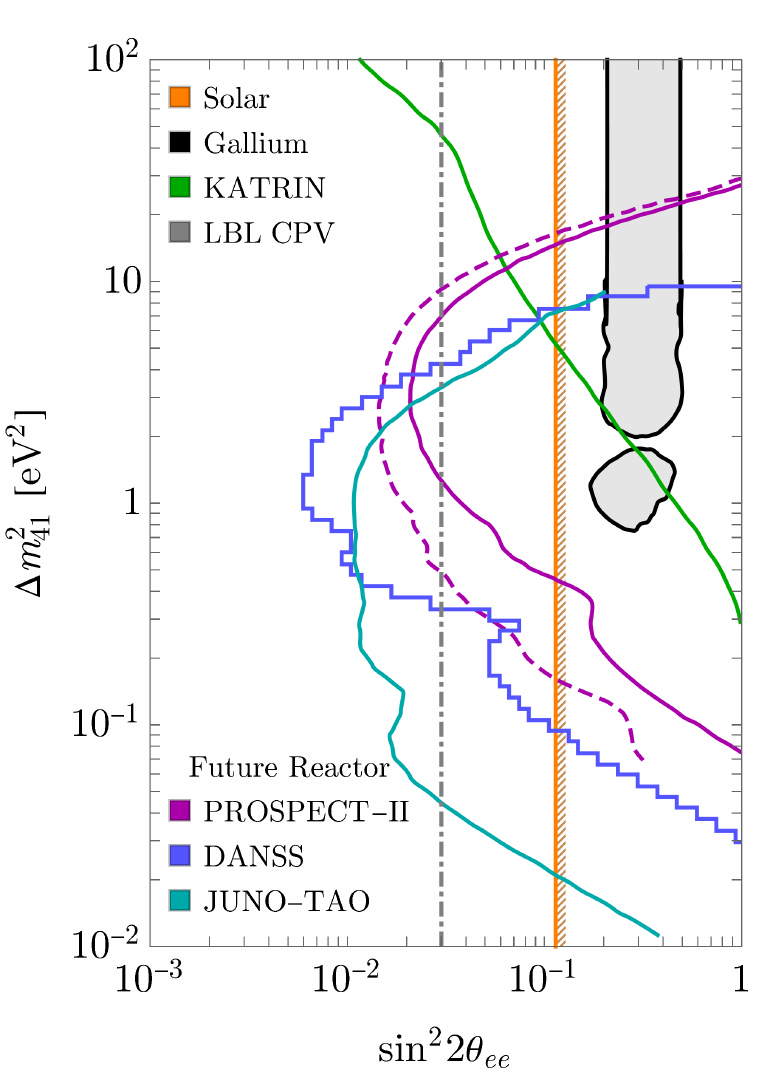}
  \caption{Left: Current constraints on a sterile neutrino from $\nu_e$/$\overline{\nu}_e$ disappearance. Color fillings represent preferences; hatching represents exclusions. The dashed, gray region is the fit to reactor rate deficits using the HM flux model \cite{Giunti:2021kab}, given for context. See text for more details.
  Right: The future sensitivities of KATRIN \cite{KATRIN:2022ith} (green; 95\% C.L.), PROSPECT-II \cite{Andriamirado:2021qjc} (purple; 90\% C.L.), DANSS (light blue; 90\% CL$_s$) and JUNO-TAO \cite{juno_tao} (cyan; 90\% CL$_s$). For PROSPECT-II, two configurations are shown: two years at an HEU core (solid), and four years at an HEU core plus two years at an LEU core (dashed). The dot-dashed gray line is the $CP$ violation disambiguation limit relevant for DUNE \cite{KayserVal}.}
  \label{fig:raa_status}
\end{figure}

If one measures the \emph{spectrum} of antineutrinos instead of the energy-integrated rate, then one can potentially observe oscillations directly. Moreover, if one measures the spectrum at two (or more) baselines, then their \emph{ratio} is largely insensitive to the details of the underlying flux model. Antineutrino spectra had been measured prior to the 2010s at, e.g., ILL~\cite{bib:ILL_nu, ILL_fix} and Bugey~\cite{bib:Bugey3_osc} -- these were considered in Ref.~\cite{bib:mention2011} -- but this program blossomed as a result of searches for nonzero $\theta_{13}$. Daya Bay, Double Chooz and RENO all employed detectors at multiple locations at the $\sim$km scale, as appropriate for $\Delta m_{31}^2$, but these were situated at too long of baselines to probe oscillations at the eV scale. However, experiments exploiting some combination of (1) multiple detectors, (2) movable detectors, or (3) segmented detectors have been constructed at short baselines -- within 25 m -- to search for eV-scale oscillations. Over the past decade, SBL searches have been performed by DANSS \cite{danss_osc}, NEOS \cite{bib:neos}, Neutrino-4 \cite{bib:neutrino4_osc, neutrino4_prd}, PROSPECT \cite{prospect_osc, prospect_prd} and STEREO \cite{stereo_2018, stereo_2019}; similar searches are ongoing at NEOS-II \cite{neos2} and SoLid \cite{solid}. Past global analyses of reactor spectral ratios \cite{GiuntiGlobal, schwetz_global, GiuntiRatio, dentler_global, conrad_sterile, Berryman:2019hme, giunti_katrin, huber_berryman} have inferred a preference for new oscillations as high as $\gtrsim3\sigma$, but a combination of more data and improved statistical treatments \cite{fc, stat_sterile, n4_comment, Coloma:2020ajw, Giunti:2021iti} suggests that this is no more than 1-2$\sigma$ \cite{giunti_mc, Berryman:2021yan}. The $2\sigma$ C.L. exclusion curve from a global fit of SBL spectral measurements \cite{Berryman:2021yan} is shown in magenta in Fig.~\ref{fig:raa_status} (left). In comparison, Daya Bay and Bugey-3 were studied jointly in Ref.~\cite{bib:prl_joint2020}; the result (90\% CL$_s$) is shown in cyan.\footnote{Neither of these experiments is considered in Ref.~\cite{Berryman:2021yan}; the figure is thus not double-counting reactor spectra information.} It is a triumph of experiment that the field has matured to the point of percent-level oscillation sensitivities over the course of roughly a decade.

It is pertinent to consider how reactors fit into the landscape of $\nu_e/\overline{\nu}_e$ disappearance studies, and of sterile neutrino searches more broadly. In Fig.~\ref{fig:raa_status} (left), we show constraints on $\sin^2 2\theta_{ee}$ from solar neutrino experiments \cite{Goldhagen:2021kxe} (orange; 2$\sigma$)\footnote{This constraint assumes the GS98 solar model \cite{Vinyoles:2016djt}; had the AGSS09 solar model \cite{Vinyoles:2016djt} been used, the resulting constraint would be modestly stronger.} and from KATRIN \cite{KATRIN:2022ith} (green; 95\% C.L.). The region preferred ($2\sigma$) by a combined analysis of gallium experiments \cite{Berryman:2021yan} is shaded in gray. In addition to SAGE and GALLEX, this includes recent results from BEST \cite{Barinov:2021asz, Barinov:2022wfh}, where a $\gtrsim5\sigma$ deficit has been reported \cite{Barinov:2021mjj}. Constraints have also been derived from $\nu_e$ scattering on $^{12}$C at KARMEN and LSND \cite{Armbruster:1998uk, LSND:2001fbw, Conrad:2011ce}, as well as from T2K \cite{Abe:2014nuo} and MicroBooNE \cite{Denton:2021czb, Arguelles:2021meu}; these have been omitted for clarity. Curiously, the solar constraint and the gallium preference are in $\gtrsim3\sigma$ tension \cite{Berryman:2021yan}; reactor spectral measurements are compatible with either, while the compatibility between gallium and reactor rates is, as described above, flux-model dependent \cite{Giunti:2021kab}. Moreover, there are no significant indications of anomalous $\nu_\mu$ disappearance~\cite{minos}; when combined with $\nu_e$ disappearance null results, this results in significant tension with the LSND and MicroBooNE anomalies \cite{bib:kopp,bib:prl_joint2020,dentler_global}. On top of all of this, eV-scale sterile neutrinos contribute to $N_\text{eff}$ and $\sum m_\nu$; cosmological observations place severe limits on nonstandard contributions to these quantities, disfavoring essentially all of the parameter space shown in Fig.~\ref{fig:raa_status} \cite{Enqvist:1991qj, Melchiorri:2008gq, Hannestad:2012ky, Archidiacono:2013xxa, Mirizzi:2013kva, Gariazzo:2013gua, Bridle:2016isd, Feng:2017nss, Knee:2018rvj, Berryman:2019nvr, Gariazzo:2019gyi, Adams:2020nue, Hagstotz:2020ukm}. This all suggests that $3+1$ oscillations do not comfortably describe the data. The question becomes: Is there a compelling conventional or BSM alternative?

%\emph{More broadly, the 3+1 scheme is under tension. Reactors fit into the bigger picture --- even if they're telling you what the new physics \emph{can't} be! Show some breadth --- particularly interesting scenarios?}

The next-simplest model one could invoke would be to introduce multiple species of sterile neutrinos. This has been studied in, e.g., Refs.~\cite{Kopp:2011qd, Conrad:2012qt, bib:kopp, conrad_sterile}; including only additional oscillation frequencies does not appreciably resolve these tensions.  
Other proposed scenarios include sterile neutrino decay~\cite{decay_machado,decay_deg}, the presence of nonstandard interactions among either the active or sterile neutrinos~\cite{nsi_marf,nsi,nsi_denton}, hidden sector couplings to neutrinos~\cite{Batell:2009di}, 
or some combination of multiple effects~\cite{decay_conrad}.  
Reactor experiments will play an essential part in a necessarily diverse global program to assessing which (if any) of these scenarios are correct.  
As noted in Sec.~\ref{sec:synergies}, they provide a clean environment in which to study oscillations, owing to (1) the flavor purity of the source; (2) the low energies, which prevent heavy states from polluting the observed signal; and (3) the relative absence of matter effects.  
If the existing SBL anomalies persist, and are confirmed at, e.g., the SBN program at Fermilab~\cite{sbn} and more robust future iterations of the BEST radiochemical experiment, then reactor experiments will continue to play an important role in discriminating between possible explanations thereof.  
The use of multiple arms of the US neutrino program to elucidate a more complex `non-vanilla' sterile sector is very well-illustrated in Ref.~\cite{decay_conrad}: in this example, which envisions a 2-component sterile sector, US short-baseline reactor data is crucial for constraining active-sterile oscillation parameters, while US short-baseline accelerator experiments are best at pinning down radiative decay phenomena experienced by the heavier sterile state.  

%BLAH: 3-4 sentence paragraph about CP-violation.
As noted below Eq.~\eqref{eq:pmns_matrix_extended}, introducing a sterile neutrino also introduces two new $CP$-violating phases, which enriches the possibilities for $CP$-violation studies at long-baseline accelerator experiments. On one hand, 3+1 oscillations that violate $CP$ may be confounded with $CP$-conserving, 3+0 oscillations \cite{Kayser}; on the other, large-amplitude, $CP$-conserving oscillations with a sterile neutrino may generate false signals of $CP$ violation at, e.g., DUNE \cite{cp_kelly}. While $P_{ee}$ in Eq.~\eqref{eq:Pee_SBL} is necessarily $CP$-conserving, the sensitivity of reactors to the existence of additional neutrinos is crucial for the disambiguation of such a signature. This potential parameter degeneracy can broken if $\sin^2 2\theta_{ee}$ can be measured at the level $\lesssim 0.03$ \cite{KayserVal}, shown by the dot-dashed line in Fig.~\ref{fig:raa_status}.

\subsection{The Future of Short-Baseline Reactor Experiments}

\begin{table*}[t!]
    \centering
    \begin{tabular}{c||c|c|c}
    \hline
    Experiment & $L$ [m] & $P_\text{th}$ [MW] & Material(s) \\ \hline \hline
    %Bugey-3~\cite{Declais:1994su} & 15, 40, 95 & 2800 & LS \\ \hline 
    DANSS~\cite{danss_osc} & $\sim11-13$ & 3100 & SS \\ \hline
    MiniCHANDLER~\cite{Haghighat:2018mve} & 25 & 2900 & SS \\ \hline
    NEOS ~\cite{bib:neos} & 24 & 2800 & LS \\ \hline
    NEOS-II~\cite{bib:neos} & 24 & 2800 & LS \\ \hline
    Neutrino-4~\cite{bib:neutrino4_osc} & $\sim6-12$ & 100 & LS \\ \hline
    PROSPECT~\cite{prospect_osc} & $\sim7-9$ & 85 & LS \\ \hline
    PROSPECT-II~\cite{PROSPECT-II} & $\sim7-9$ & 85 & LS \\ \hline
    SoLid~\cite{solid} & $\sim6-9$ & 40-100 & SS \\ \hline
    STEREO~\cite{stereo_2018} & $\sim9-11$ & 58 & LS \\ \hline
    JUNO-TAO~\cite{juno_tao} & $\sim30$ & 4600 & LS \\ \hline
    \hline
    %
    %Chooz [] & 998, 1115 & 8500 (?) & LS \\ \hline
    Daya Bay~\cite{bib:prl_reactor} & 550, 1650 & 17,400 & LS \\ \hline
    Double Chooz~\cite{dc_nature} & 400, 1050 & 8500 & LS \\ \hline
    %Palo Verde [] & Complete & 750, 890 & 11 630 & LS \\ \hline
    RENO~\cite{bib:reno_shape} & 430, 1450 & 16,800 & LS \\ \hline \hline
    JUNO~\cite{JUNO:2020xtj} & 52,500 & 26,600 & LS \\ \hline
%    KamLAND [] & & & LS \\ \hline
    \end{tabular}
    \caption{A tabulation of IBD-based reactor experiments that were either performed in roughly the last decade or will be performed in the near future.  Experiments are sorted into short, medium, and long-baseline categories.}
    \label{tab:ibd_experiments}
\end{table*}

As of 2022, at least four new short-baseline reactor neutrino detectors are in preparation or under construction, with plans to address the open questions described above. The JUNO-TAO detector, a satellite detector for JUNO, will begin taking data in 2023 at a baseline of $\sim 30$ m from a commercial power reactor in China~\cite{juno_tao}. 
The PROSPECT-II detector, a planned upgrade of the PROSPECT detector, anticipates taking a second run of data within 10 m of the HFIR reactor in the US and possibly at other sites~\cite{Andriamirado:2021qjc}.
The DANSS Collaboration is currently upgrading their detector to improve their photostatistics, and thus their energy resolution \cite{DANSS-2021}. The Neutrino-4 Collaboration is also preparing an upgrade: a combination of increasing the detector volume and introducing pulse-shape discrimination is expected to triple their statistics, though the impact on their sterile neutrino sensitivity has not yet been made public \cite{N4_talk_Fomin}.
Each of these experiments will extend sensitivity to non-standard neutrino oscillation well beyond current limits, into regions of interest for the still-unresolved gallium anomaly and the continuing tension between short-baseline accelerator results.  
These experiments are likely to provide particularly good sensitivity in the $\sim$2-20~eV$^2$ mass splitting region, where current limits on active-sterile mixing are comparatively weaker in the electron disappearance channel.  
While probing this region, JUNO-TAO, PROSPECT-II and DANSS will also be able to authoritatively address existing claims of moderate confidence-level observations of sterile neutrino oscillations at the Neutrino-4 experiment~\cite{neutrino4_prd}.
These detectors will also increase the precision of neutrino spectrum measurements, described more in Sec.~\ref{sec:source}.  

As the neutrino community seeks to resolve remaining short-baseline neutrino anomalies, reactor experiments such as PROSPECT-II and JUNO-TAO provide several points of complementarity to other approaches. As shown in Fig.~\ref{fig:raa_status}, the projected PROSPECT-II sensitivity will combine with the projected KATRIN sensitivity to fully cover the parameter space favored by the current gallium anomaly (which, as noted above, is already disfavored solar experiments) and to definitively exclude an oscillation solution to the RAA. Although the curves in Fig.~\ref{fig:raa_status} correspond most directly to a 3+1 sterile neutrino models, they illustrate the general point that reactor neutrinos explore a flavor channel (pure $\nu_e$) where there may not be input from other sources. 
They do so with relatively low cost compared to accelerator experiments, because the reactor sources are already in operation and the detector size can be on the meter-scale.  

In addition to JUNO-TAO, PROSPECT-II, DANSS and Neutrino-4, which all use IBD interactions in scintillator as the detector channel, a growing number of experiments are seeking to measure CEvNS interactions at reactor sources.  
Ongoing reactor CEvNS projects are listed in Table~\ref{tab:cevns_experiments}.  
Compared to the established IBD channel, the CEvNS signal presents a much greater experimental challenge due to high sensitivity to radiation and instrumental background.
%Compared to the established IBD channel, the CEvNS signal presents a much greater experimental challenge. 
So far, the low-energy CEvNS signal has not been detected above the large backgrounds to this approach.  
When it becomes visible, the CEvNS signal will provide information about reactor neutrino fluxes and interactions below the IBD threshold and, like IBD searches, complement accelerator- and DAR-based searches for sterile neutrino oscillations.  
These experiments are discussed in more detail in Sections~\ref{sec:bsm} and~\ref{sec:detectors}.  

% \emph{Tabulate (recent and future) experiments.}
% \emph{Future IBD experiments: PROSPECT-II, TAO. (Figure: sensitivities compared to/contrasted with RAA and spectral experiments.) Important in their own right, but also serve as a complement to the accelerator program, in particular --- high marginal benefit for a relatively low marginal cost (connect to Sec.~\ref{sec:synergies}). Explicit mention of $\delta_{CP}$ --- reactor-accelerator complementarity (1) prevents mismeasurement and (2) allows to determine new sources of $CP$ violation.}

\begin{table*}[t!]
    \centering
    \begin{tabular}{c||c|c|c|c}
    \hline
    Experiment & $L$ [m] & $P_\text{th}$ [MW] & Material(s) & Technology \\ \hline \hline
    CHILLAX \cite{chillax_magcevns_2021} & $\sim$25 & $\sim$1000 & LAr \& LXe & Dual-Phase TPC \\ \hline
    CONNIE \cite{connie, CONNIE:2021ngo} & 30 & 3800 & Si & Skipper CCDs \\ \hline
    CONUS \cite{CONUS:2020skt} & 17 & 3900 & Ge & Ionization \\ \hline 
    MINER \cite{NF10:MINER:2016igy} & $\sim2-10$ & 1 & Ge, SI, Al$_2$O$_3$ & Bolometry \\ \hline 
    NCC-1701 \cite{collar_bsm} & 8 & 2960 & Ge & Ionization \\ \hline 
    NEON \cite{Choi:2020qcj} & 24 & 2800 & NaI(Tl) & Scintillation \\ \hline 
    NEWS-G \cite{news-g_magcevns_2020} & - & - & Ne & Ionization \\ \hline 
    $\nu$GeN \cite{NF10:NUGEN:Belov_2015} & $\sim10$ & 3100 & Ge & Ionization \\ \hline 
    NUCLEUS \cite{NUCLEUS:2019igx} & - & - & CaWO$_4$ \& Al$_2$O$_3$ & Bolometry \\ \hline 
    NUXE \cite{nuxe_magcevns_2020, Ni:2021mwa} & $\sim$25 & $\sim$3000 & LXe & Ionization/Scintillation \\ \hline 
    RED-100 \cite{NF10:RED:Akimov_2017} & 19 & 3100 & LXe & Dual-Phase TPC \\ \hline 
    Ricochet \cite{ricochet, NF10:Ricochet:2021rjo} & 8.8 & 58 & Ge \& Zn & Bolometry \\ \hline 
    SBC \cite{SBC:2021yal} & 3 & 1 & LXe & Scintillation \\ \hline 
    TEXONO \cite{Kerman:2016jqp, TEXONO:2020vnv} & 28 & 2900 & Ge & Ionization \\ \hline 
    $\nu$IOLETA \cite{violeta_magcevns_2021} & 8, 12 & 2000 & Si & Skipper CCDs \\ \hline    \end{tabular}
    \caption{A tabulation of CE$\nu$NS reactor experiments, including their reactor standoff $L$, the reactor (thermal) power $P_{\text{th}}$, component material(s) and detection technology.}
    \label{tab:cevns_experiments}
\end{table*}

% \emph{CE$\nu$NS as a new player in this field --- provide a census of the (many!)~experiments that are cropping up. Experimental challenges (backgrounds, thresholds, quenching, etc.) and opportunities ($\nu$ flux below IBD threshold) --- again, complementarity to accelerator/$\pi$DAR sources.}

\subsection{Medium- and Long-Baseline Reactor Experiments}

%\emph{The narrative has focused on SBL experiments because of all the hubbub related to RAA and eV-scale sterile neutrinos, but MBL/LBL experiments are also important for studying nonstandard matter effects -- again, complementarity to the accelerator program because of low energies. Status of, e.g., NSI searches from MBL/LBL experiments --- the future, i.e., JUNO. Other nonstandard effects?}

We conclude this section by commenting on searches for nonstandard oscillations at medium- and long-baseline reactor experiments. In Fig.~\ref{fig:raa_status}, we have already noted the combined constraint from Daya Bay and Bugey-3 \cite{bib:prl_joint2020}; the constraint is dominated by Daya Bay below $\Delta m_{41}^2 \lesssim 0.3$ eV$^2$. Similar exclusions have been derived for RENO and Double Chooz~\cite{DoubleChooz:2020pnv,RENO:2020uip}. Long-baseline experiments are sensitive to smaller values of $\Delta m_{41}^2$ than those shown in Fig.~\ref{fig:raa_status}: the JUNO collaboration forecasts a sensitivity to $\sin^2 2\theta_{ee} \gtrsim 0.02$ for $3\times10^{-4} \lesssim \Delta m_{41}^2 \lesssim 2\times 10^{-3}$~\cite{juno2}. 
% We also highlight the synergies between the SBL and LBL programs in this endeavor: because TAO will serve as the near detector for JUNO, correlating observations between the two will allow for more robust findings than either would have on its own. This is nicely illustrated in the analyses presented in the context of sterile neutrinos in, e.g., Ref.~\cite{Basto-Gonzalez:2021aus}.

%Another popular class of models is those that allow for nonstandard interactions (NSI) for neutrinos. One consequence of this is to modify the matter effects that neutrinos experience as they propagate. Matter effects are essentially absent for short- and medium-baseline oscillations, and even at long baselines, these only slightly modify three-neutrino oscillations. As such, nuclear reactors are ill-suited for searches for NSI on the basis of modifications to oscillations. That said, the same new interactions may also modify antineutrino production and detection rates at reactor experiments; searches for these types of nonstandard effects will be discussed in Sec.~\ref{sec:bsm}.

Neutrino oscillations are fundamentally contingent on the coherence of the neutrino wave-packet; decoherence could dramatically change the oscillation probabilities at medium and long baselines. 
The Daya Bay Collaboration has studied these effects in Ref.~\cite{DayaBay:2016ouy} and finds that they are not significant in their existing data. This is confirmed in joint analyses of Daya Bay, RENO and KamLAND in Refs.~\cite{deGouvea:2020hfl, deGouvea:2021uvg}. 
The JUNO collaboration has benchmarked their sensitivities to several models of decoherence in Ref.~\cite{JUNO:2021ydg} (see also Ref.~\cite{deGouvea:2020hfl}); they forecast approximately one order of magnitude improvement in measuring the size of the neutrino wave-packet. 
Decoherence effects link up with sterile neutrino searches in a nontrivial way: recent work \cite{Arguelles:2022bvt} finds that these can be important in correctly assessing constraints at SBL reactor experiments for $\Delta m_{41}^2 \sim \mathcal{O}(\text{eV}^2)$. 
These findings again highlight the importance of robust reactor programs at both short and long baselines. 

We finally briefly note the capabilities of longer-baseline reactor experiments in probing a wider variety of exotic BSM scenarios.  
A variety of such studies have been performed at high-statistics medium-baseline experiments, such as $CPT$ and Lorentz-invariance violation searches at Daya Bay~\cite{DayaBay:2018fsh} and Double Chooz~\cite{DoubleChooz:2012eiq}.  
Other exotic studies, such as searches for large extra dimensions have also been proposed~\cite{Basto-Gonzalez:2021aus}.

\section{Probing Neutrino Properties and Unknown Particles with Reactors Neutrino Detectors (NF03, NF05)}
\label{sec:bsm}

\begin{comment}
\textcolor{red}{Editors: Louis Strigari, Jingke Xu,  Guillermo Fernandez Moroni}

\begin{itemize}
\item Brief overview of SM properties one can test at a reactor: magnetic/electric dipole, charge, weinberg angle, non-standard couplings, decay, others?  

\item Brief overview of other particles you can look for: DM made by the reactor (dark photons, ALPs, mCP, etc), or ambient dark matter (BDM, superheavy WIMPS, etc).  

\item Overview which future experiments might enable substantial advancements in measurements on these topics: low-threshold (CEvNS or nu-e) detectors, surface-based or underground IBD detectors, what else?  

TABLE: mapping experiment type to different BSM physics topic.
\end{itemize}
\end{comment}
\subsection{Key Takeaways}% of Sec.~\ref{sec:sbl}}

\begin{itemize}
%    \item Nuclear reactor provides the most intense neutrino source in the earth.
    \item Reactor antineutrinos, due to their low energies, are capable of scattering coherently from all nucleons in a target nucleus, which greatly enhances the expected cross-section of this CEvNS process with respect to other interaction channels.  
    \item For this reason, reactors offer unprecedented sensitivity in measuring the Standard Model CEvNS cross-sections at low momentum transfer, as well as data-model deviations indicative of a range of BSM physics processes.  
    \item A range of low-threshold detection technologies currently under active development can allow access to this new low-momentum transfer regime.  
    \item Other novel aspects of reactor-based experiments, such as their on-surface location and their proximity to large reactor-produced photon fluxes, offer promise in probing the existence of a range of hidden sector particles and interactions.
    
%    Nuclear reactors are also a very intense source of photons that can interact with the reactor structure and produce other dark sector candidates (such as axion-like particles, hidden photons, etc.) that can leave the core and be detected by low threshold sensors.  
\end{itemize}

\subsection{Reactor CE$\nu$NS and Low-Energy Processes:  Theory and Experimental Limits} 
CE$\nu$NS is a neutral-current process that arises when the momentum transfer in the neutrino-nucleus interaction is less than the inverse of the size of the nucleus. For typical nuclei, this corresponds to neutrinos with energies $E_\nu\lesssim 50$ MeV. In the SM, the interaction  is mediated by the $Z$-boson, with its vector component leading to the coherent enhancement. As a reference point, we first write the cross section in the form
\begin{align}
\frac{\text{d}\sigma}{\text{d}T}=\frac{G_F^2M}{4\pi}\bigg(1-\frac{M T}{2E_\nu^2}\bigg)Q_\text{w}^2\big[F_\text{w}(q^2)\big]^2\,,
\label{eq:SMcrosssection} 
\end{align}
where $G_F$ is the Fermi constant, $T=E_R=q^2/(2M)=E_\nu-E_\nu'$ is the nuclear recoil energy (taking values in $[0,2E_\nu^2/(M+2E_\nu)]$), $F_\text{w}(q^2)$ is the weak form factor, $M$ is the mass of the target nucleus, and $E_\nu$ ($E_\nu'$) is the energy of the incoming (outgoing) neutrino. The tree-level weak charge is defined by 
\begin{align}
Q_\text{w}=Z \big(1-4\sin^2\theta_W\big)-N\,,
\end{align}
with proton number $Z$, neutron number $N$, and weak mixing angle $\sin^2 \theta_W$. To first approximation, the weak form factor $F_\text{w}(q^2)$ depends on the nuclear density distribution of protons and neutrons. In the coherence limit $q^2\to 0$ it is normalized to $F_\text{w}(0)=1$, with the coherent enhancement of the cross section reflected by the scaling with $N^2$ via the weak charge, given the accidental suppression of the proton weak charge $Q_\text{w}^p\ll1$. Consequently, this implies that \cevns is mainly sensitive to the neutron distribution in the nucleus.     

\par Nuclear reactors have long been utilized as copious sources of electron anti-neutrinos.  
Neutrinos from reactors have been detected using the inverse beta decay reaction, $\bar{\nu}_e + p +$1.806~MeV$ \rightarrow e^+ + n$, by observing both the outgoing positron and coincident neutron. 
%There are four isotopes whose fission produce neutrinos above the inverse beta decay threshold: $^{235}U$, $^{241}P$, $^{239}P$, and $^{238}U$. The neutrino flux is determined from the power produced by the reactor.
The characteristic neutrino energy for this source is $\lesssim 1$ MeV, roughly an order of magnitude or more lower than the average energies of neutrinos produced by accelerator sources.  
Due to these low energies, the coherence condition for the recoil is largely preserved over the entire reactor energy regime, so that there is no dependence on the internal structure of the nucleus. 

%Louis, can you also talk about what new physics may be practically probed based on reactor CEvNS? - JX

In general, the presence of any BSM physics will modify the previous cross sections, thus altering the expected number of events detected via the CEvNS reaction in a detector. In a general fashion, we write the total cross section in the presence of BSM as
\begin{align}
	\frac{d\sigma}{dE_R}= \left.\frac{d\sigma}{dE_R}\right|_{\rm SM} + \left.\frac{d\sigma}{dE_R}\right|_{\rm BSM},
\end{align}
where the first term is the SM cross section for either neutrino-electron and \cevns~interactions, and the second is the modification created by the BSM interactions. 
Note that any possible interference effect that can appear according to the nature of the new mediators are included in the BSM cross section.

%%%%%%%%%%%%%%%%%%%%%%%%%%%%%%%%%%%%%%%%%
\begin{table}[t]
\caption{Contributions to the neutrino-electron and \cevns\ cross-sections for the different scenarios considered here. The $g_V, g_A$ are given by $g_V=\frac{1}{2}+2\sin^2\theta_W, g_A=\frac{1}{2}$~\cite{Cerdeno:2016sfi}. \label{tab:Int} }
    \centering
    \begin{tabular}{cccc}
        \toprule\toprule
       Interaction  & Non-zero couplings & $\left.\frac{d\sigma_{\nu e}}{dE_R}\right|_{\rm BSM}$ & $\left.\frac{d\sigma_{\rm CE\nu NS}}{dE_R}\right|_{\rm BSM}$ \\ \midrule\midrule
        Magnetic Moment & $\mu_{\nu_e}$ & $\alpha_{\rm EM}\mu_{\nu_e}^2\frac{E_\nu-E_R}{E_\nu E_R}$  & $\alpha_{\rm EM}\mu_{\nu_e}^2 Z^2\frac{E_\nu-E_R}{E_\nu E_R}\mathcal{F}^2(E_R)$  \\\midrule
       Scalar & $g_{\nu,\phi},g_{es},g_{qs}$ & $\frac{g_{\nu,\phi}^2 g_{es}^2E_Rm_e^2}{4\pi E_\nu^2(2E_Rm_e+m_\phi^2)^2}$  & $\frac{Q_S^2m_N^2E_R g_{\nu\phi}^2g_{qs}^2}{4\pi E_\nu^2(2E_Rm_N+m_\phi^2)^2}$  \\\midrule
       Pseudoscalar&  $g_{\nu,\phi},g_{ep},g_{qp}$  & $\frac{g_{\nu,\phi}^2 g_{ep}^2E_R^2m_e}{8\pi E_\nu^2(2E_Rm_e+m_\phi^2)^2}$  & $0$  \\ \midrule
       \multirow{2}*{Vector} &\multirow{2}*{ $g_{\nu Z^\prime},g_{ev},g_{qv}$} & $\frac{\sqrt{2} G_F m_e g_V g_{\nu Z^\prime} g_{ev}}{\pi (2E_Rm_e+m_{Z^\prime}^2)}$  & $-\frac{G_F m_N Q_V^{\rm SM} Q_V^\prime (2E_\nu^2-E_Rm_N)}{2\sqrt{2}\pi E_\nu^2(2E_Rm_N+m_{Z^\prime}^2)}$  \\  
 &  & $+\frac{g_{\nu Z^\prime}^2 g_{ev}^2m_e}{2\pi (2E_Rm_e+m_{Z^\prime}^2)^2}$ & $+\frac{Q_V^{\prime 2} m_N (2E_\nu^2-E_Rm_N)}{4\pi E_\nu^2(2E_Rm_N+m_{Z^\prime}^2)^2}$\\ \midrule
        \multirow{3}*{} &  & $-\frac{\sqrt{2} G_F m_e g_A g_{\nu Z^\prime} g_{ea}}{\pi (2E_Rm_e+m_{Z^\prime}^2)}$ & $\frac{G_F m_N Q_A Q_A^\prime (2E_\nu^2+E_Rm_N)}{2\sqrt{2}\pi E_\nu^2(2E_Rm_N+m_{Z^\prime}^2)}$ \\ 
		Axial & $g_{\nu Z^\prime},g_{ea},g_{qa}$ & $+\frac{g_{\nu Z^\prime}^2 g_{ea}^2m_e}{2\pi (2E_Rm_e+m_{Z^\prime}^2)^2}$ & $-\frac{G_F m_N Q_V^{\rm SM}Q_A^\prime E_\nu E_R}{2\sqrt{2}\pi E_\nu^2(2E_Rm_N+m_{Z^\prime}^2)}$\\
		&&& $+\frac{Q_A^{\prime 2} m_N (2E_\nu^2+E_Rm_N)}{4\pi E_\nu^2(2E_Rm_N+m_{Z^\prime}^2)^2}$\\\midrule
        \bottomrule
    \end{tabular}
\end{table}

The new physics can be enhanced by light mediators. It could be the photon coupling through electromagnetic properties of the neutrino, or additional mediators having couplings to neutrinos, charged leptons and quarks. In the spirit of simplified models, we assume a Lagrangian at low energies which includes terms for the new interactions with the SM fermions without specifying the gauge invariant models at high energies as in \cite{fernandezmoroni2021physics}. 
% \begin{align}
% 	\mathscr{L}=\,&\mathscr{L}_{\rm SM} + (g_{\nu\phi}\phi\overline{\nu}_R\nu_L+{\rm h.c.}) + g_{\nu Z^\prime}\overline{\nu}_L\gamma^\mu\nu_LZ_\mu^\prime\notag\\
% 	&+g_{ls}\phi\overline{\ell}\ell+g_{qs}\phi\overline{q}q -ig_{lp}\phi\overline{\ell}\gamma^5\ell-ig_{qp}\phi\overline{q}\gamma^5q\notag\\
% 	&+g_{lv}\overline{\ell}\gamma^\mu \ell Z_\mu^\prime+g_{qs}\overline{q}\gamma^\mu qZ_\mu^\prime +g_{la}\overline{\ell}\gamma^\mu\gamma^5 \ell Z_\mu^\prime+g_{qa}\overline{q}\gamma^\mu\gamma^5 qZ_\mu^\prime.
% \end{align}
% We denote a new scalar or vector boson by $\phi$ or $Z'$, respectively, and the $g$'s are all dimensionless couplings.
For each scenario, the modification of both neutrino-electron and \cevns\ cross sections will have a specific shape, possibly including interference effects. In Table~\ref{tab:Int} we summarize and compile the distinct BSM contributions to the neutrino-electron and \cevns~cross sections for each light mediator scenario, together with the non-zero couplings relevant in each case.  
%BLAH: Can we give more examples of experiments/analyses that have probed these couplings and give a laundry list here: TEXONO, Krasnoyarsk, Reines, Giunti?

In the specific case of \cevns, there is an additional step; we need to translate the interactions from the quark to the nucleon level. The coherence factors related to the specific mediator are given by (see e.g. Refs.~\cite{Alarcon:2011zs, Alarcon:2012nr, DelNobile:2013sia, Hill:2014yxa, Cerdeno:2016sfi})
\begin{subequations}
	\begin{align}
		Q_V^\prime &= 3 (N+Z) g_{\nu Z^\prime} g_{qv},\\
		Q_A^\prime &= 0.3 S_N g_{\nu Z^\prime} g_{qa},\\
		Q_A &= 1.3 S_N,\\
		Q_S &= 14(N+Z) + 1.1Z,
	\end{align}
\end{subequations}
corresponding to the vector, axial, SM axial, and scalar currents, with nuclear spin $S_N$, and neutrino-$Z^\prime$ and quark vector couplings $g_{\nu Z^\prime}$ and $g_{qv}$, respectively. 

\begin{figure}[h!]
    \centering
    \includegraphics[width = 1\textwidth]{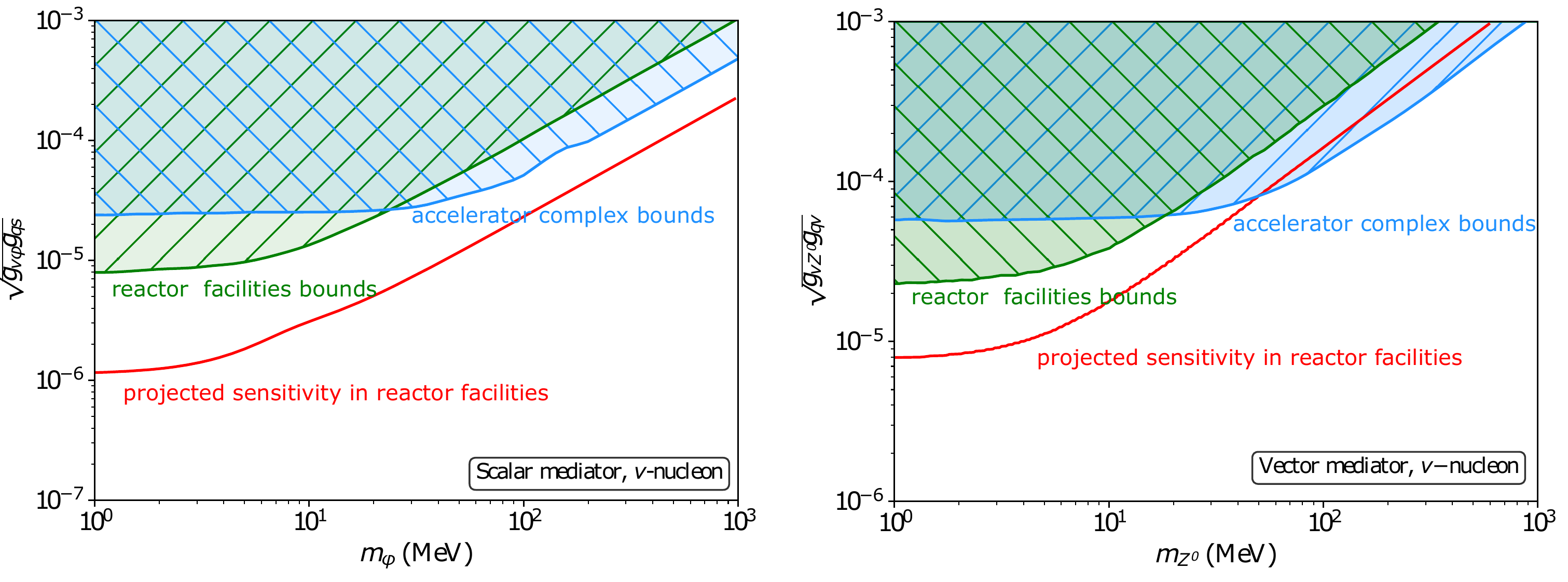}
    \caption{Current bounds and projected sensitivity bounds for new neutrino interactions with nucleons through a scalar mediator (left) and vector mediator (right). Plots show with different colors the parameter space ruled out using neutrinos from accelerator complex and neutrinos from nuclear reactor facilities. Figures taken from \cite{fernandezmoroni2021physics}. }
    \label{fig:sens_bsm_nuclear}
\end{figure}

Figure ~\ref{fig:sens_bsm_nuclear} shows the sensitivity at 90\%~C.L. of a new light scalar mediator coupling to neutrinos and quarks (left panel) and the sensitivity to a light vector mediator coupled to neutrinos and quarks (right panel), for current experiments using accelerator neutrinos (blue area) and reactor neutrinos (green area)~\cite{connie_bsm,collar_bsm}.  
New sensor technologies aiming to detect CEvNS at nuclear reactors have thresholds low enough to reap the benefits of the large neutrino flux of the reactor and access these new physics models.  
Both graphs show a better sensitivity for mediator masses below 20 MeV in reactor-based CEvNS experiments.  
The projected sensitivity for a 10 kg experiments using Skipper CCD~\cite{Tiffenberg:2017aac} with silicon as the target material is also shown in both plots. The sensor allows for a energy threshold of a few eV of the equivalent ionization energy.  
A wide range of coupling constants is unconstrained in the parameter space for masses for light mediators~\cite{fernandezmoroni2021physics}.  
Since the interaction cross sections scale with the fourth power of the coupling parameter (y-axes in the plots), the increase in sensibility of this new search is several orders of magnitude of the existing limits.  

The combination of three aspects -- the cross-section enhancement for nuclear interaction for the reactor neutrino energies, the very low energy threshold of new technologies to observe faint depositions, and the reactor being the most intense neutrino flux on earth -- make the proposed technique a unique tool to search for dark sector candidates in new regimes.

\subsection{Experimental Requirements For Reactor CE$\nu$NS Detection}

\begin{figure}[h!]
    \centering
    \includegraphics[width = 0.49 \textwidth]{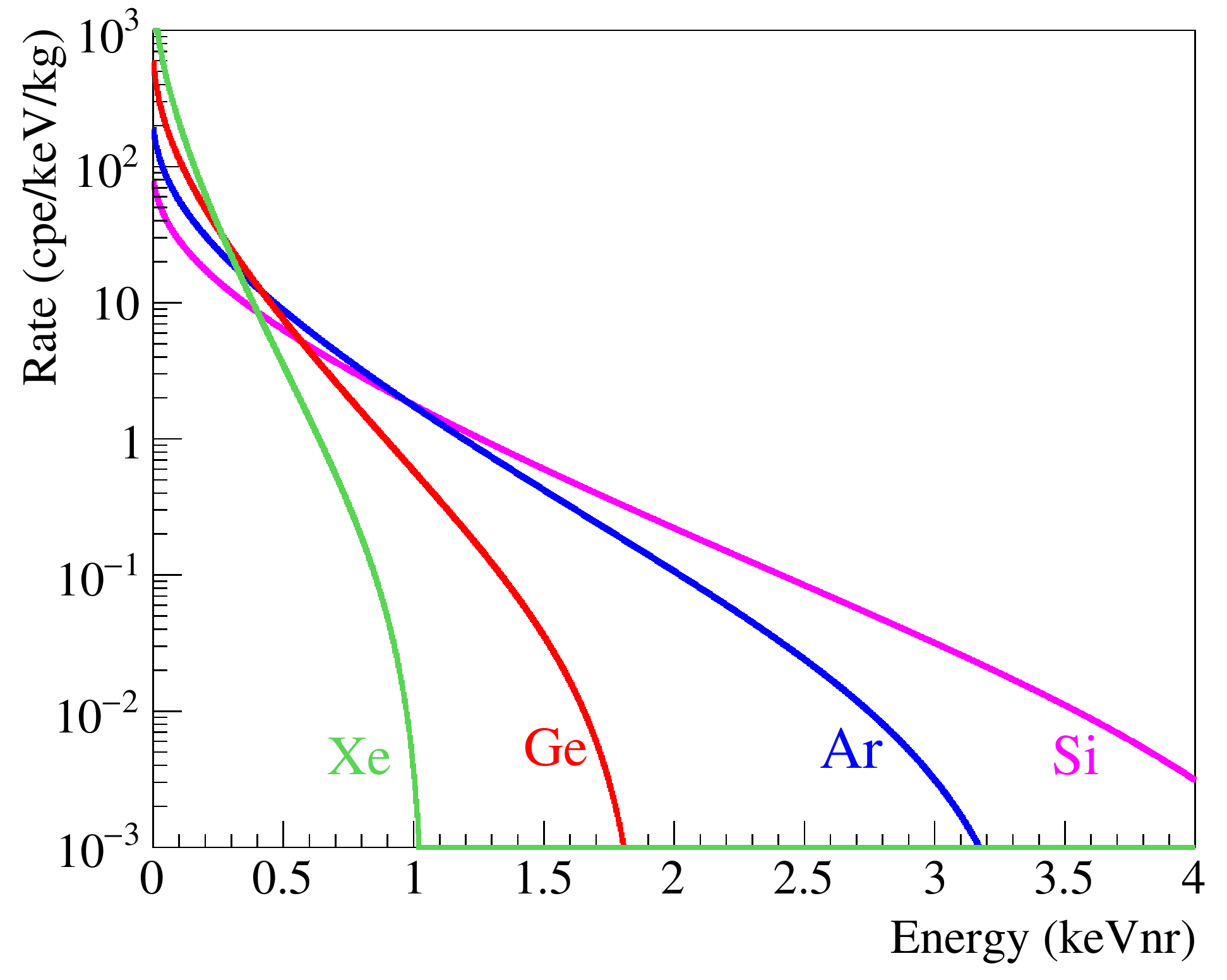}
    \includegraphics[width = 0.49 \textwidth]{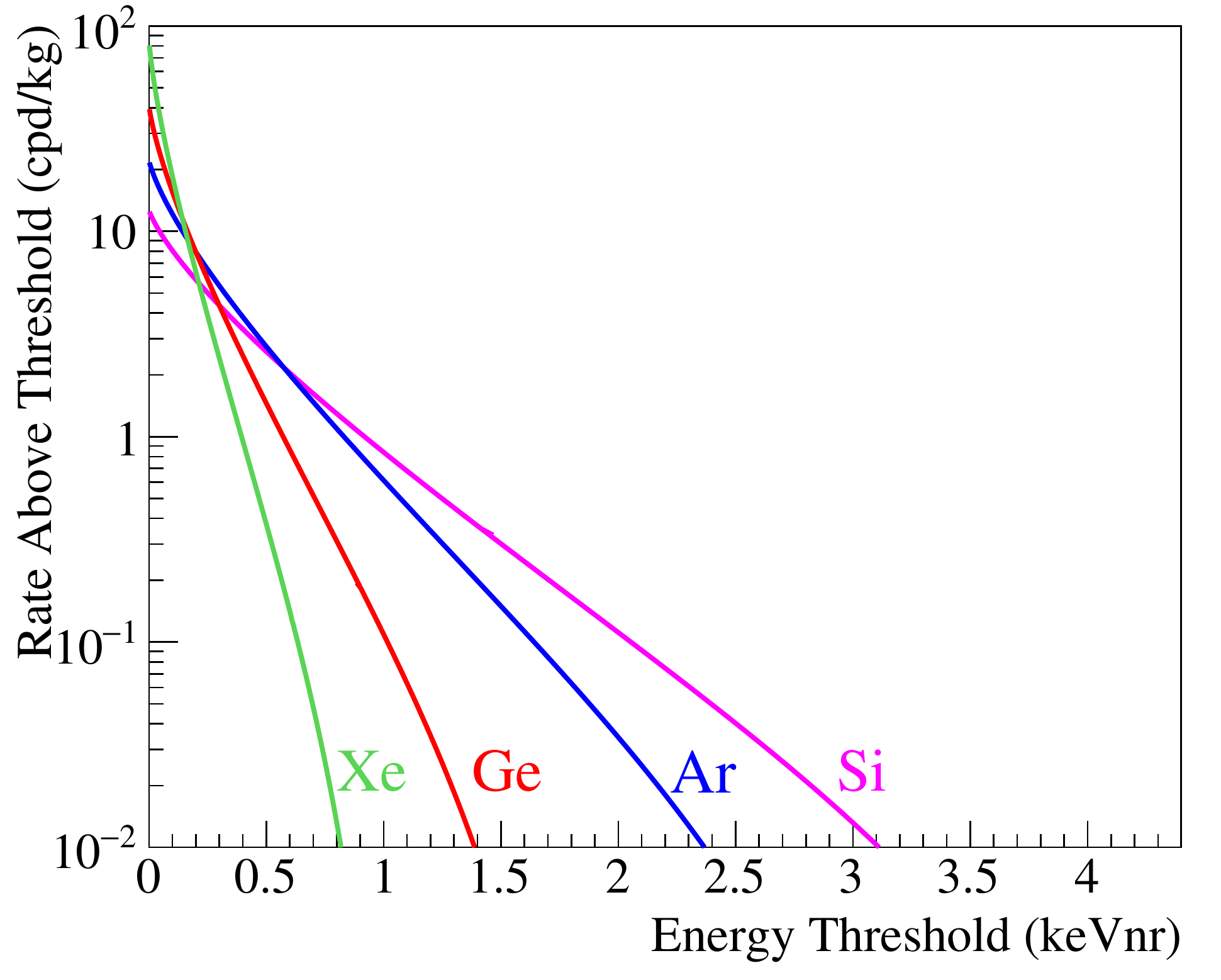}
    \caption{\textbf{Left:} The expected reactor CEvNS energy spectra in a Si/Ar/Ge/Xe target, with the assumption of 1kg target mass and 25m standoff distance from a 1GW reactor core; reactor antineutrino spectrum is taken from \cite{vogel_review}. \textbf{Right:} Integrated CEvNS event rate in 1 kg of Si/Ar/Ge/Xe as a threshold of detector energy threshold, with the same assumption on reactor parameters as for the left figure. }
    \label{fig:cevns_spectra}
\end{figure}

The maximum nuclear recoil energy resulted from CEvNS interactions can be approximated as $2k^2/M$, and is usually at the keV level or below for reactor antineutrinos with a characteristic energy of $\lesssim$ 1MeV. As illustrated in Fig.~\ref{fig:cevns_spectra}, with a Si/Ar/Ge/Xe target, 90\% of the reactor CEvNS signals will have an energy below 0.8/0.6/0.3/0.2 keV. In addition, the majority of energy transferred from the antineutrino to a nucleus is dissipated as heat. As a result, for detector technologies that measure scintillation and/or ionization signatures from particle interactions, only a small fraction of nuclear recoil energy is observable. In Si~\cite{Chavarria2016_SiRecoil}, Ge~\cite{Jones1975_GeRecoil, Scholz2016_GeRecoil,Bonhomme:2022lcz} and Xe~\cite{Lenardo2019_XeRecoil} the reduction factors of measurable energy (or quenching factors) have been measured down to $\sim$0.3 keV, with typical suppression values around 6--10 in this energy regime. This quenching effect, in addition to the low nuclear recoil energy, makes the detection of reactor CEvNS signals challenging. 

Thanks to the progress of direct detection dark matter experiments in the last few decades, low-threshold detectors sensitive to keV-level nuclear recoils have been developed~\cite{XENON1T2019_S2Only, CDMS2020_HVDetector, DAMIC2020_DM, SENSEI2020_DM}. A typical dark matter experiment focuses on  nuclear recoils from a few keV to tens of keV, and thus the detection of reactor CEvNS requires the detector thresholds to be further reduced. Several experimental efforts have been launched to advance the low-energy sensitivity of detector technologies including Si and Ge ionization detectors~\cite{SENSEI2020_DM, CDMS2020_HVDetector, CONUS:2020skt, CONNIE:2021ngo}, liquid argon and xenon scintillation/ionization detectors~\cite{XENON1T2019_S2Only, DarkSide2018_S2Only, chillax_magcevns_2021, nuxe_magcevns_2020, NF10:RED:Akimov_2017, SBC:2021yal}, and cryogenic bolometers~\cite{CRESST2019_DM, NF10:MINER:2016igy, NF10:NUCLEUS:Strauss:2017cuu, NF10:Ricochet:2021rjo}. Up to date, energy thresholds in the range of tens of eV to hundreds of eV have been demonstrated in bolometers and ionization detectors.

Coincidentally, at a detection threshold of $\sim$200eV, reactor CEvNS experiments using different targets are expected to observe comparable event rates per target mass (Fig.~\ref{fig:cevns_spectra}). Because of the near-exponential shape of the CEvNS spectra, an experiment with a 50 eV threshold will be able to collect 5--10 times more CEvNS events than those with 200 eV thresholds, demonstrating the need to develop lower-threshold detector technologies. 
Further, due to the low expected event rate of neutrino interactions, a detector also needs to have a large active mass to obtain sufficient statistics to study possible BSM physics associated with CEvNS. 
Currently available low-threshold detectors such as Skipper CCDs are limited to active masses at a hundred-gram level~\cite{SENSEI2020_DM, NF10:CONNIE:2021}, while high-mass detectors such as liquid argon and xenon Time Projection Chambers (TPCs) are limited to an energy threshold of hundreds of eV~\cite{LUX2021_S2Only, XENON1T2019_S2Only, DarkSide2018_S2Only}. Ongoing R\&D efforts are currently pursuing substantial improvements in these directions \cite{DAMIC-M,Oscura,darkside_2021,Lenardo_2019}. More R\&Ds is needed in the next decade to perform a first definitive experimental observation of reactor CEvNS.

In addition to detector threshold and active mass, another important aspect to consider in reactor CEvNS detection experiments is the excess backgrounds observed in the low energy regions of different detectors~\cite{EXCESS2021_summary, EXCESS2021, LUX2020_Ebackground, XENON1T2019_S2Only, DarkSide2018_S2Only}, which operate at very different temperatures and have different signal readout schemes. 
Such backgrounds often manifest themselves as a fast rising event rate as the energy approaches the detector threshold, and can vary drastically in rate, temporal behavior, and other characteristics. 
In ionization detectors these backgrounds are suspected to arise from the trap and delayed release of electrons~\cite{LUX2020_Ebackground, DarkSide2018_S2Only} or low-energy interactions near the active volume~\cite{EXCESS2021_summary}, and in bolometers they are sometimes hypothesized to originate from crystal stress or accumulation of energy in the active volume~\cite{EXCESS2021, EXCESS2021_summary}. Much remains unstudied for these experiments to enjoy meaningful nuclear recoil sensitivities in the reactor CEvNS energy regime. 

%Although the CEvNS process cannot be used to unfold the incoming neutrino energy at the event level, the observed interaction rate and spectrum contain valuable statistical information regarding the reactor neutrino energy spectrum.  
%Most importantly, because the CEvNS interaction is threshold-less, a low-mass target like Si and Ar can in principle capture interactions by neutrinos below the IBD energy threshold of 1.8 MeV. 
%With enough statistics and accuracy, this information can help test different formulations of antineutrino production in a reactor. 
%In addition, because the CEvNS process is flavor-blind, a short baseline reactor CEvNS experiment can be capable of studying the mixing of electron antineutrinos with a sterile neutrino, and by operating a single detector at different standoff distances an experiment will cancel out various systematics from the flux prediction and detector effects. 
%Last but not least, the CEvNS energy spectra can also contain information about BSM physics such as non-zero neutrino magnetic moment, which can modify the event rate in the lowest-energy region of CEvNS event spectra. 

%\section{IBD} 
%TBD 

\subsection{Exotic particle searches at nuclear reactors} 

Other novel aspects of reactor-based experiments, such as their on-surface location and their proximity to large reactor-produced photon fluxes, can be leveraged to probe the existence of a range of hidden sector particles and interactions.  
Below, we illustrate with a few examples.  

\begin{figure}[h!]
    \centering
    \includegraphics[width = 0.6\textwidth]{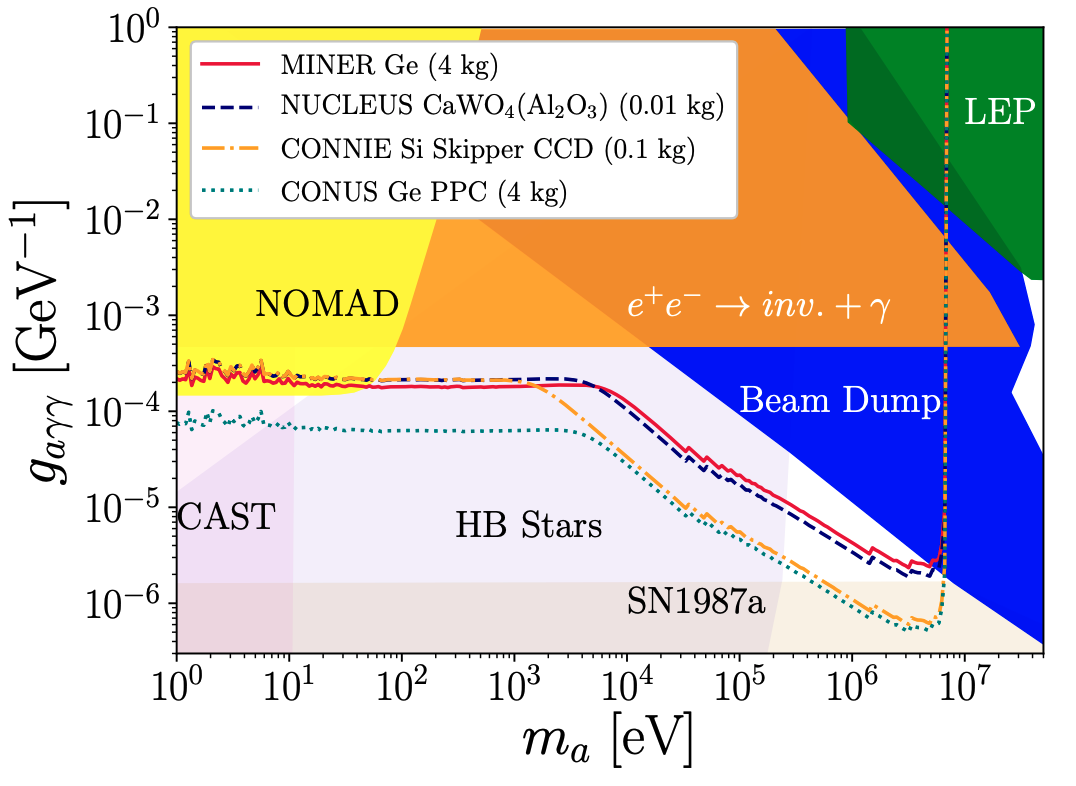}
    \caption{Comparison of sensitivity of axion like particles searches at nuclear reactor compared with excising bounds. Figures taken from \cite{Dent_2020}.}
    \label{fig:sens_axion_like}
\end{figure}

Nuclear reactors are also an intense source of photons and neutrons, which can interact with the materials of the reactor structure or spontaneously transform to produce hidden sector candidates that could escape from the reactor core and reach a nearby detector. 
This new framework of production and detection at nuclear facilities has been studied due to the large production rates obtained in reactors and the availability of new  technologies to detect them \cite{Sierra2021,Dent_2020,danilov_2019}. 
As an example  of the sensitivity of this technique Fig~\ref{fig:sens_axion_like} (from \cite{Dent_2020}) shows the reach in the search for axion-like particles for different low threshold sensor technologies in nuclear reactors (different color lines) compared to other astrophysically derived constraints (shaded areas). The plot shows the parameter space of axion-like particles coupling to photons (coupling constant in the  \textit{y} axes and particle mass in the \textit{x} axes. As the plot shows, the new technique shows unprecedented sensitivity to regions that cannot be accessed by other experiments for axion masses around 1 MeV (the so called cosmological triangle).  
These detectors can also similarly detect other indirectly electron- or photon-coupled hidden sector particles generated in the core, such as millicharged particles~\cite{TEXONO:2018nir}.  
Neutron-sensitive detectors, such as those used in reactor IBD experiments, are highly capable of probing neutron-coupled hidden sector particles; a search setting world-leading limits on hidden neutrons was recently reported by the STEREO experiment~\cite{Almazan:2021fvo}.  

Most short-baseline reactor experiments are located on-surface and lack a substantial amount of overburden.  
While disadvantageous from the perspective of increases in neutrino-like cosmic backgrounds, it also provides unique advantages for the detection of high-cross-section cosmogenic dark matter particles that would be otherwise attenuated before reaching a detector~\cite{Cappiello:2019qsw}
The PROSPECT reactor antineutrino experiment has used its overburden-free, PSD-capable IBD detector to perform a sensitive search for single proton recoils from interactions of boosted dark matter in the sub-GeV mass regime~\cite{prospect_dm}.  
Similar on-surface reactor-located rare event searches may also be applicable for pursuit of other BSM particle types, such as multiply-interacting massive particles (MIMPS)~\cite{Bramante:2018qbc,Bramante:2018tos} or macroscopic dark matter~\cite{Bai:2019ogh}.

\section{Improving Reactor and Nuclear Physics Knowledge Through Neutrino Measurements and Modelling (NF09)}
\label{sec:source}
%\textcolor{red}{Editors: Xianyi Zhang, Tom Langford}

\subsection{Key Takeaways}
\begin{itemize}
    \item Precise knowledge of the total magnitude and energy spectrum of reactor antineutrino emissions is a vital ingredient in performing some future neutrino physics measurements.  
%    model dependent particle research and background estimation in HEP experiments.
    \item Recent neutrino experiments have been very successful in advancing the state of knowledge of reactor antineutrino emissions, most notably by uncovering the reactor flux and spectrum anomalies. % isotopic and
    \item The increased precision of reactor neutrino measurements has had a broader science impact by spurring investments and improvements in non-neutrino nuclear physics measurements, nuclear data, and reactor antineutrino modelling.  
    \item Next-generation IBD and non-IBD experiments are poised to improve their reactor flux and spectrum measurement precision beyond the associated modelling uncertainties, enabling data-driven improvements to reactor and nuclear physics.
\end{itemize}

\subsection{Reactor Neutrino Flux and Spectrum Measurements}

% \begin{itemize}
%     \item Theta-13 measurements of flux and spectra at multi-Rx LEU sites (DYB, DC, RENO)
%     \item Time-evolution extraction of p9 vs u5 nuebar yield
%     \item SBL measurements of flux and spectra at HEU and LEU
%     \item Unfolding of "Prompt" energy into antineutrino energy spectra
%     \item Joint analyses to enhance physics reach beyond a single experiment
% \end{itemize}
Antineutrino emissions from LEU and HEU reactors have been precisely measured by a range of IBD detection experiments covering baselines from roughly 7~m to 2000~m. 
While some experiments have measured emissions from HEU reactors, which burn only $^{235}$U, most others have sampled LEU reactors, whose neutrino emissions are contributed by the primary fissile isotopes ($^{235}$U, $^{238}$U, $^{239}$Pu, and $^{241}$Pu) according to their  fission fractions at a specific point in the reactor's burnup cycle.  
These measurements enable accurate evaluation of antineutrino yields and spectra per fission from the primary fissile isotopes, as well as providing cross-checks for antineutrino flux predictions made from nuclear databases and beta-spectra conversions.

\begin{table}[t]
    \centering
    \begin{tabular}{c|cccc|c}
    \toprule\toprule
    Experiment  &   $f_{235}$ &   $f_{238}$ &   $f_{239}$ &   $f_{241}$ &   Measurements\\
    \midrule
    Bugey-3 &   0.614 & 0.074 & 0.274 & 0.03   & flux/spect \\
    Bugey-4 &   0.614 & 0.074 & 0.274 & 0.03   & flux \\
    Daya Bay &  0.630-0.511 & 0.075-0.077 & 0.253-0.345 & 0.042-0.068  &   flux/evol/spect  \\ 
    RENO    &  0.62-0.527 & 0.072-0.074 & 0.262-0.333 & 0.046-0.066 &   flux/evol/spect  \\
    Double CHOOZ    &  0.520 & 0.087 & 0.333 & 0.060 &   flux/spect   \\
    \midrule
    ILL    & 1  & 0  & 0  & 0  &   flux    \\
    Savannah River & 1  & 0  & 0  & 0  &   flux  \\
    STEREO    & 1  & 0  & 0  & 0  &   flux/spect    \\
    PROSPECT    & 1  & 0  & 0  & 0  &   spect    \\
    \bottomrule
    \end{tabular}
    \caption{Examples of IBD experiments' measurement of reactor neutrino flux, spectrum, and evolution, with different reactor compositions.}
    \label{tab:pastibdexps}
\end{table}

Experiments listed in Table~\ref{tab:pastibdexps} measured the IBD detection rate from various reactors with organic scintillator targets.  
Using precise knowledge of the rate of reactor fission in the core and the IBD detection efficiency~(see Refs.~\cite{bib:cpc_reactor,bib:prd_rate} for details), IBD rate measurements can be converted to a measure of IBD yield, or antineutrino flux times the well-known IBD cross-section~\cite{Vogel:1999zy}.  
Time-averaged IBD yield measurements made by most experiments provided a first global picture of a family of uncorrelated or modestly correlated  data points from different baselines and fissile isotope compositions~\cite{bib:mention2011,bib:chao}.  
Among the example experiments in Table~\ref{tab:pastibdexps}, Bugey-4, Daya Bay, RENO and Double CHOOZ measured the IBD rate from corresponding reactors with experimental uncertainties of $1.4\%$, $1.5\%$, $2.1\%$, and $1.0\%$, respectively~\cite{bib:B4,bib:prd_rate,reno_evol,dc_nature}.
The examples on HEU produced IBD rate includes the ILL, Savannah River, and STEREO measurements with uncertainty of $9.1\%$, $2.9\%$, and $2.5\%$, respectively~\cite{bib:ILL_nu,bib:srp,stereo_rate}.  
From this dataset, IBD yields of \uFive~could be tightly constrained using HEU measurements, while constraints on the yields of the remaining isotopes remained quite loose~\cite{Giunti}.  

\begin{figure}[!h]
  \centering
  \includegraphics[width=0.5\textwidth]{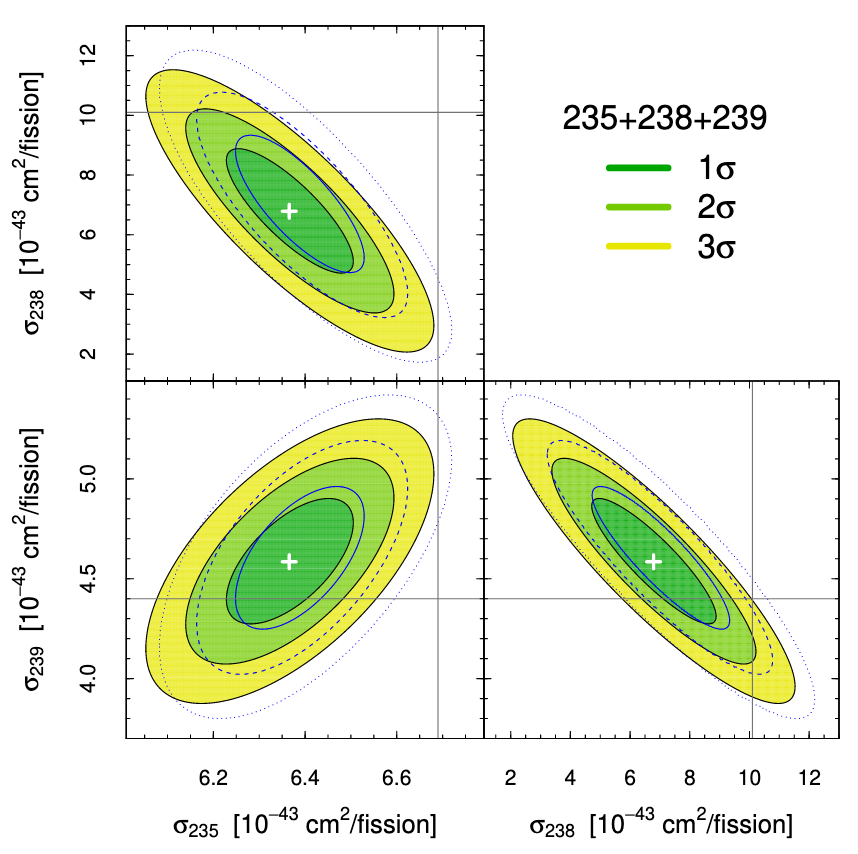}
  \caption{Allowed regions for isotopic IBD yields of \uFive, \pNine, and \uEight~provided by a fit of time-integrated and `flux evolution' IBD yield datasets.  For this fit, sterile neutrino oscillations are assumed to be negligible.  From Ref~\cite{giunti_diagnose}.  }
  \label{fig:globalibdX}
\end{figure}

Beyond time-integrated yields, the Daya Bay and RENO experiments more recently reported IBD yields measured at various points in their reactors' fuel cycles with the same reactor-detector configuration~\cite{bib:prl_evol,reno_evol}, yielding a set of highly-correlated data points capable of substantially improving direct knowledge of \pNine~and \uEight~yields~\cite{Giunti2,surukuchi_flux,giunti_diagnose}.  
Best-fit isotopic IBD yields provided by time-integrated and so-called `flux evolution' datasets are shown in Figure~\ref{fig:globalibdX}.  
%The fission fraction difference between experiments and their evolution with burnt fuel offers a path to evaluate neutrino flux from specific fissile isotopes, mostly among $^{235}$U and $^{235}$Pu, the two most abundant fissile isotopes in LEU reactors. 
%Analysis of neutrino production or IBD rate from different fissile isotopes were done in Ref~\cite{PhysRevD.99.073005}. %and others.

As overviewed in Table~\ref{tab:pastibdexps}, many of these reactor experiments have also reported the differential spectrum of IBD positron energies detected per fission, while others have further unfolded this IBD positron spectrum into an interacting antineutrino energy spectrum per fission.  
Meaningful measurements of this type require detectors with positron energy resolutions roughly of order 20\% or better.  
While first high-statistics absolute spectrum measurements at LEU reactors first became available in the mid-1990s~\cite{bib:Bugey3}, available precision greatly improved with the $\theta_{13}$ experiments of the 2010's~\cite{bib:reno_shape,dc_bump,bib:prl_reactor}.  
Precision HEU spectra only become available very recently with the PROSPECT and STEREO experiments~\cite{prospect_spec,stereo_shape}.  
The interacting neutrino spectrum per fission for \uFive~and~\pNine~was reported by Daya Bay measuring spectra at different points in its reactors' fuel cycles~\cite{bib:prl_235239,Adey:2021rty}.  
Measured \uFive~isotopic spectra have been demonstrated to be generally consistent between Daya Bay, PROSPECT, and STEREO~\cite{bib:prosDBjoint,bib:prosSTEREOjoint}.  

\subsection{Modeling Reactor Antineutrino Emissions}

Two complementary methods are available for modelling the per-experiment or isotopic IBD yields and spectra per fission reported in the previous section~\cite{vogel_review}.  
The first is the `summation' or `\textit{ab-initio}' method in which the flux and spectra are directly calculated from tabulated fission yields and branching ratios. 
This method uses nuclear databases, such as JEFF~\cite{bib:JEFF33}, to account the fission yields, as well as data on beta-unstable isotopes from ENSDF databases~\cite{bib:ENSDF} to sum the theoretical beta spectra of hundreds of fission products and thousands of beta branches.  
Uncertainties in the summation method are contributed by missing information of beta-unstable isotopes and uncertainties of beta decay branching and fission product yields.  
Until very recently, tabulations also did not account for correlations in fission yield and decay uncertainties between isotopes and branches, meaning that uncertainty envelopes, even when provided, are ill-defined. 
%The lack of tabulation of correlations in fission yield measurement uncertainties, as well as the missing information of specific decaying isotopes leading to large uncertainties.
Recently, cataloguing of fission yield correlations~\cite{Matthews2021,FIORITO2014331} and addition of improved decay data using total absorption spectroscopy (TAGS) techniques~\cite{GREENWOOD199795,PhysRevLett.105.202501,tas_lots,tas_few,tas_br,tas_nb,tas_nb2,tas_rb,tas_rbbr,tas_rbi} has provided the promise of reducing and better understanding summation uncertainties.  
%Several approaches~\cite{} of searching  also offers paths to better understand the summed uncertainty of this method. 

The second method, generally considered to be more precise, performs the conversion of measured aggregate post-fission beta decay spectra into antineutrino spectra through the fitting of a limited number of individual beta branches~\cite{bib:huber,bib:mention2011}.  
The universally used aggregate beta decay datasets underlying this method were measured by neutron-induced fission of $^{235}$U~\cite{SCHRECKENBACH1985325}, $^{239}$Pu~\cite{VONFEILITZSCH1982162}, and $^{241}$Pu~\cite{HAHN1989365} at the ILL reactor. 
Beta-branches are fitted to the cumulative beta spectra such that the sum of branches is the best-fit to measured beta spectrum.
This data-driven approach has the advantage of being immune to uncertainties from unknown or unmeasured beta decay spectra.
However, the fitted branches do not fully represent the $\sim$1000 fission-produced beta branches actually present in the spectrum.  
Theoretical corrections, including forbidden transitions~\cite{hayes_first,Hayen:2019eop} and weak magnetism corrections~\cite{bib:HayesMag}, which add additional uncertainties. 
Flux prediction of neutrinos from~$^{238}$U, and other non-fissile isotopes in reactor facilities, still rely on other experimental data or nuclear database summation.  

\begin{figure}[!h]
    \centering
    \includegraphics[width=0.65\textwidth]{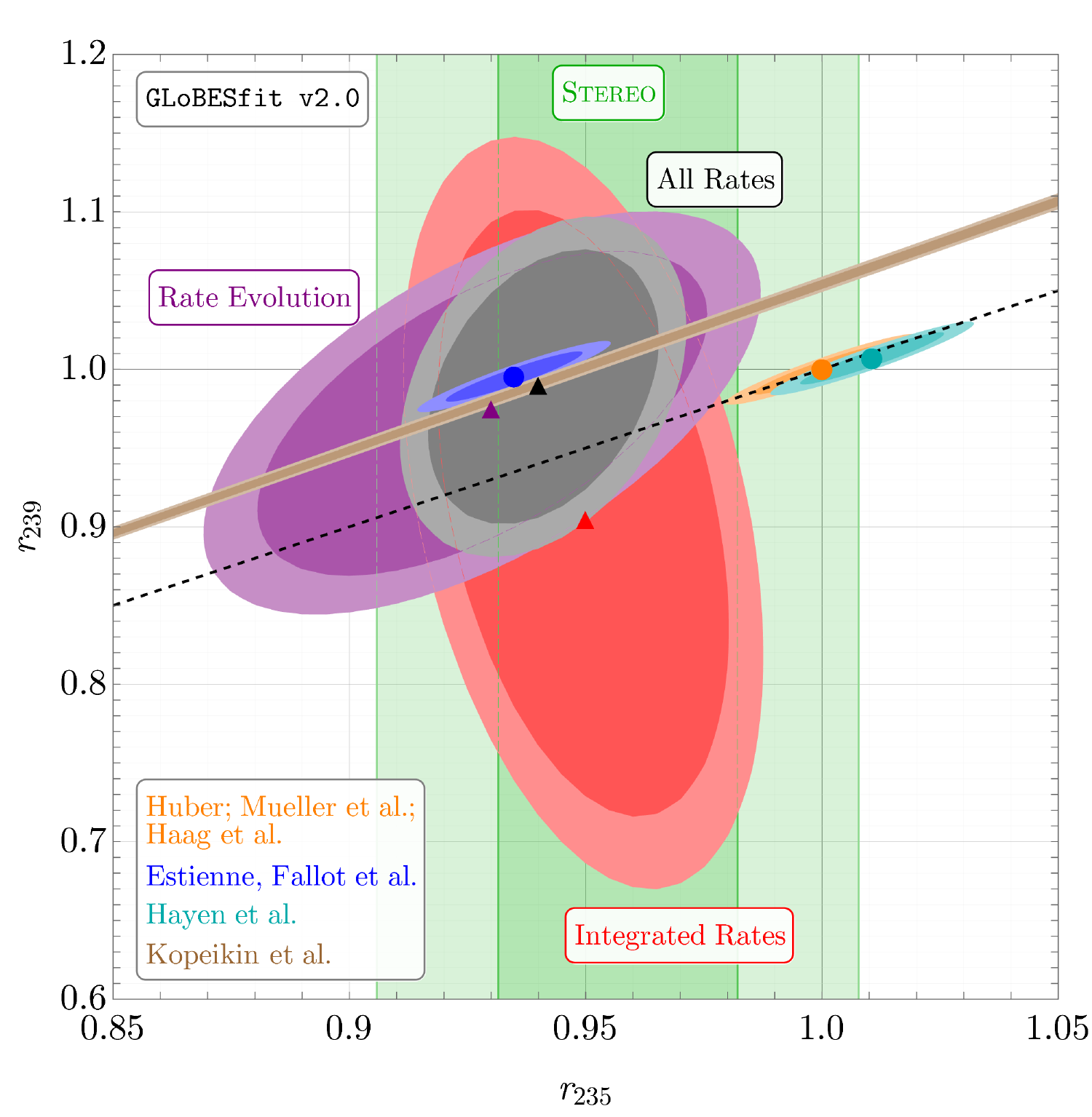}
    \caption{The 95\% C.L. (dark) and 99\% C.L. (light) contours in $r_{235}$--$r_{239}$ plane for integrated rate (red), fuel evolution (purple) and all reactor experiments (black), where $r_{X}$ is the ratio of the flux predicted/measured for isotope $X$ over its HM prediction. The result from STEREO \cite{stereo_rate} is shown in green; the bands represent the $1\sigma$ (dark) and $2\sigma$ (light) regions for one degree of freedom. The orange, blue and cyan ellipses represent the expectations from the HM, EF and HKSS flux models, respectively; $1\sigma$ ($2\sigma$) is shown in dark (light) shades. The brown bands represent the $1\sigma$ (dark) and $2\sigma$ (light) determination of the $^{239}$Pu/$^{235}$U ratio from the Kurchatov Institute \cite{Kopeikin:2021rnb, kopeikin2021}. The black, dashed line represents the line along which $r_{235}=r_{239}$. The triangles represent the best-fit values for the three fits, and the circles show the central values for the flux models. Figure and caption adapted from Ref.~\cite{huber_berryman}.}
    \label{fig:r235r239}
\end{figure}

These two methods have complementary, largely uncorrelated uncertainties, and efforts have been taken in recent years to compare their outcomes.  
While the conversion and summation spectral predictions had been initially thought to be in conflict~\cite{bib:dwyer,hayes_shoulder}, more recent studies using up-to-date database and decay information have found striking spectral shape agreement between prediction methods~\cite{bib:fallot2}.  
On the other hand, all recent studies have found discrepancies between the methods' predicted energy-integrated fluxes, both in overall magnitude and in the relative offset between  \uFive~and~\pNine~yields~\cite{hayes_evol,bib:fallot2,huber_berryman}.  
Flux model offsets are illustrated in Figure~\ref{fig:r235r239} as the difference between blue and orange/cyan circle data points.  

%A major step forward has been made by the use of total-absorption gamma spectroscopy (TAGS) to directly measure the branching ratios of many key isotopes needed for summation models. 
%In comparison to previous measurements which relied on HPGe detectors, TAGS-based measurements are less susceptible to the Pandemonium effect and produce more reliable ground-state and excited-state feeding information.
%Correlations of fission product yields are also studied with data driven and MC methods to reduce the uncertainty from nuclear databases.

\subsection{Data-Model Discrepancies}
%\begin{itemize}
%    \item 5MeV Bump (not improved by any recent data/model)
%    \item Yield Time-evolution: u5 yield looks to be mis-modelled, p9 %looks fine
%    \item SBL HEU flux: u5 yield looks to be mis-modelled
%\end{itemize}

With improvements in reliability of the models and precision of IBD yield and spectrum measurement in the last decade, a variety of data-model discrepancies have emerged.  
First, the global average of measured reactor neutrino fluxes were found to be offset with respect to the more-precise conversion prediction~\cite{bib:mention2011} -- the `reactor antineutrino anomaly' described in some detail in Section~\ref{sec:sbl}.  
This discrepancy is visible as the diagonal offset between the red and orange ellipses in Figure~\ref{fig:r235r239}.  
%compared to most of predictions, as shown in figure~\ref{fig:globalreactorflux}.
%This deficit formed the foundation of the RAA and spurred the possibility of sterile neutrinos, launching a generation of SBL reactor measurements searching for oscillation effects.
More recently, the flux evolution datasets from Daya Bay and RENO have elucidated that, absent neutrino oscillations, this flux anomaly can be more accurately interpreted as an offset in measured and predicted \uFive~IBD yields, visible as a horizontal offset between the purple and orange ellipses in Figure~\ref{fig:r235r239}. Moreover, the consonance between flux evolution datasets, summation predictions, and recent conversion predictions using new fission beta measurements~\cite{Kopeikin:2021rnb} indicates that ILL-measured beta spectrum inputs to the conversion approach may be largely to blame for IBD yield data-model discrepancies.  
Historical reactor decay heat measurements have also been recently investigated towards this end~\cite{sonzogni_heat}.  

% Both IBD rate measurements with different reactors and with fuel evolution have found, without the sterile neutrino oscillation, $^{235}$U contributes to the discrepancy with a ratio to the HM of $\sim95\%$, while 

\begin{figure}[!h]
    \centering
    \includegraphics[width=0.9\textwidth]{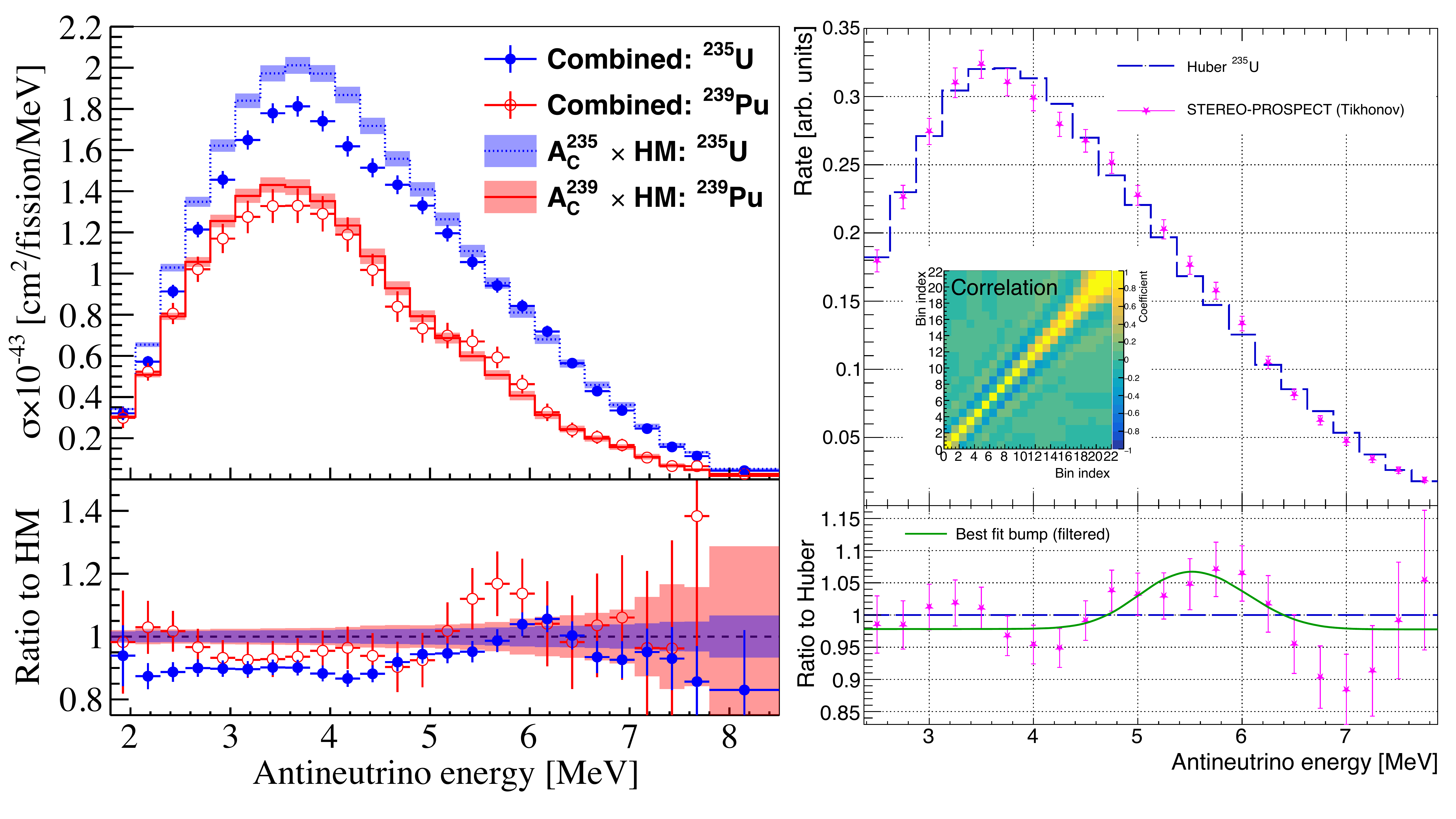}
    \caption{Joint unfolded interacting antineutrino energy spectrum of $^{235}$U and $^{239}$Pu from Daya Bay and PROSPECT (left) and of $^{235}$U from STEREO and PROSPECT (right).  Comparisons to the Huber-Mueller model are given in both cases.  From~\cite{bib:prosDBjoint} and~\cite{bib:prosSTEREOjoint}.}
    \label{fig:unfoldedspectra}
\end{figure}

Recent measurements of the antineutrino energy spectra at LEU and HEU reactors also demonstrate discrepancies between data and predictions. 
As demonstrated in Figure~\ref{fig:unfoldedspectra}, there is most notably an excess of events observed at approximately 5~MeV which is not matched by theoretical models.  
This so-called `bump' has been the focus of much interest in the neutrino as well as the nuclear physics community, since there are only a small number of high-$Q$ isotopes which contribute the majority of neutrinos in this region~\cite{sonzogni_insights}.  
While this spectral deviation was first observed at LEU reactors, short-baseline measurements by PROSPECT and STEREO have observed a similar-sized effect in the spectrum of $^{235}$U, indicating a common mis-modelling of the interacting antineutrino energy spectrum of multiple fissile isotopes~\cite{prospect_prd,stereo_shape,bib:prosSTEREOjoint}.  
This spectral data-model discrepancy appears to be common across all prediction types, even after the introduction of updated fission yield and nuclear structure datasets~\cite{bib:fallot2}.  
The universality of this problem indicates an issue with an input common to both prediction techniques, such as the assumed theoretical shape of the beta spectra used in both calculations~\cite{hayes_shape}.  

%Joint analysis of unfolded neutrino spectra by Daya Bay-PROSPECT~\cite{DYB_PROS}, STEREO-PROSPECT~\cite{STR_PROS} indicated the local discrepancy between 5-6~MeV antineutrino spectra is preferred to result from data-model disagreement in $^{235}$U, with local discrepancies at $4.2\sigma$ and $2.5\sigma$ respectively. 

%\begin{figure}[!h]
%  \centering
%  \includegraphics[width=0.49\textwidth]{./figures/Global_flux_different1.png}
%  \includegraphics[width=0.49\textwidth]{./figures/Global_flux_different2.png}
%  \caption{Global average of reactor neutrino flux. Figure and caption from Ref~\cite{huber_berryman}.}
%  \label{fig:jointspectra}
%\end{figure}

\subsection{Future Improvements in Understanding Isotopic Neutrino Emissions}

A range of ongoing and future experimental IBD-based efforts offer the promise of improving the precision of isotopic antineutrino flux and spectrum measurements.  
Most recently, the NEOS-II experiment was deployed in Sep, 2018, and just completed a $\sim500$-day reactor-on data taking period encompassing the entire fuel cycle of a single 2.8~GW~commercial LEU core at the YoungKwang Hanbit nuclear power plant.  
The experiment aims to measure the IBD rate and energy spectrum of this reactor core at 24~m baseline and perform an analysis of antineutrino spectrum and flux evolution.  
While its IBD statistics are unlikely to approach those provided by Daya Bay, its single-core measurement enables it to observe a broader range of reactor fuel content, potentially enabling isotopic measurements comparable to Daya Bay and RENO.  
Plausible gains in isotopic IBD yield measurement precision achievable in a single-core LEU experiment are overviewed in Ref.~\cite{surukuchi_flux}.  

\begin{figure}
    \centering
    \includegraphics[width=0.51\textwidth]{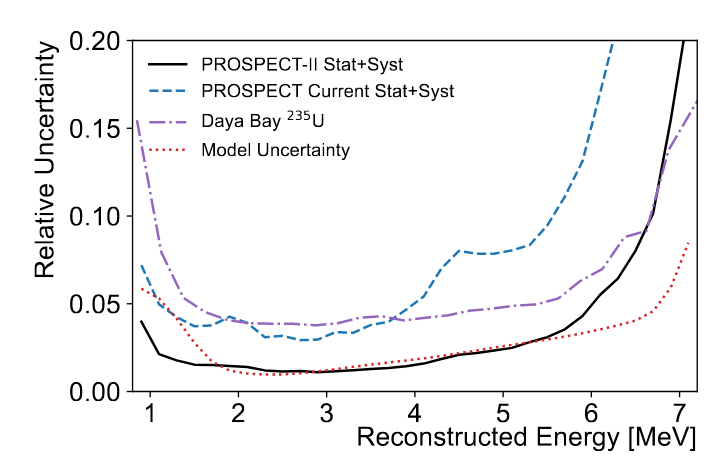}
    \includegraphics[width=0.48\textwidth]{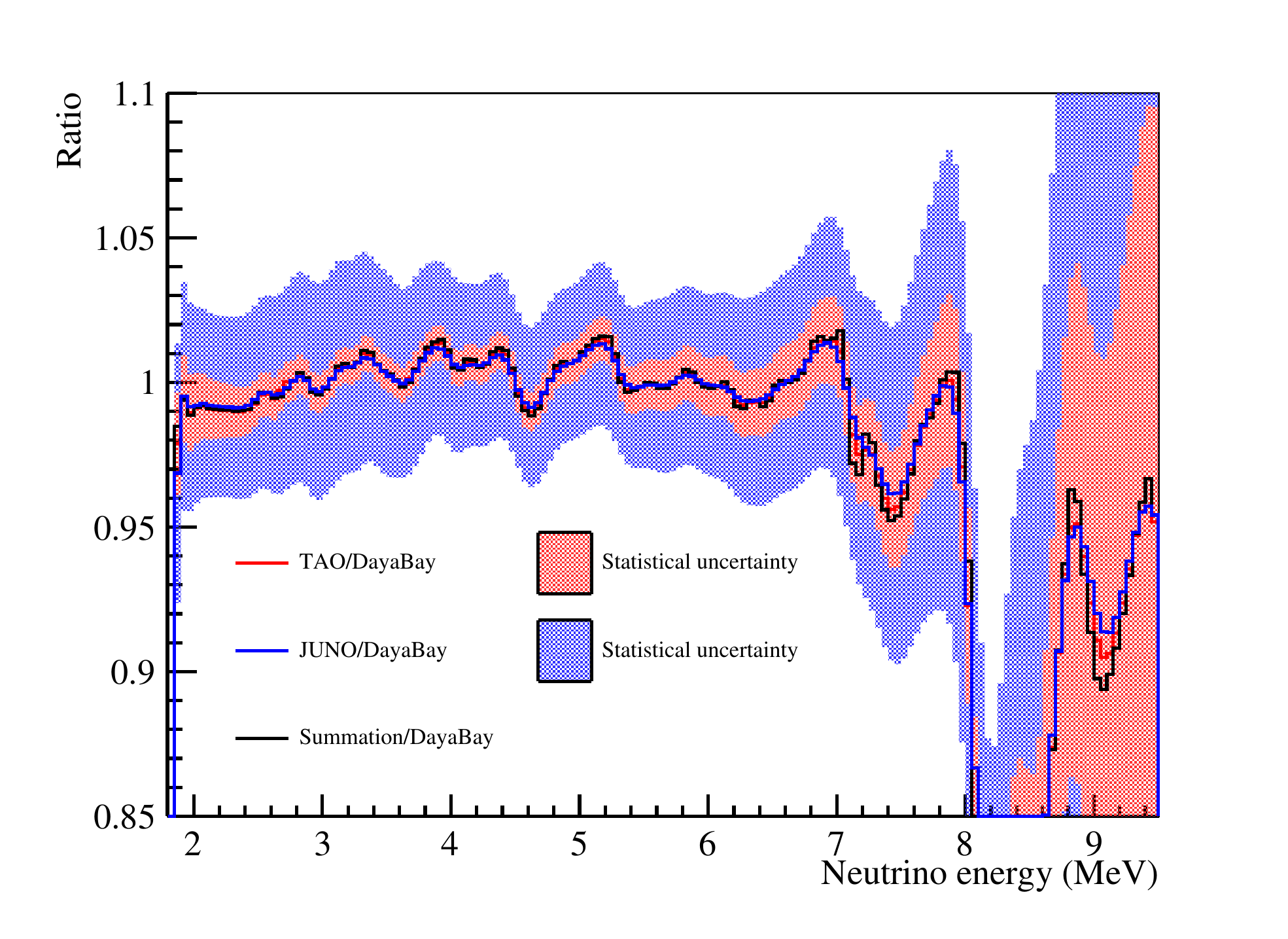}
    \caption{Left: PROSPECT-II $^{235}$U spectrum measurement uncertainties after two years of data-taking.  From~\cite{Andriamirado:2021qjc}.  Right: Comparison of projected JUNO-TAO and JUNO measurements and uncertainties with Daya Bay measurements, assuming that the true LEU reactor spectrum measured by JUNO-TAO and JUNO is given by Ref.~\cite{bib:fallot2}; JUNO-TAO's sensitivity to fine structure in the LEU reactor antineutrino spectrum is clearly illustrated.  From Ref.~\cite{juno_tao}.}
    \label{fig:tao_spectrum}
\end{figure}

Beyond this, a pair of proposed future high-precision short-baseline reactor experiments aim to build on recent successes utilizing neutrinos to enhance understanding of nuclear data.  
The PROSPECT collaboration has proposed a follow-up measurement with an improved detector called PROSPECT-II to be deployed at 7-9~m from the High Flux Isotope Reactor at Oak Ridge National Lab~\cite{PROSPECT-II}. 
The proposed run plan will increase its acquired IBD dataset by more than a factor of five over PROSPECT's first run, alongside an increased signal-to-background ratio.  
Additionally, PROSPECT plans a new measurement of the absolute flux of neutrinos from $^{235}$U reaching a $\sim$2.5\% precision primarily limited by knowledge of the HFIR reactor core's thermal power.  
These measurements will provide an important test of theoretical models in a simple system primarily composed of a single fissile isotope, \uFive.  
The expected \uFive~spectrum measurement uncertainties of PROSPECT-II uncertainty are shown in figure~\ref{fig:tao_spectrum}: its \uFive\ precision will substantially exceed Daya Bay, and will rival that of the theoretical models.  
Subsequent deployment of PROSPECT-II at an LEU reactor would allow correlated flux measurements between core types, further enhancing knowledge of individual isotopic contributions, again outlined in Ref.~\cite{surukuchi_flux}.  

In southern China, a high-resolution ($<$2\%/$\sqrt{E~\mathrm{(MeV)}}$) satellite detector for the JUNO project, called JUNO-TAO, is in the development phase and will be deployed at $\sim$25~m from one LEU reactor at the Taishan nuclear power plant~\cite{juno_tao}.  
JUNO-TAO will collect a large (millions) IBD dataset with excellent energy resolution over multiple fuel cycles, which should enable searches for sub-structure in the neutrino spectrum from individual beta-decays, as shown in Figure~\ref{fig:tao_spectrum}.  
When analyzed in combination with a high-precision HEU experiment, such as that provided by PROSPECT-II, these datasets should enable major improvements in knowledge of the antineutrino spectrum produced after \pNine~and~\uEight~fissions.  

%FIGURE: forecast isotopic spectrum measurement precision from different experiments for all isotopes (PROSPECT, JUNO-TAO, etc)

% \begin{figure}
%     \centering
%     \includegraphics{}
%     \caption{Caption}
%     \label{fig:my_label}
% \end{figure}

%TABLE: forecasted isotopic yield measurement precision for different future experiments

%\subsection{Improving Modeling and Constraints With Non-IBD Experiments} 
%\begin{itemize}
%    \item TAGS measurements
%    \item Reactor fissile isotope beta measurements (Kopeikin, ILL measurements)
%    \item Beta shape measurement?
%    \item Fission yield measurement?
%    \item cevns for low energy
%\end{itemize}

Data-model discrepancies have been authoritatively demonstrated by recent high-precision reactor antineutrino measurements.  
A resolution of this picture will likely require not just improvements in IBD datasets, but also the advancement of a variety of non-IBD nuclear physics and neutrino datasets.  
An overview of relevant recommendations for improving non-IBD datasets can be found in Refs.~\cite{bib:IAEA,bib:WoNDRAM}.  

On the conversion prediction side, recent Russian measurements of aggregate beta spectrum/yield ratios have cast doubt on the accuracy of the original ILL datasets~\cite{kopeikin2021,Kopeikin:2021rnb}.  
To authoritatively resolve this issue, a high-precision aggregate beta spectrum measurement using modern neutron facilities and measurement techniques should be performed for all major and minor fission isotopes; such a measurement should be achievable at a number of US-based neutron facilities.  
The robustness of both conversion and summation predictions could be enhanced via measurement of beta spectrum shapes for a few forbidden beta decay transitions of high-$Q$ isotopes that provide a dominant contribution to the high-energy reactor antineutrino spectrum.  
Such a measurement would verify this key theoretical input to both calculations.  
For summation predictions, continuation of total absorption gamma spectroscopy (TAGS) measurements should be carried out to further minimize the incidence of Pandemonium-affected data in the nuclear data. 
%However, the most likely problematic isotopes have already been addressed in recent measurements, meaning that future improvements may have minimal impact on summation outputs.   

Up to this point, direct antineutrino measurements have been unable to test the accuracy of summation modelling below the 1.8~MeV proton IBD interaction threshold. High-precision measurements of recoil spectra from the threshold-free CEvNS interaction at reactors offer the promise of addressing this current weakness in the global antineutrino flux picture.  
%reactors can provide non-IBD based data with high statistics for reactor modelling. 
%CEvNS events below the IBD threshold are valuable to compare with historical models and pinpoint the isotopic contribution to the reactor neutrino discrepancy.

\section{Priorities for Improving Reactor Antineutrino Detection (NF10)}
\label{sec:detectors}

\begin{comment}
\textcolor{red}{Editors: Christian Roca, Stefan Schoppmann}
%\textcolor{red}{In Josh Klein's white paper?}
\subsubsection{Initial notes:}
\begin{itemize}

\item Again overview the currently-discussed detection technolgies: primary low-threshold (CEvNS and nu-e) and IBD-based detection

\item Discuss pathways towards realizing reactor CEvNS detection

\item Discuss scope of possible improvements to SBL \nuebar detection

\item Discuss prospects

\item For each of these items should stress synergies with in tech development (background rejection; scintillator development, threshold reduction, calibration challenges), and thus with differing physics sub-fields (DM, 0nubb, applications, ...)

TABLE: Outlining overlap between RxNu technology in question and other sub-field in question.  

\end{itemize}
\end{comment}

\subsection{Key Takeaways}
\begin{itemize}
    \item A broad range of detection technologies are required to cover the full range of physics topics accessible via detection of reactor neutrinos using coherent-neutrino nucleus scattering.
    \item Small short baseline and large medium baseline reactor neutrino experiments require improvements in particle identification, light collection, and/or target composition to achieve future fundamental and applied physics goals.  
    \item Technology used in reactor neutrino physics overlaps with those used in direct dark matter searches and in a range of neutrino physics topics, such as long-baseline beam neutrino oscillations, neutrinoless double beta decay, solar neutrino physics, geoneutrino detection, and more.  
\end{itemize}

\subsection{Reactor Antineutrino Detection Technologies}

The reactor neutrino sub-field has been particularly prolific within the broader scope of neutrino physics in recent years.  
However, persistent tensions between the results of multiple short baseline experiments, together with the yet to-be realized detection of \cevns using reactor neutrinos, are strong reasons to continue improving on current techniques and developing new enabling technologies.
To ensure a broad range for known and unknown physics with reactor neutrinos, it is necessary that new experiments and development efforts cover a wide range of detection principles. In this Section, we provide a condensed description of the many current initiatives in  pursuit of low-threshold and/or MeV-scale reactor antineutrino detection.

\subsection{Very Low Energy Detection}
Coherent scattering of neutrinos off nuclei has become a growing field of interest in reactor neutrino physics and neutrino physics in general. 
For coherent scattering, the neutrino energy transfer occurs with the entire nucleus rather than a single nucleon, meaning that energy transfer has to be very low. In addition, a large fraction of transferred energy is released as heat or lost due to quenching effects.  
While coherent scattering was already discovered at high energy spallation neutron sources, fully coherent scattering would happen only at low reactor energies and thus very sensitive new detector technologies are required.  
These detectors need to offer a low threshold and low noise levels.  
In addition, those detectors require a thick shielding and overburden as they are running close to a continuous reactor, as opposed to have an accelerator-based timing reference signal to suppress background.

The use of low-threshold detectors for performing novel non-standard physics searches was described previously in Sections~\ref{sec:sbl} and~\ref{sec:bsm}.  
Since coherent scattering detectors in reactor neutrino physics are sensitive to very low energies, implemented technologies also offer promise beyond reactor neutrino detection.  
For example, such a technology would also be useful in probing the scattering of low-mass dark matter.  First results have already been delivered on these topics~\cite{NF10:CONUS:2021dwh}.  
In the following, different types of low energy detectors in the context of coherent neutrino nucleus scattering are discussed.

\textbf{P-type High Purity Germanium Detectors.} \hspace{0.5cm} 
These detectors belong to the class of ionization detectors. Opposed to conventional n-type point contact technology, p-type point contact permits high purity Ge detectors to bypass the characteristic limited charge collection efficiency and degraded energy resolution. This results in reduced capacitance while offering a large detector volume of about 1\,kg per detector unit. The small value of the capacitance results in low electronic noise and allows to lower the detection threshold to values between 200 and 300\,eVee. A mechanical cooling is commonly used and shielding is either employed via sandwiches of lead, copper, polyethylene or active vetoing through scintillation crystals. There are four major experiments at the commissioning or data-taking stage that could reveal a positive detection of reactor neutrino \cevns in the near future: CONUS~\cite{NF10:CONUS:Bonet:2020ntx}, NuGEN~\cite{NF10:NUGEN:Belov_2015}, TEXONO~\cite{NF10:TEXONO:2014eky} and the NCC-1701 vessel at Dresden-II nuclear reactor \cite{collar_bsm}.

\textbf{Skipper Si CCD.} \hspace{0.5cm} In the most general sense, the interaction principle of Charge-Coupled Device (CCD) is based on the photoelectric effect, where incident photons are absorbed in a silicon substrate generating as a consequence one or more electron-hole pairs. In conventional scientific CCDs, low-frequency readout noise results in variations in the measured charge per pixel creating a fundamental limitation on precise single-photon counting. Some initiatives like CONNIE \cite{NF10:CONNIE:2016} have been applying  CCDs to neutrino detection for many years, providing an upper limit for reactor \cevns event rate.

However, in the recent years, a new noise-reduction technique has emerged based in the use of a floating gate output stage to perform repeated charge measurements for each pixel. This multiple readout technique was implemented in the form of a Skipper-CCD achieving ultra-low readout noise that stood several orders of magnitude below values obtained with conventional CCD detection \cite{NF10:CCD:2017}. The application of this novel technology is expected to bring unprecedented detection precision down to the eV energy scale. CONNIE recently upgraded to Skipper-CCDs \cite{NF10:CONNIE:2021} showing preliminary stable and very low values of readout noise. Another new initiative called $\nu$IOLETA has taken the chance to join the efforts for building a kg-scale experiment based on Skipper-CCDs projecting a 90\% C.L observation of \cevns in 1.5 days of data taking.

Besides allowing high precision measurements of the SM at low energies, Skipper-CCD might enable a unique exploration of any physics hiding beyond that. Light-boson mediated interactions, neutrino magnetic moment or dark matter searches are strong candidates to be investigated.

\textbf{Noble Element Detectors.} \hspace{0.5cm} Noble element detectors, especially liquid xenon (LXe) and liquid argon (LAr) detectors, have been developed during the last decade mainly for direct dark matter searches. One of their main advantages is the extremely low detection threshold, a feature that makes noble element technology an excellent candidate to observe \cevns. By means of time projection chambers filled with the aforementioned noble elements, low energy interactions like these have been sought by analyzing ionization signals, but to date the sensitivity in the few-electron region has been compromised by backgrounds. The most recognizable effort trying to observe \cevns using this technology is the RED collaboration~\cite{NF10:RED:Akimov_2017}. This experiment uses a dual phase xenon detector of 100\,kg fiducial volume. Ionization electrons created in the liquid phase are extracted through electric fields and amplified in the gaseous phase. The scintillation light of about 30 photons per electron in the gaseous phase is then detected by photosensors. This experiment has achieved a low background rate down to 4 ionization electrons while operating at the surface level. R\&D efforts to reduce the single-and-few electrons background in noble liquid detectors are being pursued in the NUXE program~\cite{nuxe_magcevns_2020,Ni:2021mwa}, which plans to use a 30~kg LXe active target to detect reactor neutrino CEvNS events with signals down to single ionization electrons.  

Synergies with dark matter searches using similar technologies exist. More concretely, the observation of \cevns using noble gases will provide valuable input for a precise signal and background modeling for next-generation LAr and LXe based dark matter experiments. Finally, it could present a new way to monitor the nuclear fuel cycle using neutrinos for nuclear safeguarding applications.

\textbf{Bolometers.}  \hspace{0.5cm} Bolometers are designed as heat detectors and measure phonons created by nuclear recoils. Operating at mK temperatures, these detectors are able to achieve very low thresholds down to 20\,eV. Three collaborations, NUCLEUS~\cite{NF10:NUCLEUS:Strauss:2017cuu,NUCLEUS:2019igx}, Ricochet~\cite{NF10:Ricochet:2021rjo}, and MINER~\cite{NF10:MINER:2016igy}, are following this strategy. NUCLEUS uses $\text{CaWO}_4$ and $\text{Al}_2\text{O}_3$ crystals, while Ricochet and MINER use Ge/Zn and Ge/Si targets, respectively. To ensure a reasonable energy resolution, the detector crystals in use have to be kept small, in the order of 10\,g. An exception is MINER with a detector at the order of 1\,kg, since they detect charged particles through phonons created by the charged particles in high field regions of the detector. Their detector therefore belongs rather to the class of ionization detectors. Besides MINER, also Ricochet can exploit ionization and heat signals. This allows them, from the comparison of these signals, to perform particle identification and therefore background rejection. 

\textbf{Crystal Scintillator Detectors.} \hspace{0.5cm} An alternative form of scintillation-based detectors revolves around crystal scintillators. One of the main advantages of this technology is the high yield of photons produced by scintillator crystals while producing low amounts of background. The crystals are also relatively accessible and permit for the use of large and relatively inexpensive pieces. The NEON collaboration~\cite{NF10:NEON:2020Jaejin} uses short 15\,kg NaI crystals to improve the light collection efficiency. Crystals are read out by PMTs on both ends. They achieve a 220\,eV energy threshold. An active liquid scintillator veto is surrounding the target crystal. Data taking has started from December 2020 which includes 1 month reactor-off period.

\textbf{Color Center Passive Detectors.} \hspace{0.5cm} Crystal defects have been identified as candidate for the detection of low energy nuclear recoils. Recently, it was proposed to use materials where these defects act as color centers and to use modern microscopic techniques, specifically selective plane illumination microscopy, to image individual radiation induced color centers in bulk volumes~\cite{Cogswell:2021qlq}. This technology could provide passive detectors for reactor CEvNS, both for basic science and also nuclear security applications.

\subsection{IBD Detection Technology Improvements}
Unlike the previously discussed very low energy detection of coherent neutrino nucleus scattering, reactor neutrino detection via inverse beta decay is well established. To improve the scalability and/or background rejection performance of IBD detectors, novel detector media are currently being investigated.  These developments may improve the achievable physics precision of IBD-based detectors and increase their capability or versatility as reactor monitoring instruments.  

\textbf{$^{6}$Li-doped Organic Scintillators.} \hspace{0.5cm} 
The study of reactor antineutrinos has traditionally pivoted around organic scintillators. Among common requirements like high scintillation light yield and good optical transmission, organic scintillators need to provide excellent particle identification for fast neutrons and neutron captures in order to properly identify IBD interactions. To successfully fulfill these criteria organic scintillator compounds can be mixed with PSD-capabilities in mind and then doped with a neutron-catcher isotope like $^6$Li. \\
Liquid PSD-capable scintillators with $^6$Li-doping (LiLS) have already been produced and used in experiments like PROSPECT at the ton-scale \cite{PROSPECT-I}. Besides its PSD-capabilies, LiLS production is easily scalable which permits larger proton-rich targets with long-term stability at standard temperatures. A complementary alternative to LiLS that permits readily transportation and flexibility comes from $^6$Li-doped plastic scintillators (LiPS). While historically plastics have been found to exhibit much poorer discrimination properties, in the recent years significant progress has been made in synthesizing stable PSD-capable plastic scintillators ~\cite{NF10:Zaitseva:2102PSD}, even with dissolved $^6$Li \cite{NF10:Zaitseva:2013LiPS}. Some initiatives like the ROADSTR near-field working group ~\cite{ROADSTR:2021loi} and SANDD \cite{NF10:SANDD:2019Li} are currently developing novel prototypes for readily mobile reactor antineutrino detectors using PSD-capable LiPS.

\textbf{Water-based Liquid Scintillators.} \hspace{0.5cm} 
Monolithic optical detectors have a long history of success in neutrino physics via IBD or ES, from water Cherenkov detectors to liquid scintillator detectors. As new experiments push current limits into previously unexplored regions of phase space, a priority are advanced detection techniques for particle identification and background rejection. A promising new approach is given by exploiting Cherenkov and scintillation signals simultaneously using water-based organic liquid scintillators, i.e water is loaded with $\sim10$\,\% liquid scintillator~\cite{NF10:Yeh,NF10:Caravaca:2020lfs}. This technology is foreseen in the Eos experiment and could be deployed in planned experiments for reactor monitoring like AIT-NEO~\cite{NF10:WATCHMAN}. There are also potential synergies with future kilo-ton experiments like Theia~\cite{NF10:Theia} which will have a rich physics program including topics in high-energy, nuclear, geo, and astrophysics such as neutrino mass ordering, CP-violation in the leptonic sector, solar neutrinos, diffuse supernova neutrinos, neutrinos from supernova bursts, neutrinos from the Earth’s crust, nucleon decay, and neutrinoless double beta decay with sensitivity towards normal neutrino mass ordering.

Powerful aspects are the particle identification (PID) capabilities offered from the Cherenkov/scintillation ratio~\cite{NF10:Caravaca:2016fjg}. This PID has the potential to significantly improve signal/background discrimination of alpha/beta and beta/gamma particles and arises from two sources: the time profile of scintillation light emitted in response to a recoiling proton may differ from electron-like events due to quenching effects and the ratio of Cherenkov to scintillation light will differ between heavier and lighter particles. Additionally, recent developments have demonstrated the capability to identify neutron/gamma particles through the pulse shape discrimination of the scintillation light ~\cite{Ford:PSD_WBLS}. 

\textbf{Mixed and Slow Liquid Scintillators.} \hspace{0.5cm} 
Alternative approaches to improve discrimination power via the time profile of scintillation light exists. This can be achieved by using compound scintillators blended from two or more scintillator components. In addition, varying the concentration of fluors allows to slow down the scintillation pulse time profile. This allows in particular to distinguish between nuclear and electronic recoils. The recoil protons excite more triplet states of the solvent molecules than electrons or positrons, therefore leading to a different magnitude of quenching. These triplet states have longer decay times increasing the charge ratio in the tail of the scintillation pulse. Blended scintillators were successfully exploited for PID in the past~\cite{NF10:Buck:2018cac,NF10:Kim:2015pba}.

\textbf{Opaque Scintillator.} \hspace{0.5cm} 
LiquidO is a detection approach relying on opaque scintillators that represents a departure from conventional scintillation detectors. The main principle resides in stochastically confining light around the production point by reducing the scattering length of photons to below the cm level, while keeping the absorption length high enough to ensure a good light output~\cite{NF10:LiquidO:2019}. The localised detection of trapped photons provides imaging of topological energy depositions that translates into superior event-by-event identification and position reconstruction. In order to capture the confined light, the detector is traversed by a tight array of optical wave-shifting fibers that collect the light at the interaction point and transport it to photodetectors, typically SiPMs, located at the end of each fiber. While LiquidO can have multiple fiber orientations running simultaneously to reach 3D imaging, it is possible to use timing, if the resolution is good enough, to infer the projected position along the fiber. The LiquidO detection technique is not limited to the use of scintillation. In fact, Cherenkov light can and has been detected this way. However, the use of scintillation is key for low energy neutrino detection. In addition to its imaging capabilities, the opaque medium of LiquidO offers unique opportunities for heavy loading (in the order of 10\% or more), as the lack of a transparency requirement relaxes the constraints on the optical model.

An experimental proof-of-concept was successfully run in 2018, called Micro-LiquidO, with an active volume of 0.2~L. Its successor, called Mini-LiquidO, is currently in operation and completing data taking with a volume of 7.5~L. The first opaque scintillating medium used in both LiquidO detectors was NoWaSH~\cite{NF10:Buck:2019}, an admixture from LAB and PPO as the scintillator and paraffin wax to provide the opacity. This compound has displayed below-cm scattering lengths while keeping a high profile of photons per MeV. Above $40\degree$C it mixes homogeneously with ease, while becoming highly viscous when cooling to room temperature. Preliminary studies of NoWaSH also support the possibility of metal loading into the admixture, a feature much needed for different physics goals. Other solutions for possible opaque scintillators exist~\cite{NF10:Wagner:2018ajx} and are in the early stages of R\&D within the LiquidO scientific consortium.

In the context of reactor antineutrino IBD detection, LiquidO could have the ability to separate positrons from electrons and gammas on an event-by-event basis, enabling a major improvement of the signal-to-background ratio and reducing the reliance on overburden. LiquidO’s native muon-tracking capability with sub-cm precision is also expected to enable a tight control of cosmogenically produced backgrounds. A full 5~ton reactor neutrino program detector has been funded by the EIC-Pathfinder-2021 European program and will start construction in early 2023. LiquidO technology is also actively being considered for the detection of solar neutrinos using indium, geoneutrinos, accelerator neutrinos, and double beta decay~\cite{NF10:LiquidO:2019}. 

\textbf{Gd-doped Water} \hspace{0.5cm} 
Gadolinium-doping has long been recognized as a key advance in the context of enhancing sensitivity to neutrons and thus antineutrinos in IBD detectors. The Super-Kamiokande gadolinium upgrade~\cite{Super-Kamiokande:2021the} reflects the importance of this technological enhancement for fundamental neutrino physics at the MeV scale. Similarly, Gd-doping presents the opportunity to improve sensitivity to reactors in large-scale detectors, especially for mid-to-far-field monitoring and exclusion applications.  Detectors such as the proposed kiloton-scale AIT-NEO detector~\cite{NF10:WATCHMAN}  will permit exploration of further enhancements to the sensitivity of gadolinum-doped water detectors in both domains.  For example the use of smaller and/or faster photosensors offers the prospect of improved vertex resolution compared to SK-Gd, with beneficial effects on fiducialiization and background rejection.  The scale of the detector also permits detailed experimental validation of the performance of technologies such as wavelength shifting plates, and new methods for in-situ characterization of water attenuation in doped media. Reconstruction of supernova directionality through the electron scatter channel may be achievable by tagging IBD events using the gadolinium dopant.  AIT-NEO can also be used to study the combined benefits of the essential gadolinium dopant with  those coming from alternative media such as water-based liquid scintillator.

\subsection{Synergies}

Given the technology overlap between reactor neutrinos and other rare event detection fields like dark matter or neutrinoless double-beta decay allow for interesting synergies that could be exploited. We discuss here some of these synergies, leaving the broader picture of potential applications to the next section.

\begin{itemize}
    \item High Purity Ge detectors: low threshold detection allows for $0\nu \beta\beta$ decay, gamma and x-ray spectroscopy.
    \item Plastic Scintillators: their flexibility could be practical for reactor monitoring purposes through readily mobile neutrino detectors (see Sec. \ref{sec:applications}).
    \item Skipper CCD: nuclear spectroscopy, massive multiplexed optical/near-infrared cosmic surveys to study the dark sector, direct DM searches.
    \item Bolometers: their sensitivity to nuclear recoil make them ideal for dark matter/axion searches or probing the structure of nuclei.
    \item Noble liquids: accurate signal and background modeling for the next generation of dark matter experiments.
    \item Water-based scintillators: strong PID capabilities and broad energy range would allow multi-disciplinary research, including BSM physics like $0\nu \beta\beta$ decay or proton decay.
    \item Opaque scintillators: background suppression and flexible doping allow for multiple types of neutrino research, like $0\nu \beta\beta$ or solar neutrinos.
\end{itemize}

\section{Applications of Reactor Neutrinos (NF07)}
\label{sec:applications}

%\begin{comment}
%\textcolor{red}{Editors: Tomi Akindele, Andrew Conant}
\subsection{Key Takeaways}
\begin{itemize}
    \item Neutrino measurements for fundamental physics and nonproliferation applications require  strongly overlapping technology and workforce capabilities.
    \item There are strong synergies between the future scientific goals, nuclear data needs, and technology pathways of both fields.
    \item Stakeholders for both fundamental and applied neutrino physics programs would benefit from coordination of investments in reactor-based experiments and demonstrations.
\end{itemize}

\subsection{Antineutrino Applications Overview}
Measurement of antineutrinos can provide information about the operation of a nuclear reactor as well as addressing important science goals for the Neutrino Physics community.  
Application of reactor antineutrino detection technology, the development of which has largely been motivated by the pursuit of fundamental scientific discovery, enables remote  monitoring of nuclear reactors which has the potential to address nuclear energy and security problems.
Conversely, engagement and support from these application communities could provide additional impetus for investments in improving reactor antineutrino flux predictions and detection technology, benefiting scientific efforts at such facilities. 

Here, we describe mutual benefits that the scientific and application communities could enjoy from strong and enduring engagement.

\subsection{Potential Societal Benefits from the Application of Neutrino Detection}

The preceding sections of this white paper describe the scientific case for using reactor neutrinos to help understand properties of the Standard Model and BSM physics. In addition, neutrinos from reactors can be leveraged to gain information on the reactor itself or nuclear science at large. Currently, there are over 400 commercial and 200 research reactors operating worldwide, the former of which produce approximately 10\% of the globe's electricity. Nuclear reactors present a viable clean energy source which can help combat the effects of climate change, and more reactors are projected to be constructed every year. However, concern over the misuse of nuclear technologies and materials is one of the several impediments to the widespread adoption of this power source. Antineutrino detection can potentially support the safe and peaceful use of nuclear energy as a non-intrusive, remote measurement method to increase confidence and transparency by verifying that reactors are being used in a manner consistent with their declared purpose. Furthermore,  reactor neutrino emissions are inherently coupled to underlying nuclear data such as fission yields and the characteristics of short-lived beta-unstable isotopes, and neutrino measurements can help to improve our knowledge of these parameters~\cite{bib:WoNDRAM}.

Neutrino physicists have proposed several uses of neutrino detection to monitor reactors and spent nuclear fuel, as well as completing several demonstrations close to reactors~\cite{Bernstein:2019hix}. 
%Reactor antineutrino measurements have demonstrated the capability to monitor reactors in a variety of scenarios. 
The monitoring concepts proposed can roughly be grouped into two categories, near-field and far-field. Near-field monitoring concepts typically involve ton-scale detectors that are located within $\sim 100$~m of a reactor core. Ideally, such a system would be able to operate with limited overburden to provide more flexibility and avoid the need for modifications to a facility or the fortunate circumstance of an existing deployment location that provides substantial cosmic background attenuation. Near-field systems can potentially determine reactor status (on/off), power level, and fissile content. 
Far-field  monitoring concepts typically involve below-ground detectors of hundred ton to several hundred kiloton scale, located well beyond the facility boundary ($\sim 2-200$~km from the reactor core). Demonstration of these concepts has been encouraged by an enduring NNSA strategic goal to demonstrate the capability to remotely monitor nuclear reactors using antineutrinos. Potential benefits include a reduction in intrusiveness from the perspective of the country being monitored,  the elimination of any potential for interference with facility operations,  and the ability to exclude the possibility of reactor operations over radii as large as $\sim 200$~km. Cost and practicality must also be carefully assessed due to the rapid fall-off in antineutrino flux with increasing standoff, which of necessity implies larger detector sizes for timely signal accumulation and increasing overburden for background suppression.
%Far-field monitoring concepts typically involve below-ground detectors of  100 ton or greater scale located beyond the boundary of a facility ($\sim$ 2km or greater from the reactor core). Given the rapid fall-off in neutrino flux with distance, motivations for operating in the far-field include  elimination of any potential for interference with facility operation or the desire to exclude the possibility of reactor operations in a geographical area (radius of $\lesssim100$~km).

% Near-field to identify reactor operational status and changes in fissile inventory in the near field ($<$ 1000m baselines) and the presence of operating reactors in the far field ($\sim100$km baselines).

The Nu~Tools study~\cite{Akindele:2021sbh}, commissioned by the National Nuclear Security Administration (NNSA) Office of Defense Nuclear Nonproliferation Research \& Development (DNN R\&D), identified  current and future areas of utility for neutrinos in nuclear energy and security via end-user engagement through over 40 interviews with potential stakeholders. The study identified several promising applications in which neutrino technology can service nuclear energy and security needs. The two most promising use case applications identified by the study were advanced reactor safeguards and future nuclear deals. Some forthcoming advanced reactor design are not amenable to existing safeguards techniques and neutrino detection might be able to play a role. Applications to safeguards of spent fuel and nuclear accident response show some promise, although further study is needed. Notably, application of neutrino detection to the current fleet of nuclear reactors operated under safeguards overseen by the International Atomic Energy Agency was not found to be promising. 

\subsection{Overlaps between Applications and High Energy Physics Opportunities}
The pursuit of fundamental discovery often motivates technology development that then enables new applications; however, in the context of neutrino physics there is the opportunity for the converse to also occur. For example, detectors with  the ability to deploy above-ground and packaged for mobility have been identified as attractive by potential antineutrino monitoring end-users. As described in~\cite{Bernstein:2019hix}, applications-focused technology R\&D in this direction informed the successful design efforts of the current generation of short baseline sterile neutrino searches at reactors which must also operate with limited overburden. Looking towards future possibilities, the deployment of monitoring detectors at different reactor types could provide information to further constrain and improve flux and spectrum predictions. To give a specific example, an antineutrino-based power diagnostic for the forthcoming Versatile Test Reactor could support the materials science mission of that facility, while also measuring the antineutrino emissions from a reactor with exotic fuel types and fast neutron spectra \cite{Hill_AAP}. As discussed elsewhere in this whitepaper, precision flux and spectrum predictions could enable BSM physics searches using  reactor neutrinos.

Mobile detectors able to measure spectra with common systematic uncertainties would be especially beneficial and have evident appeal for applications. Finally, if antineutrino detection is adopted as a means to monitor reactors at 10-100km standoff, this may present an opportunity for oscillation measurements at unique baselines that could reduce uncertainties on neutrino mixing parameters. Furthermore, such detectors could contribute to the Supernova early warning system.

\subsection{Overlaps With Technology Development}
Fundamental and applied neutrino science can both benefit from advances in detection technology. 
Cooperation on common goals and techniques can enable new physics probes and expand the application space of neutrino detection. 
Methods to reduce backgrounds and improve the energy resolution and efficiency of detectors utilizing Inverse Beta Decay (IBD) would improve the sensitivity of short baseline sterile neutrino searches and application observables.
Detection of coherent elastic neutrino-nucleus scattering (CEvNS) at reactors is challenging, but if achieved with low enough threshold could provide unique measurements below the IBD threshold for applications as well as a rich physics program.  Finally, directional neutrino detection would be another advance with strong mutual benefits. This capability would mitigate declared reactor backgrounds for far-field detection of undeclared facilities, enhance searches for the diffuse supernova background and geoneutrinos, and provide the source direction for optical observations of a core collapse supernova. 
Improvements in light collection, photo-detection, and fast inexpensive readout systems would also be of benefit to both science and applications.

\subsection{Workforce Development Pipeline and Non-traditional Career Paths}
The training early career HEP physicists receive in graduate programs related to neutrino physics experiments greatly benefits applied antineutrino research for nuclear safeguards. 
Alternatively, the relatively small scale of most application efforts often provides students and postdocs the opportunity to contribute to all aspects of an experimental particle physics project.
Such opportunities ameliorate the ability of early career physicists to develop their own research programs and maintain a workforce pipeline.
Institutions that focus on nuclear energy and safeguards technology provide an additional career option that enables physicists to continue to develop HEP relevant technical skills and make important societal contributions.

\subsection{Realizing Synergies between Neutrino Physics and Neutrino Applications}
As recommended by the Nu~Tools study, stakeholders for both fundamental neutrino physics and applications would benefit from taking advantage of these overlaps and coordinating investments for detectors deployed at reactors \cite{Akindele:2021sbh}. 
Additionally, researchers in both fields would benefit from coordination within the community to identify key overlaps between goals through workshops, community engagements, and attending targeted conference series.
Such organization could reduce redundancies in parallel technology development efforts and foster stronger interactions between experts across the wide range of neutrino science and its applications. 
%\end{comment}

% \section{Conclusions}
% \label{sec:Conclusions}

\clearpage

\bibliographystyle{iopart-num-long}
\bibliography{main}

\end{document}